\documentclass[fleqn,usenatbib]{mnras}

\usepackage{newtxtext,newtxmath}
\usepackage[T1]{fontenc}
\usepackage{ae,aecompl}

\usepackage{graphicx}	% Including figure files
\usepackage{amsmath}	% Advanced maths commands
\usepackage[dvipsnames]{xcolor}
\usepackage{multirow}
\usepackage{delarray}
\usepackage{color}
\usepackage{here}
\usepackage{float}
\usepackage{tensor}
\usepackage{xspace}
\usepackage{orcidlink}
\usepackage[multiple]{footmisc}

\newcommand{\mach}{\mathcal{M}}
\newcommand{\xvect}{\mathbf{x}}

\newcommand{\be}{\begin{equation}} \newcommand{\ee}{\end{equation}}

\newcommand{\SFE}{\mathrm{SFE}}
\newcommand{\solarmass}{\mathrm{M}_{\rm \sun}}
\newcommand{\msun}{\solarmass}

\newcommand{\Msonic}{M_{\rm sonic}}

\newcommand{\MJeans}{M_{\rm Jeans}}
\newcommand{\MSR}{\Lambda_\mathrm{MSR}}
\newcommand{\MSO}{\Lambda_\mathrm{MSO}}
\newcommand{\alphath}{\alpha_{\mathrm{th}}}
\newcommand{\alphaturb}{\alpha_{\mathrm{turb}}}
\newcommand{\alphasys}{\alpha_{\mathrm{sys}}}
\newcommand{\alphag}{\alpha^\prime}

\newcommand{\alphaB}{\alpha_{\mathrm{B}}}

\newcommand{\pc}{\mathrm{pc}}

\newcommand{\vvector}{\mathbf{v}}

\newcommand{\appropto}{\mathrel{\vcenter{
  \offinterlineskip\halign{\hfil$##$\cr
    \propto\cr\noalign{\kern2pt}\sim\cr\noalign{\kern-2pt}}}}}

\newcommand{\myquote}[1]{``#1''}

\newcommand{\blue}[1]{\textbf{\textcolor{blue}{#1}}}

\usepackage[normalem]{ulem}

\newcommand{\add}[1]{{\blue{#1}}}
\defcitealias{grudic_starforge_methods}{Paper I}
\defcitealias{guszejnov_starforge_jets}{Paper II}
\defcitealias{grudic_starforge_m2e4}{Paper II}
\title[STARFORGE: Cluster assembly and mass segregation]{Cluster assembly and the origin of mass segregation in the STARFORGE simulations}

\author[]{
D\'avid Guszejnov\orcidlink{0000-0001-5541-3150}$^{1}$\thanks{guszejnov@utexas.edu},
Carleen Markey\orcidlink{0000-0003-0629-8840}$^{1,2,3}$,
Stella S. R. Offner\orcidlink{0000-0003-1252-9916}$^{1}$,
Michael Y. Grudi\'{c}\orcidlink{0000-0002-1655-5604}$^{4}$,
\newauthor
Claude-Andr{\'e} Faucher-Gigu{\`e}re\orcidlink{0000-0002-4900-6628}$^{4}$,
Anna L. Rosen\orcidlink{0000-0003-4423-0660}$^{5}$
Philip F. Hopkins\orcidlink{0000-0003-3729-1684}$^{6}$,
\\
$^{1}$Department of Astronomy, University of Texas at Austin, TX 78712, USA \\
$^{2}$Department of Physics and Astronomy, 525 Northwestern Avenue, Purdue University, West Lafayette, IN 47907, USA\\
$^{3}$Department of Physics $|$ Carnegie Mellon University, 5000 Forbes Avenue, Pittsburgh, PA 15213, USA \\
$^{4}${CIERA and Department of Physics and Astronomy, Northwestern University, 2145 Sheridan Road, Evanston, IL 60208, USA}\\
$^{5}$Center for Astrophysics $|$ Harvard \& Smithsonian, 60 Garden St, Cambridge, MA 02138, USA \\
$^{6}$TAPIR, Mailcode 350-17, California Institute of Technology, Pasadena, CA 91125, USA \\
}

\date{\today \vspace{-0.6cm}}

\pubyear{2022}

\begin{document}
\label{firstpage}
\pagerange{\pageref{firstpage}--\pageref{lastpage}}
\maketitle

% Abstract of the paper
\begin{abstract}
Stars form in dense, clustered environments, where feedback from newly formed stars eventually ejects the gas, terminating star formation and leaving behind one or more star clusters. Using the STARFORGE simulations, it is possible to simulate this process in its entirety  within a molecular cloud, while explicitly evolving the gas radiation and magnetic fields and following the formation of individual, low-mass stars. We find that individual star-formation sites merge to form ever larger structures, while still accreting gas. Thus clusters are assembled through a series of mergers. During the cluster assembly process a small fraction of stars are ejected from their clusters; we find no significant difference between the mass distribution of the ejected stellar population and that of stars inside clusters. The star-formation sites that are the building blocks of clusters start out mass segregated with one or a few massive stars at their center. As they merge the newly formed clusters maintain this feature, causing them to have mass-segregated substructures without themselves being centrally condensed. The merged clusters relax to a centrally condensed mass segregated configuration through dynamical interactions between their members, but this process does not finish before feedback expels the remaining gas from the cluster. In the simulated runs the gas-free clusters then become unbound and break up.
We find that turbulent driving and a periodic cloud geometry can significantly reduce clustering and prevent gas expulsion. Meanwhile, the initial surface density and level of turbulence have little qualitative effect on cluster evolution, despite the significantly different star formation histories. %SO Ideally include a final big picture takeaway, but not sure what to suggest. 
\end{abstract}

% Select between one and six entries from the list of approved keywords.
% Don't make up new ones.
\begin{keywords}
galaxies: star clusters: general -- stars: formation --  stars: kinematics and dynamics -- stars: luminosity function, mass function  
\end{keywords}

%%%%%%%%%%%%%%%%%%%%%%%%%%%%%%%%%%%%%%%%%%%%%%%%%%

%%%%%%%%%%%%%%%%% BODY OF PAPER %%%%%%%%%%%%%%%%%%
 \section{Introduction}\label{sec:intro}
 
Stars predominantly form in dense clusters of hundreds to a few $10^5$ stars \citep{clustering_lada, bressert_2010}, making cluster formation a key part of the star formation process. Newly formed clusters can dissolve due to gas ejection resulting from stellar feedback, internal relaxation, dynamical friction and tidal fields (\citealt{krumholz_2019_cluster_review}), making the present day observable clusters the surviving members of the original population. Observed bound clusters have historically been categorized as \emph{open clusters} and \emph{globular clusters} depending on their location and age, but emerging evidence suggests that these two classes are not different with regards to their formation and internal dynamics but instead experience a different cosmological history (see e.g., \citealt{Kruijssen_2014_GC_formation} and the review of \citealt{krumholz_2019_cluster_review}). Unbound clusters are often referred to as \emph{stellar associations} and are typically found at sites of recent star formation \citep{Gouliermis_2018_unbound_stars_hierarchical_structure}. 
%\SO{Add associations.} 

The relatively low number of observed clusters compared to the abundance of star formation sites suggests that most (non-massive) star formation sites create only short-lived clusters \citep{clustering_lada}.
%\alr{I believe this fosuces more on the local neighborhood, which does not exhibit massive star forming sites - e.g.,clusters/associations that originate from low-density GMCs, if so I suggest stating that explicitly},
Longer-lived bound clusters must require specific star formation histories and initial conditions (see \citealt{Krumholz_McKee_2020_bound_cluster_formation} for details). The exact formation mechanism of clusters within star-forming molecular clouds is not known, despite intense theoretical and observational effort. However, recent observations (e.g., \citealt{bressert_2010,Gouliermis_2018_unbound_stars_hierarchical_structure}) support the idea of hierarchical star formation, where stars form in regions of various densities, prescribed by the underlying hierarchy of ISM structure (e.g., along filaments). Simulations of small star-forming clouds have reproduced this scenario and formed bound star clusters through hierarchical assembly, where small sub-clusters merge with their neighbors, eventually forming a massive bound structure (e.g., \citealt{bonnell:2003.hierarchical, grudic_2017, vazquez_semadeni_2017_hierarchical_star_cluster_formation}).

A key step in the cluster formation process is the onset of stellar feedback that  first stops the accretion of individual stars then expels the gas from the cluster. Exactly how this \emph{gas expulsion} happens has dramatic effects on the future evolution of the cloud \citep{krause2020}. Violent gas expulsion leads to the quick dissolution of the cluster (i.e., \myquote{infant mortality}, see \citealt{Hills_1980_cluster_analytic, clustering_lada, Baumgardt_Kroupa_2007_cluster_gas_expulsion,Fall_2010_feedback_cluster_mass_function}), however highly substructured clusters may survive even instantaneous gas expulsion \citep{Farias_2018_gas_expulsion_substructure}.  Recent hydrodynamical simulations have also found indications of gravitational feedback from gas expulsion \citep{Geen_2018_cloud_scale_SF,ZamoraAviles_2019_gravitational_feedback_cluster_dispersal}, such that asymmetry in the expelled gas shell produces a net gravitational force %acceleration 
on the stars. \textit{Gaia} measurements have identified several clusters undergoing gas expulsion, which appear to be expanding \citep{Kuhn_2019_gaia_young_cluster_kinematics}. 

%\SO{I think you should focus on the key open problems on cluster? How do clusters form (summarize different models)?}
%\SO{Also please cite Krause et al. 2020 review. Some useful stuff on gas expulsion in S5, which should be helpful for later discussion and context with past work. } % https://arxiv.org/pdf/2005.00801.pdf
%\alr{Discuss star cluster disruption studies, i.e. the debate whether if star clusters are disrupted during their infancy due to stellar feedback or due to environmental effects (see works by Fall, Bastien, Chevance, Kruijssen etc.). I have some text on this you can leverage/rewrite if you'd like}.

The stellar distribution also provides insights into the initial conditions and past cluster evolution. Many observed star clusters exhibit \emph{mass segregation}, whereby massive stars are concentrated in the centers of clusters  \citep{Hillenbrand_Hartmann_1998_ONC_structure,Kirk_Myers_2011_mass_segregaton}. 
%\SO{More detail: Whether mass segregation is produced by nature, i.e., massive stars preferentially form towards cluster centers, or nurture, where dynamics causes massive stars to migrate to the center, is debated.}
 Mass-segregation may be a natural outcome of the star formation process, such that clusters are %inherently 
 born segregated (e.g., \citealt{Mckee_tan_2003_turbulent_core,Bonnell_Born_2006_compatitive_accretion}). In this scenario massive stars form at the locations with the highest density gas, such that mass segregation is \emph{primordial}. Alternatively, star clusters may not be initially non-segregated but become so due to dynamical interactions \citep{Spitzer_69_cluster_nbody_mass_segregation} that cause massive stars to sink to the bottom of the potential well, i.e., the cluster center. Numerical investigations have been limited by the dynamic range of star formation simulations as the simulation must track the formation of individual stars and model their motions over the cluster relaxation timescale.  Thus, works investigating the origin of mass segregation have been constrained to modeling small clusters \citep[e.g.,][]{Kirk_2014_small_cluster_formation} and clusters without self-consistent gas treatment \citep[e.g.,][]{parker_2014}. %SO I mean to cite this reference https://ui.adsabs.harvard.edu/abs/2014MNRAS.438..620P/abstract

In this paper we present radiation-magneto-hydrodynamic (RMHD) simulations from the STAR FORmation in Gaseous Environments (STARFORGE) project\footnote{\url{http://www.starforge.space}}. These simulations follow the evolution of turbulent and magnetized giant molecular clouds (GMCs) from the onset of star formation until it is disrupted by stellar feedback, while also following the formation of individual stars above the H burning limit (for details see \citealt{grudic_starforge_methods} and \citealt{guszejnov_starforge_jets}, henceforth referred to as \citetalias{grudic_starforge_methods} and \citetalias{guszejnov_starforge_jets}). Note that the hydrodynamic simulations previously used to study star cluster formation in the literature had smaller dynamic ranges so they were either restricted to simulating a small clump \citep[e.g.,][]{Kirk_2014_small_cluster_formation} or did not follow individual low-mass stars \citep[e.g.,][]{Geen_2018_cloud_scale_SF,ZamoraAviles_2019_gravitational_feedback_cluster_dispersal}. The STARFORGE simulations follow the the assembly of star clusters %as soon as any star formation starts and follow it until 
through gas dispersal, which is modeled self-consistently by including all major feedback processes (i.e., protostellar jets, stellar radiation and winds, supernovae). This allows us to determine whether mass-segregation is primordial and explore the role mergers play in cluster assembly. 
Note that in this work we focus on the stellar clustering in the simulations. For a detailed analysis on the cloud evolution, star formation history, and stellar mass spectrum see the companion paper \cite{grudic_starforge_m2e4} (henceforth referred to as \citetalias{grudic_starforge_m2e4})% and \cite{guszejnov_starforge_imf}. %\alr{Given the large \# of papers that will follow, it might be too much to list them alll as Paper \#. Does FIRE do this?}

We briefly summarize the STARFORGE simulations in \S\ref{sec:starforge}, and refer the reader to \citetalias{grudic_starforge_methods} for more details on the numerical capabilities of STARFORGE. Our cluster identification methods are detailed in \S\ref{sec:cluster_identification} with special attention to the time dependent nature of the cluster assignment problem. We present our results for the fiducial cloud parameter simulation in \S\ref{sec:fiducial_results}, describing the evolution of cluster properties in \S\ref{sec:cluster_property_evol}, mass-segregation in \S\ref{sec:cluster_mass-segregation}, and the mass function of stellar populations inside and outside clusters in \S\ref{sec:cluster_IMF}. In \S\ref{sec:variation_results} we investigate how variations in the initial cloud conditions, including the initial surface density, velocity dispersion, geometry and turbulent driving affect the cluster formation process. We discuss the implications of these results and the related caveats in \S\ref{sec:discussion}.  Finally, we summarize our results and conclusions in \S\ref{sec:conclusion}.

%literature
% \citet{bressert_2010} %mapping SF in 500 pc, most stars form in clusters
% \citet{Allison_2009_MST} %MST intro
% \citet{Kruijssen_2014_GC_formation} GC formation similar to other clusters

% \citet{clustering_lada} %review
% \citet{Portegies_Zwart_2010_young_massive_clusters}%review
% \citet{krumholz_2019_cluster_review} %review
% \citet{Krumholz_McKee_2020_bound_cluster_formation} %review of cluster formation theories
% \citet{grudic_2017} Mike's hierarchical assembly cluster paper
% \citet{grudic_2020_cluster_formation} Mike's cluster formation paper, concentrating on bound fraction and cluster masses etc. as a function of IC

%\citet{Moeckel_Bate_2010_posthydro_nbody_cluster} running an Nbody sim after hydro to see how clutsers evolve, looks at segregation

%\citet{Kirk_2014_small_cluster_formation} formation of small clusters in isoT simulations, looks at MSR and shows that subclusters have 1-2 most masisve stars at center

% %%%%%%%%%%%%%%%%%%%%%%%%%%%%%%%%%%%%%%%%%%%%%%%%%% 

 \section{Numerical Methods}\label{sec:methods}
 
  \subsection{The STARFORGE simulations}\label{sec:starforge}
  
  For this work we utilize simulations from the STARFORGE project, which are run with the {\small GIZMO} code\footnote{\url{http://www.tapir.caltech.edu/~phopkins/Site/GIZMO.html}}. A full description and presentation of the STARFORGE methods including a variety of tests and algorithm details are given in \citetalias{grudic_starforge_methods}. We only briefly summarize the key points here. Readers familiar with the STARFORGE simulation methods should skip ahead to \S\ref{sec:cluster_identification}.
  
  \subsubsection{Physics}\label{sec:physics}
  %\mike{we end up writing ``see Paper I for details" a lot here}

  We simulate star-forming clouds with the {\small GIZMO} code \citep{hopkins2015_gizmo}, using the Lagrangian meshless finite-mass (MFM) method for magnetohydrodynamics \citep{hopkins_gizmo_mhd}, assuming ideal MHD (with the constrained gradient scheme of \citealt{Hopkins_2016_divb_cleaning} to ensure that $\nabla\cdot \mathbf{B}= 0$ to high numerical precision.
  
  Gravity is solved with an improved version of the Barnes-Hut tree method from \citet{Springel_2005_gadget} with high-order integration of sink particle trajectories to accurately follow multiple sink systems. Force softening is fully adaptive for gas cells \citep{price_monaghan_softening, hopkins2015_gizmo}. Accreting sink particles (stars) have a fixed $18\,\rm AU$ kernel radius. We adopt the sink formation and accretion algorithm from \citet{Bate_1995_accretion}, while accurately accounting for thermal, magnetic, kinetic and gravitational energies and angular momentum. As such we are able to follow the formation and evolution of binaries and multiples with separations larger than the softening length. 
  
  Sink particles represent individual stars. Once they form they follow the protostellar evolution model from \citet{Offner_2009_radiative_sim}, which is also used in the {\small ORION} code.
  %\alr{I would add that this model is based on the protostellar models of Hosokawa+2009}.
  In this model the protostar is treated as a collapsing polytrope: the collapse is divided into distinct phases during which the qualitative behavior changes. These phases are \myquote{pre-collapse}, \myquote{no burning}, \myquote{core deuterium burning at fixed temperature}, \myquote{core deuterium burning at variable temperature}, \myquote{shell deuterium burning} and \myquote{zero age main sequence}. %\alr{This just may be my preference, but I would suggest to include zero age main sequence, since the stars are only given the radii and L from the Tout+1996 formula and does not take MS evolution into account}
  This module dynamically evolves stellar properties (e.g., radius, accretion and internal luminosities) throughout the simulation. For details see Appendix B of \citet{Offner_2009_radiative_sim} and \citetalias{grudic_starforge_methods}. %\alr{Question: do you have any massive stars that go WR? If so, I'd suggest including that here because it sounds like you only take into account the protostellar and MS properties. Which is fine for the low-mass stars, but not the most massive ones and it's those stars that generate the most feedback, esp. during the WR phase and SNe.}\DG{They are mentioned later in the wind prescription}

  \myquote{Non-isothermal} or \myquote{cooling} STARFORGE runs utilize the radiative cooling and thermo-chemistry module presented in \citet{hopkins2017_fire2} that contains detailed metallicity-dependent cooling and heating physics from $T=10-10^{10}\,$K, including recombination, thermal bremsstrahlung, metal lines (following \citealt{Wiersma2009_cooling}), molecular lines, fine structure (following \citealt{CLOUDY}) and dust collisional processes. The cooling module self-consistently solves for the internal energy and ionization state of the gas (see Appendix B of \citealt{hopkins2017_fire2}). The gas adiabatic index is calculated from a fit to density based on the results of \citet{Vaidya_2015_EOS}. The runs in this paper explicitly treat radiation (RHD runs), unlike \citetalias{guszejnov_starforge_jets}. This means co-evolving the gas, dust, and radiation temperature self-consistently as in \citet{Hopkins_FIRE_radiation_2018}, including the stellar luminosity in various bands accounting for photon transport, absorption and emission using dust opacity. We use a first-moment or M1 \citep{levermore_1984_M1} RHD solver with a reduced-speed-of-light (RSOL) of $30\,\mathrm{km/s}$ and transport photons in 5 distinct bands (IR, optical/NIR, NUV, FUV, and ionizing). Our treatment automatically handles the trapping of cooling radiation in the optically-thick limit. In addition to local sources (i.e., stars) an external heating source is added representing the interstellar radiation field (ISRF) and a temperature floor of $T_\mathrm{floor}=2.7\,\mathrm{K}$ (corresponding to the cosmic microwave background temperature) is enforced.

% Note that we compare simulations with two different treatment of radiation transport
%   \begin{itemize}
%       \item Our \myquote{ApproxRad} simulations (same runs as in \citetalias{guszejnov_starforge_jets}) do not explicitly evolve radiation-hydrodynamics (RHD). Instead they assume a constant dust temperature of $T_\mathrm{dust}=10\,\mathrm{K}$ along with a temperature floor of $T_\mathrm{floor}=10\,\mathrm{K}$. They attempt to approximately capture the transition between optically thick and optically thin cooling regimes following \citet{rafikov2007} and modeling each gas cell as a plane-parallel atmosphere with with optical depth to escape integrated using the {\small TreeCol} algorithm \citep{treecol}.
%       \item The \myquote{RHD} simulations co-evolve the gas, dust, and radiation temperature self-consistently as in \citet{Hopkins_FIRE_radiation_2018}, including the stellar luminosity in various bands accounting for photon transport, absorption and emission using dust opacity. We use a first-moment or M1 \citep{levermore_1984_M1} RHD solver with a reduced-speed-of-light (RSOL) of $30\,\mathrm{km/s}$ transport photons in 5 distinct bands (IR, optical, NUV, FUV, ionizing, see \citetalias{grudic_starforge_methods} for details). Our treatment automatically handles the trapping of cooling radiation in the optically-thick limit. In addition to local sources (i.e. stars) an external heating source is added representing the interstellar radiation field (ISRF) and a temperature floor of $T_\mathrm{floor}=2.7\,\mathrm{K}$ is enforced.  
%   \end{itemize} 
  
  As shown in \citetalias{guszejnov_starforge_jets}, protostellar jets represent a crucial feedback mechanism as they dramatically reduce stellar masses that is achieved not just by launching some of the accreted material, but also by perturbing the accretion flow around the star. We model their effects by having sink particles launch a fixed fraction of the accreted material along their rotational axis with the Keplerian velocity at the protostellar radius. See \citetalias{grudic_starforge_methods} for details on the numerical implementation.
  
  In addition to their radiative feedback, massive main-sequence stars inject a significant amount of mass, energy, and momentum into their surroundings through stellar winds. We calculate the mass-loss rates based on a prescription given in \citet{grudic_starforge_methods}, motivated by \citet{smith_2014_winds}, and wind velocities per \citet{lamers_1995_wind_vesc}. Winds are implemented either through local mass, momentum and energy injection or direct gas cell spawning, depending on whether the free-expansion radius can be resolved. To account for Wolf-Rayet (WR) stars that dominate the wind energy and momentum budget we use a simple prescription where the mass loss rate of $M>20\,\msun$ stars is increased at the end of their lifetime using the WR lifetime prescription of \citet{Meynet_Maeder_2005_WR_lifetime}.
  
  Finally, massive stars end their life as a supernova (SN). In the simulation all $>8\,\msun$ stars are eligible to become a supernova at the end of their lifetime, for which the minimum is set as 3 Myr. SNe lead to an isotropic ejection of all mass with a total energy of $E_\mathrm{SN}=10^{44}\,\mathrm{J}=10^{51}\,\mathrm{erg}$, which is implemented through direct gas cell spawning. %\alr{What energy do you inject for each SN? I'm guessing $10^{51}$ ergs, if so you should state that here. See my comment above, you should mention the WR stage since that's likely when you form the wind bubbles. Also, do you include winds? I thought you did but you don't mention it in this section, if not then please ignore my comments related to that.}
  
  The simulations in this paper include all of the physical processes detailed above. %\alr{This is just a pet peeve of mine, so please ignore if you don't agree. I suggest adding this sentence to the previous paragraph, extremely short paragraphs just don't look right to me. (I guess I was taught in grade school that a paragraph should have at least 3 sentences.)}

\subsubsection{Initial Conditions \& Parameters of Clouds}\label{sec:initial_conditions}
    
  We generate our initial conditions (ICs) using {\small MakeCloud} \citep{Grudic_MakeCloud_2021}, identically to \citetalias{guszejnov_starforge_jets}. Unless otherwise specified our runs utilize \emph{\myquote{Sphere} ICs}, meaning that we initialize a spherical cloud (radius $R_\mathrm{cloud}$ and mass $M_\mathrm{0}$) with uniform density, surrounded by diffuse gas with a density contrast of 1000. The cloud is placed at the center of a periodic $10 R_\mathrm{cloud}$ box. The initial velocity field is a Gaussian random field with power spectrum $E_k\propto k^{-2}$ \citep{ostriker_2001_mhd} compromised of a natural mixture of compressive and solenoidal modes, scaled to the value prescribed by the $\alpha_{\rm turb}\equiv 5 \sigma^2 R_\mathrm{cloud}/(3 G M_0)$ turbulent virial parameter where $\sigma$ is the 3D gas velocity dispersion. The initial clouds have a uniform $B_z$ magnetic field whose strength is set by the $\mu$ normalized mass-to-flux ratio \citep{Mouschovias_Spitzer_1976_magnetic_collapse}. There is no external driving in these simulations. Note that the initial temperature is effectively set by the gas-dust mixture quickly reaching equilibrium with the interstellar radiation field (ISRF), for which we assume solar neighborhood conditions \citep{draine.ism}. %Note that in simulations without explicitly evolved radiation (\myquote{ApproxRad}) the gas is initialized at $T=10\,\kelvin$, while in \myquote{RHD} runs the initial temperature is set by the gas-dust mixture quickly reaching equilibrium with the ISRF, for which we assume solar neighborhood conditions.

  We also run simulations using \emph{\myquote{Box} ICs}, similar to the driven boxes used in e.g., \citet{li:2004.mhd.turb.sf,Federrath_2014_jets,Cunningham_2018_feedback}. These are initialized as a constant density, zero velocity periodic cubic box with the same temperature prescription as \myquote{Sphere} ICs. This periodic box is then \myquote{stirred} using the driving algorithm from by \citealt{federrath_sim_compare_2010,bauerspringel2012}. This involves a spectrum of $E_k\propto k^{-2}$ of driving modes in Fourier space at wavelengths 1/2 - 1 the box size, with an appropriate decay time for driving mode correlations ($t_{\mathrm{decay}}\sim t_{\mathrm{cross}}$). This stirring is initially performed without gravity for five global freefall times $\left(t_{\mathrm{ff}}\equiv\sqrt{\frac{3 \pi}{32 G \rho_0}}\right)$, to achieve saturated MHD turbulence. The normalization of the driving spectrum is set so that in equilibrium the gas in the box has a turbulent velocity dispersion that gives the desired $\mach$ and $\alpha_{\rm turb}$, same as in the Sphere runs. We use purely solenoidal driving, which remains active throughout the simulation after gravity is switched on. %\alr{Why purely solenoidal? A antural mixture is usually taken to be 2/3 solenoidal, 1/3 compressive or vice versa (I forget the exact order).}. \DG{I don't recall the reason why we changed it so, but it made little difference before}
  We take the box side length $L_\mathrm{box}$ to give a box of equal volume to the associated {\it Sphere} cloud model. An important difference between the \textit{Sphere} and \textit{Box} runs is that in the case of driven boxes the magnetic field is enhanced by a turbulent dynamo \citep{federrath_2014_dynamo} and saturates at about $\alphaB\sim 0.1$ (i.e., 10\% relative magnetic energy to gravitational, see \citealt{Guszejnov_isoT_MHD}), %\alr{I suggest writing out what $\alpha_B$ is and same for the rest of the defining terms so that the reader doesn't have to look it up in other papers}, 
  so for Box runs the \myquote{pre-stirring} magnetic field strength (defined by $\mu$) does not directly specify the actual initial magnetic field strength when gravity is turned on (however the \myquote{pre-stirring} flux in the box will still affect the large-scale geometry of the magnetic field). 

\begin{figure*}
\begin {center}
\includegraphics[width=0.33\linewidth]{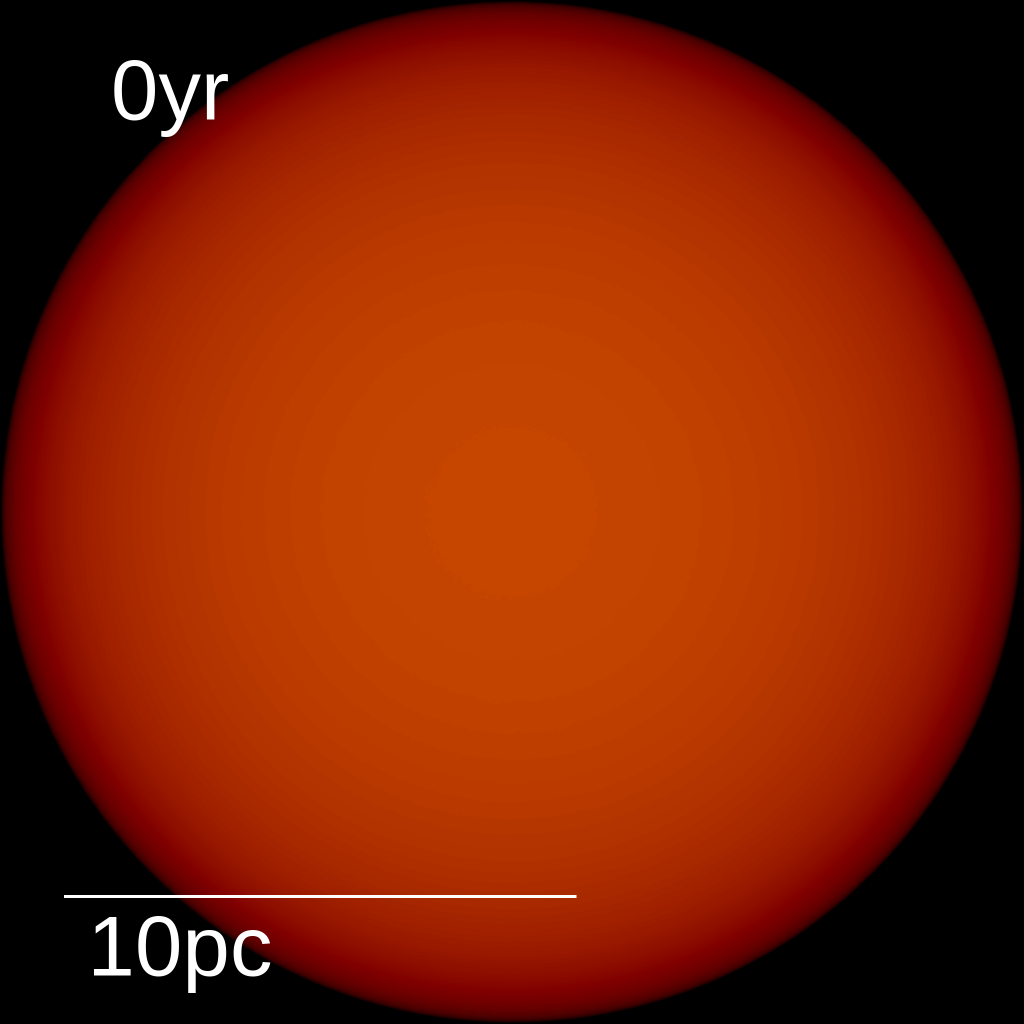}
\includegraphics[width=0.33\linewidth]{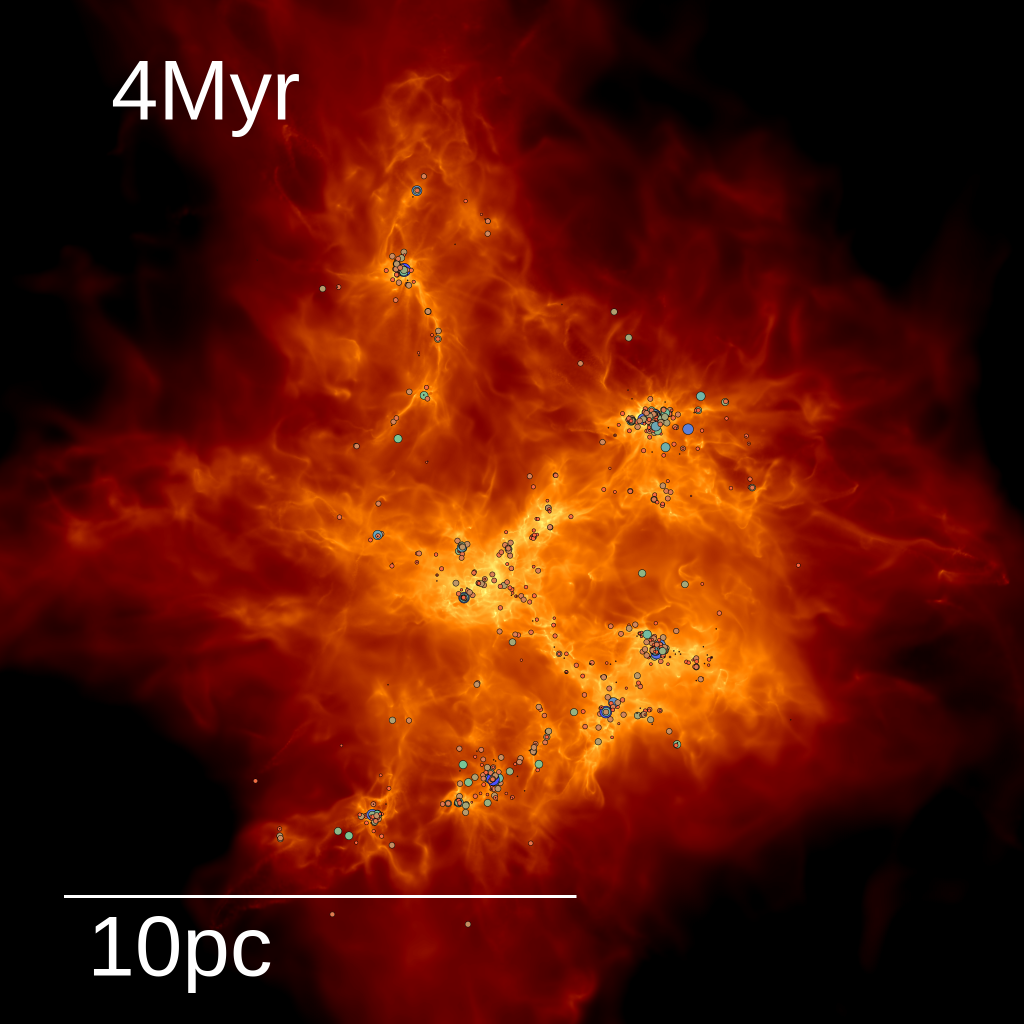}
\includegraphics[width=0.33\linewidth]{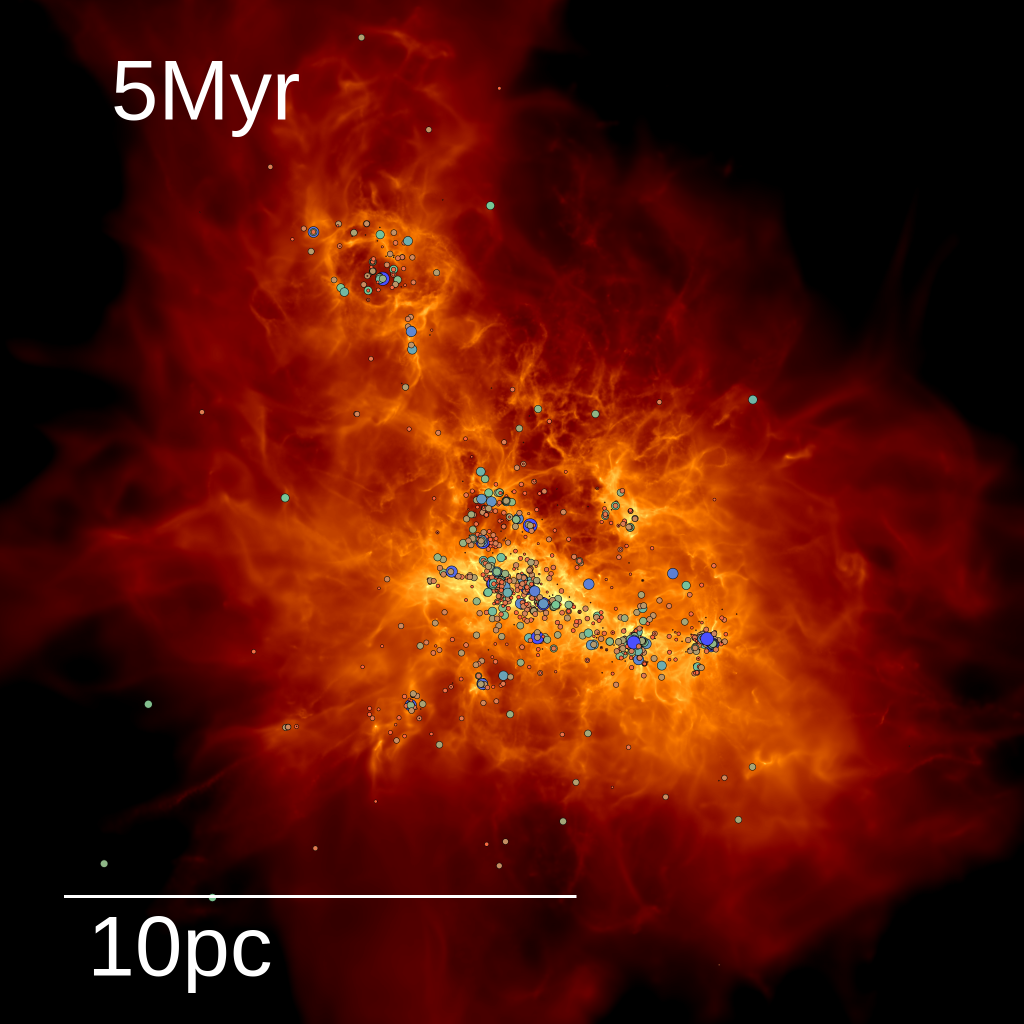}\\
\includegraphics[width=0.33\linewidth]{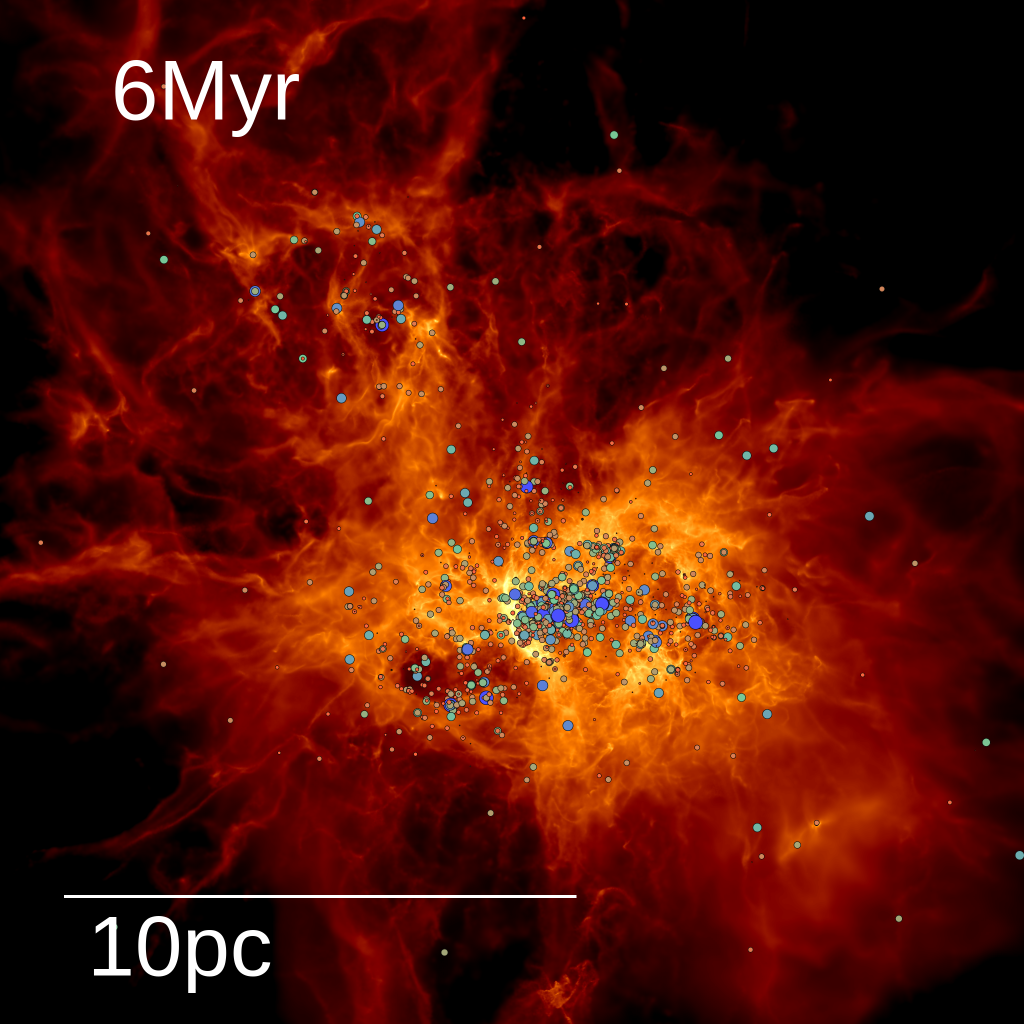}
\includegraphics[width=0.33\linewidth]{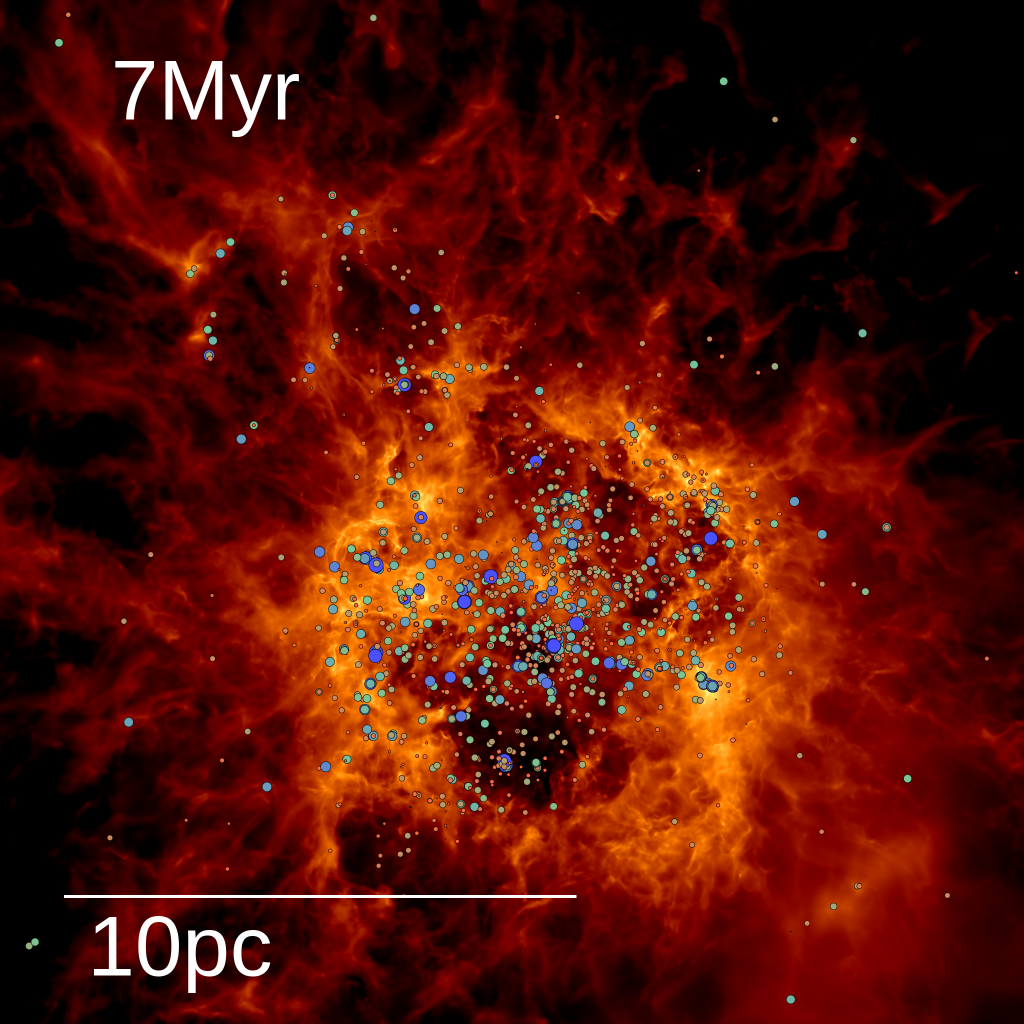}
\includegraphics[width=0.33\linewidth]{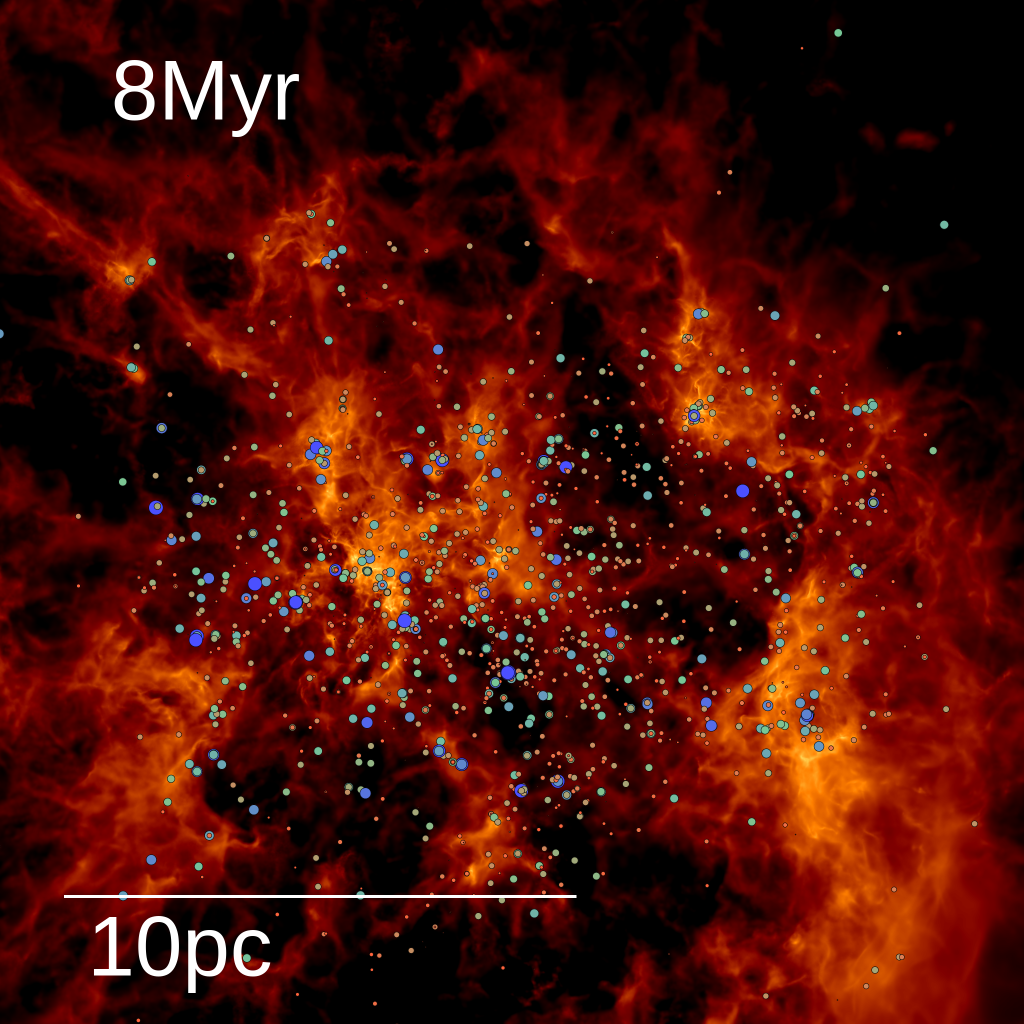}\\
\caption{Surface density maps for \textbf{M2e4\_C\_M\_J\_RT\_W} (our fiducial run), which is an $M_0=2\times 10^4\,\msun$ mass cloud resolved with $M_0/\Delta m=2\times 10^7$ initial gas cells (see Table \ref{tab:IC_phys}), at different times, from the beginning of the simulation until cloud disruption. The color scale is logarithmic and the circles represent sink particles (stars) that form in high-density regions where fragmentation can no longer be resolved. The size of the circles is increasing with mass as well as their color changing from red ($M\sim0.1\,\msun$), through green ($M\sim1\,\msun$) to blue ($M\sim10\,\msun$). This simulation resolves a dynamic range from $\sim\!\mathrm{20\,pc}$ down $\sim\!\mathrm{30\,AU}$ and is run until stellar feedback quenches star formation and disrupts the cloud.}
\label{fig:M2e4_series}
\vspace{-0.5cm}
\end {center}
\end{figure*}

  Table \ref{tab:IC_phys} shows the target parameters for the runs we present in this paper. The input parameters are the cloud mass $M_0$, size $R_0$, turbulent virial parameter $\alphaturb$ and normalized magnetic mass-to-flux ratio $\mu$ (note that initial temperature is set by the ISRF). Similar to \citetalias{guszejnov_starforge_jets} we set up our clouds to lie along a mass-size relation similar to observed GMCs in the Milky Way (e.g. \citealt{larson_law,lada:2020.sigma.gmc}, specifically assuming $\Sigma\equiv M_\mathrm{0}/ \uppi R_\mathrm{cloud}^2 = 63 \msun\,\mathrm{pc}^{-2}$), except for our one model with $10\times$ higher surface density. These clouds are marginally bound ($\alphaturb=2$, except for variation models) and start out in thermal equilibrium with the ISRF. For the initial magnetization we assume $-E_\mathrm{mag}/E_\mathrm{grav}=0.01$, which translates to $\mu=4.2$. The initial gas metallicity is assumed to be equal to the solar value. The STARFORGE simulations we use have a mass resolution of $\Delta m=10^{-3}\,\msun$, making the mass function incomplete for brown dwarfs ($M<0.08\,\msun$), which are thus omitted from our analysis (see \citetalias{grudic_starforge_methods} for convergence tests). For Sphere runs the simulations are run until stellar feedback quenches star formation and subsequently disrupts the cloud (see Figure \ref{fig:M2e4_series}). In case of the Box runs the periodic boundary conditions trap both radiation and cloud material, so the run is terminated when the box becomes saturated by stellar radiation.

\begin{table*}
    \setlength\tabcolsep{2.0pt} %compress table
	\centering
	% Phyiscs table
	\begin{tabular}{ | c | c | c | c | c | c |c |c | }
	\hline
	\textbf{Physics label}  & Thermodynamics & MHD & Protostellar Jets & Stellar Radiation & Stellar Winds \& SNe \\
	\hline
% 	\textbf{I\_M} & Ideal (M) & Isothermal (I) & Not included  \\
% 	\hline
% 	\textbf{C\_M} & Ideal (M) & ApproxRad (C) & Not included   \\
% 	\hline
% 	\textbf{C\_M\_J} & Ideal (M) & Non-isothermal, ApproxRad (C) & Included (J) & Not included & Not included  \\
% 	\textbf{C\_M\_J\_RT} & Ideal (M) & Non-isothermal, RHD (C) & Included (J) & Included (RT) & Not included   \\
	\textbf{C\_M\_J\_RT\_W}  & Non-isothermal, RHD (C) & Ideal (M) & Included (J) & Included (RT) & Included (W)   \\ 
	\hline
    \end{tabular}
    % IC table
	\begin{tabular}{|cccccc|ccccccccc|c c|}
		 \multicolumn{1}{c}{}&
		 \multicolumn{6}{c}{\bf Input Parameters} &
		 \multicolumn{8}{c}{\bf Derived Parameters}&
		 \multicolumn{2}{c}{\textbf{Resolution}} \\
		\hline
		\bf Cloud label & $M_0$ [$\msun$] & $R_{\mathrm{cloud}}$ [pc] & $L_{\mathrm{box}}$ [pc] & $\alphaturb$ & $\mu$ & $\sigma$ [km/s]  &  $\alphath$ & $\alpha$ & $\mach_{\rm A} $ & $\beta$ & $\alphaB$ & $\frac{\MJeans}{M_0}$ & $\frac{\Msonic}{M_0}$ & $\frac{M_{\Phi}}{M_0}$ &  $M_0/\Delta m$ &  $\Delta x_\mathrm{J}$ [AU] \\
		%\hline
		%\multicolumn{18}{|c|}{\bf MW cloud analogues} \\
		\hline
		%M2e4
		\bf M2e4 & $2\times 10^4$ & 10 & 16 & 2 & 4.2
		& 3.2 & 0.008 & 2.03 & 10 &  0.78 & 0.02 & 
		$3\times 10^{-3}$ & $7 \times 10^{-5}$ & 0.1 &  $2\times10^{7}$ & 30 \\
		\hline
		%M2e4_R3
		\bf M2e4\_R3 & $2\times 10^4$ & 3 &  & 2 & 4.2
		& 5.8 & 0.008 & 2.02 & 10 &  0.23 & 0.02 & 
		$5\times 10^{-4}$ & $7 \times 10^{-6}$ & 0.1 &  $2\times10^{7}$ & 30 \\
		\hline
		%M2e4_a1
		\bf M2e4\_a1 & $2\times 10^4$ & 10 &  & 1 & 4.2
		& 2.3 & 0.008 & 1.03 & 10 &  0.78 & 0.02 & 
		$3\times 10^{-3}$ & $4 \times 10^{-5}$ & 0.1 &  $2\times10^{7}$ & 30 \\
		\hline
		%M2e4_a4
		\bf M2e4\_a4 & $2\times 10^4$ & 10 &  & 4 & 4.2
		& 4.5 & 0.008 & 4.03 & 10 &  0.78 & 0.02 & 
		$3\times 10^{-3}$ & $1 \times 10^{-4}$ & 0.1 &  $2\times10^{7}$ & 30 \\
		\hline
	\end{tabular}
        \vspace{-0.1cm}
 \caption{Simulations used in this paper described with STARFORGE label conventions. \textit{Top}: Physics modules included, see \S\ref{sec:physics} and \citetalias{grudic_starforge_methods} for details on the individual physics modules. \textit{Bottom}: Initial conditions of clouds used in our runs, with $M_0$, $R_{\mathrm{cloud}}$, $\alphaturb$ and $\mu$ being the initial cloud mass, size, virial parameter, mass to magnetic flux ratio respectively (note that in runs that explicitly evolve RHD the initial gas-dust temperature is set by the ISRF). We also report the initial 3D turbulent velocity dispersion $\sigma$, thermal virial parameter $\alphath$, total virial parameter $\alpha$, Alfv\'{e}n Mach number $\mach_{\rm A}$, plasma $\beta$, magnetic virial parameter $\alphaB$, as well as the relative Jeans, sonic and magnetic mass scales (see \S2 in \citealt{Guszejnov_isoT_MHD} for definitions)  %\alr{This is my preference so feel free to ignore, but if you can easily include the def here I suggest to do so so that the reader doesn't have to flip through a bunch of papers}. 
 Note that the parameters in this table apply to both \textit{Box} and \textit{Sphere} runs as they are set up to have identical initial global parameters, with  $L_{\mathrm{box}}$ being the box size for \textit{Box} runs %\alr{I'm a little confused about L vs R for the box run, don't you set the full box to initially have a uniform density? So wouldn't R=L in the Box case? If so, I suggest removing the value for R for the box run since the sizes are different.}\DG{The volume of the box = volume of the sphere, so $L>R$} 
 and $R_{\mathrm{cloud}}$ being the cloud radius for \textit{Sphere} runs. Note that \textit{Box} runs have slightly different initial parameters (e.g., Mach number, virial parameter) due to the non-exact scaling of the driving, so the values shown here are the target values.}
 \label{tab:IC_phys}\vspace{-0.5cm}
\end{table*}

  \subsection{Cluster identification }\label{sec:cluster_identification}

  Despite almost a century of study there is no one accepted definition of what a star cluster is, as the \myquote{classical} picture of an isolated, bound, centralized group of stars is not applicable to most observations \citep{krumholz_2019_cluster_review}. Previous work in the literature defined star clusters using an absolute density threshold \citep{clustering_lada}, relative density contrast \citep{McKee_Parravano_Hollenbach_2015_stars_clusters}, boundedness \citep{Portegies_Zwart_2010_young_massive_clusters}, Bayesian decomposition into ellipsoids \citep{Kuhn_2014_cluster_bayesian_decomp} and numerous other techniques (see \citealt{Schmeja_2011_cluster_identification} for examples). Due to the lack of consensus in the literature, we choose a cluster definition that is both simple and robust for time series data (see \S\ref{sec:cleaning}). We identify star clusters using the DBSCAN (Density-based spatial clustering of applications with noise, \citealt{DBSCAN}) clustering algorithm from the {\small scikit-learn} Python library \citep{scikit_learn}, similar to \citet{Wall_2020_AMUSE_star_cluster_feedback}. DBSCAN assigns group membership using the following algorithm:
  \begin{enumerate}
      \item Any star above the H burning limit ($>0.08\,\msun$) whose $N_\mathrm{min}$ closest neighbors are within $\lambda$ distance is considered a \myquote{core particle}.
      \item All connected core particles and any particles within $\lambda$ distance are considered to be part of the same cluster. Particles not assigned to clusters are considered to be \myquote{noise.}
  \end{enumerate}
We apply DBSCAN to the 3D spatial positions of the stars, and we adopt $N_\mathrm{min}=10$ and $\lambda=1\,\pc$, %\CM{what's the justification? I went back and looked at my research note to see if I specified (I did note) I think that decision was based on clusters being ~1pc wide(?) and 10 being an experimentally chosen minimum where more led to no clusters and less led to the identification of the sub-clusters},\DG{It is somewhat arbitrary, since there is no definition for what a cluster is. Changing it to higher numbers would not qualitatively change the results (same hierarchical picture). Going lower would prolong the phase where 1 cluster=1 site, but ultimately it would give the same results.}
which effectively serves as our cluster definition. We find that changing $N_\mathrm{min}$ has no qualitative effects on our results. Reducing $\lambda=1\,\pc$ reduces the size and mass of newly formed clusters,  increases the overall number of clusters and delays mergers, however we find the evolution of cluster properties for the largest clusters to be similar.

Note that we also experimented with other, more advanced clustering methods that do not require a specified spatial scale, e.g., HDBSCAN \citep{HDBSCAN}. Algorithms like HDBSCAN identify the clustering scales from the data, thus providing results that are not biased by the somewhat arbitrary choice of clustering scale in DBSCAN. While HDBSCAN has been successfully applied to observed young clusters \citep{Kerr_2021_clustering_hdbscan}, we find that it can create confusing cluster assignments if applied to time series data. This is because the definition of what counts as a cluster in HDBSCAN is determined by the current configuration of stars, which can lead to the algorithm non-physically splitting up and merging clusters between different snapshots of a simulation. We also experiment with applying the clustering algorithm to the full 6D phase space data instead of only the 3D spatial positions, similar to the procedure applied to observational data. Doing so, however, requires a mapping from velocity to spatial scales (i.e., a phase-space metric, see \citealt{behroozi:2013.rockstar} for an example). After experimenting with several different methods (e.g. assume a linewidth-size relation, \myquote{pre-cluster} in 3D and find velocity dispersion within clusters), we ultimately find no clear advantage to using velocity data, as their main role in observations is to filter out \myquote{interloper} field stars, which are not present in our simulations.  

\subsubsection{Cluster tracking}\label{sec:cleaning}
This work aims to follow the formation and evolution of clusters, which creates a unique challenge that observations do not face, namely that cluster assignments and evolution need to be meaningful and continuous over multiple snapshots. To address this issue, we apply a series of cleaning operations after the initial cluster assignments, according to the following algorithm:
\begin{enumerate}
    \item Assign initial cluster memberships for stars in each independent snapshot using DBSCAN.
    \item Identify clusters persisting through multiple snapshots. For each cluster X in snapshot $i$ we follow the steps\add{:}
    \begin{enumerate}
        \item Find all clusters $Y$ in snapshot $i-1$ for which $X$ contains at least half of the stars of $Y$. From these the one that contains the largest fraction of stars from $X$ is considered to be the past version of $X$.
        \item If no such cluster $Y$ exists we look over older snapshots ($<i-1$, going backwards in time) and look for a cluster $Y$ where $X$ and $Y$ mutually contain at least half of the stars of the other, and consider that to be the past version of $X$.
        \item If no past version was identified we declare cluster $X$ to be a newly-formed cluster.
    \end{enumerate}
    \item Create a \myquote{cluster label history} for each star and then apply the following cleaning operations with a characteristic timescale of $t_\mathrm{clean}=100\,\mathrm{kyr}$, which effectively sets a lower limit for the cluster lifetime:
    \begin{enumerate}
        \item Remove short-lived ($<t_\mathrm{clean}$) clusters. If their stars belonged to another (not short-lived) cluster directly before joining this cluster, they keep their original assignment. This fixes a problem that arises when the clustering algorithm temporarily splits part of a cluster and then merges it back after a few snapshots. Note that short-lived cluster splits are rare when using DBSCAN and such clusters contain only a small fraction of the stars, but removing them is necessary to reduce nonphysical discontinuities in the properties of larger clusters.
        \item Inspect each star's cluster membership history and remove intermittent label assignments. If a star that initially belongs to cluster $X$ is assigned to cluster $Y$ and then back to $X$ within $t_\mathrm{clean}$ (i.e., $Y$ in a sequence of $X,X,Y,X,X$), then all $Y$ assignments are changed to $X$. This removes \myquote{flip-flopping} cluster assignments. We then remove any assignments that last a very short time ($t_\mathrm{clean}/2$). This is similar to the previous operation, but does not take into account the final label (i.e., $Y$ a sequence of $X,X,Y,Z,Z$). With these two steps we eliminate transient clusters and flip-flopping from ambiguous assignments during cluster mergers.
    \end{enumerate} 
    \item We repeat the second step and re-assign cluster IDs using the cleaned cluster label histories. This corrects errors during the original assignment, e.g., a large cluster temporarily splitting into several smaller ones.
\end{enumerate}

\subsection{Cluster properties and definitions}\label{sec:cluster_properties_defs}

To describe the star clusters in our simulations we introduce several physical quantities. We define the \emph{cluster radius} (also known as \myquote{mean-square radius} or \myquote{Spitzer radius}, see \citealt{Spitzer_1958_clusters}), as
\begin{equation}
R^2\equiv \langle ||\Delta \xvect||^2 \rangle,
\label{eq:cluster_R}
\end{equation}
where $\langle...\rangle$ denotes averaging over cluster members and $\Delta \mathbf{x}$ is the distance of a member star from the center of mass of the cluster. We also define the \emph{half-mass radius}, $R_\mathrm{50}$,  as the radius around the center of mass that encloses half the cluster mass. We define the \emph{3D cluster velocity dispersion} as 
\begin{equation}
\sigma_\mathrm{3D}^2 \equiv \langle ||\Delta \vvector||^2 \rangle,
\label{eq:cluster_sigma}
\end{equation}
where $\Delta \vvector$ is the relative velocity of a member star to the center of mass of the cluster. %\cut{ and $\langle...\rangle$ averaging over cluster members as above..} %SO Defined above, don't need to repeat 

To characterize the cluster boundedness we use the \emph{virial parameter}
\begin{equation}
\alpha \equiv \frac{2 E_{\rm kin}}{-E_{\rm grav}},
\label{eq:alpha}
\end{equation}
where $E_{\rm kin}$ and $E_{\rm grav}$ are the total kinetic and gravitational binding energy of the stars within the cluster. Note that hard binaries are common, but naively including their binding energy when trying to determine the overall boundedness of the cluster (i.e., virial parameter) yields misleading results, as these binaries essentially interact with the rest of the cluster as if they were a single point particle\footnote{Observational estimates of the virial parameter likely also suffer from biases introduced by binaries, see \cite{Gieles_2010_cluster_dispersion_binaries}.}. Thus, it is instructive to define the \emph{system virial parameter} where we merge binary and multiple systems (identified using the same algorithm as \citealt{bate_2009_rad_importance} and \citealt{guszejnov_correlation}) 
\begin{equation}
\alphasys \equiv \frac{2 E_{\rm kin, sys}}{-E_{\rm grav, sys}}.
\label{eq:alpha_sys}
\end{equation}
Here\add{,} $E_{\rm kin,sys}$ and $E_{\rm grav,sys}$ are the total kinetic and gravitational binding energy of the cluster after we replaced binary/triple/quadruple systems with their centers of mass. We can similarly define a \textit{3D system velocity dispersion} within the cluster
\begin{equation}
\sigma_\mathrm{3D, sys}^2 \equiv \langle ||\Delta \vvector_\mathrm{sys}||^2 \rangle_\mathrm{sys},
\label{eq:cluster_sigma_sys}
\end{equation}
where $\langle...\rangle_\mathrm{sys}$ is averaging over systems within the cluster. Note that close binaries are often unresolved in observed clusters, making the velocity dispersion inferred by observations closer to  $\sigma_\mathrm{3D, sys}^2$ than $\sigma_\mathrm{3D}^2$.

Note that these definitions take neither gas cells nor sink particles outside the cluster into account. Considering that clusters inevitably form in areas with dense gas, the contribution of gas to the initial boundedness is significant. As a crude estimate we calculate the amount of gas within the spatial extent of the cluster ($R$ from Eq. \ref{eq:cluster_R}) and calculate its contribution to the gravitational energy of the cluster members ($E_{\rm grav, gas}$) by assuming that this mass is distributed homogeneously within the cluster. This leads to the $\alphag$ and $\alphag_\mathrm{sys}$ virial parameters\add{:}
\begin{eqnarray}
\alphag \equiv \frac{2 E_{\rm kin}}{-E_{\rm grav}-E_{\rm grav, gas}}\label{eq:alphag},\\
\alphag_\mathrm{sys} \equiv \frac{2 E_{\rm kin, sys}}{-E_{\rm grav, sys}-E_{\rm grav, gas, sys}}.
\label{eq:alphag_sys}
\end{eqnarray}
Note that by definition $\alphag\leq \alpha$, and it only becomes equal at later times when most of the gas has been expelled from the cluster. These estimates are within a factor of few of the values returned by directly calculating the contributions from gas within the cluster. 

In our simulations we find that clusters tend to expand after gas expulsion. The clustering algorithm (\S\ref{sec:cluster_identification}) often breaks these expanding clusters into separate smaller clusters. In order to quantify the cluster expansion we introduce the \emph{mass-weighted mean radial velocity}
\be
\bar{v}_\mathrm{rad}  = \frac{\sum m v_\mathrm{rad}}{\sum m},
\label{eq:radial_v}
\ee
where $v_\mathrm{rad}$ is the radial velocity of a star relative to the cluster center of mass and the summation is over all cluster members.

%To account for this we introduce the modified $\alphag$ virial parameter
% \be
% \alphag \equiv \frac{2 E_{\rm kin}}{-E^\prime_{\rm grav}}.
% \ee
% where $E^\prime_{\rm grav}$ is the total gravitational energy of the sink particles that is calculated by taking into account all gas cells and sink particles in the simulation (i.e., the gravitational potential the sink particle actually experiences in the simulation). Note that $\alphag\leq \alpha$ and it only becomes equal at later times when most of the gas has been expelled from the cluster.

\subsubsection{Mass segregation}\label{sec:methods_mass_segregtiom}

Observed clusters exhibit \emph{mass segregation}, i.e., massive stars are ``distributed differently" than lower mass stars \citep{krumholz_2019_cluster_review}. This often means that they are concentrated at the minimum of the gravitational potential, i.e., the dense center of the cluster  \citep{Hillenbrand_Hartmann_1998_ONC_structure}. Many studies adopt this more specific criterion to define mass segregation. There are several methods in the literature to characterize this phenomenon relying on the cluster density profiles (e.g., \citealt{Hillenbrand_1997_ONC_population}) or characteristic radial distance \citep{Gouliermis_2009_cluster_segregation} of stars in various mass bins. Alternatively, one can also calculate the slope of the mass function of stars at different radii from the cluster center \citep{de_Grijs_2002_LMC_mass_segregation}. These methods, however, are sensitive to the choice of mass bins and annuli \citep{Gouliermis_2004_LMC_mass_segregation_methods}  %\mike{could imagine a bin-free version of this where the mass function slope is a parametrized function of radius and you you get the posterior distribution for the dataset of masses and radii}
and to the precise determination of the cluster center. An alternative metric that is insensitive to these is to construct a Minimum Spanning Tree (MST), the shortest graph connecting all stars without closed loops. Comparing the characteristic MST edge length between massive stars and randomly chosen stars can quantify the level of mass segregation in the cluster (see e.g. \citealt{Cartwright_Whitworth_2004_cluster_metrics} and \citealt{Allison_2009_MST}). 

In this work we consider two separate mass segregation metrics for clusters. The first one is based on the definition of \citet{Allison_2009_MST}, which quantifies the degree of mass segregation using the \emph{mass segregation ratio (MSR)} 
\begin{equation}
\MSR \equiv\frac{\langle l_\mathrm{norm} \rangle_\mathrm{MC}}{l_\mathrm{massive}},
\label{eq:lambda_msr}
\end{equation}
where $l_\mathrm{massive}$ is the \emph{mean} edge length of the MST between massive stars only. We define massive stars for the remainder of this paper as any star above $\mathrm{5\,\msun}$. %\alr{Why 5 instead of 8?}\DG{Was an arbitrary choice, AFAIK there is no hard limit that observers use when defining mass segregation, e.g. \cite{Allison_2009_MST} just defines it for x\% higheest mass stars}
Meanwhile, $l_\mathrm{norm}$ is the \emph{mean} edge length for $N_\mathrm{massive}$ randomly chosen stars, where $N_\mathrm{massive}$ is the number of massive stars. The $\langle...\rangle_\mathrm{MC}$ operation denotes constructing $N_\mathrm{sets}=500$ random sets and averaging over them, so $\langle l_\mathrm{norm} \rangle_\mathrm{MC}$ is the mean of the median MST edge lengths from $N_\mathrm{sets}$ of random realizations. Since this metric is only meaningful if at least several massive stars exist, we require $N_\mathrm{massive}\geq 5$ for it to be defined.

A significant drawback of the MST based method is that it requires at least several massive stars to be already present in the cluster, while both observations \citep{Kirk_Myers_2011_mass_segregaton} and simulations \citep{Kirk_2014_small_cluster_formation} find that even small groupings of stars with a single massive star exhibit signs of mass segregation. We find that our clusters, like observed young clusters \citep{Kirk_Myers_2011_mass_segregaton,Kerr_2021_clustering_hdbscan}, are initially highly structured, where the stellar distribution follows the hierarchical distribution of the star-forming gas. In this case, the MST method does not detect mass segregation as it only exists within smaller sub-groups of stars, i.e., if the cluster consists of several mass-segregated subclusters. 

\begin{figure}
\begin {center}
\includegraphics[width=0.95\linewidth]{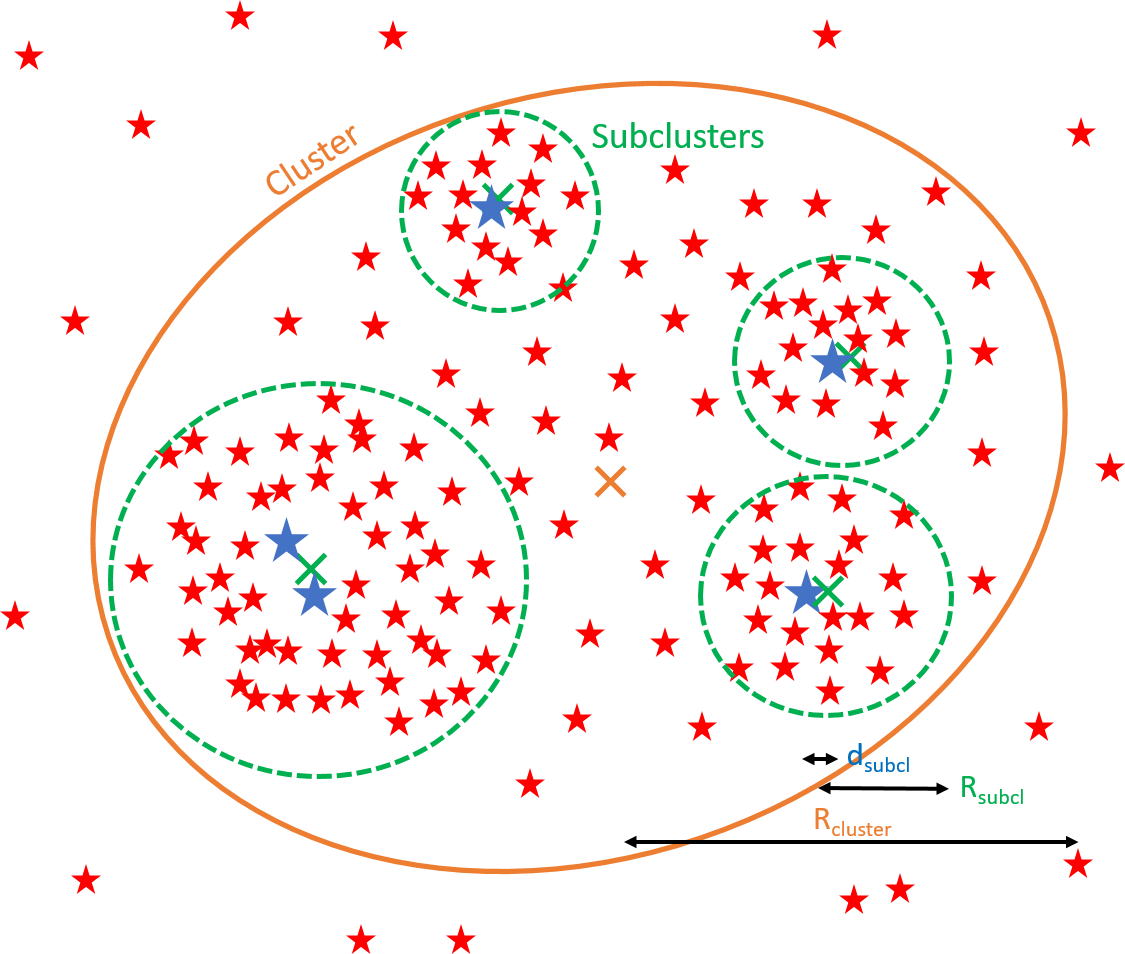}
\caption{Cartoon illustration of a young cluster with mass segregated subclusters, where low and high-mass stars are marked as red and blue respectively, showing also the centers of both the cluster and the subclusters denoted with cross markers. Due to the subclusters having only 1-2 massive stars, the mean edge length of the MST (Eq. \ref{eq:lambda_msr}) would yield $\MSR\sim R_\mathrm{cluster}/R_\mathrm{cluster}=1$, ignoring mass segregation on the subcluster level. Meanwhile, the mass segregation offset (see Eq. \ref{eq:mso}) would be $\MSO\sim R_\mathrm{subc}/d_\mathrm{subc} \gg 1$. }
\label{fig:MSO_subcluster}
\vspace{-0.5cm}
\end {center}
\end{figure}

To address this issue we identify coherent groups of stars, \emph{subclusters}, within each cluster. We define these subclusters as centrally condensed stellar over-densities and divide every cluster into one or more subclusters. We identify subclusters in each of our clusters by applying the Variational Bayesian Gaussian Mixture cluster identifying algorithm \citep{Attias_2000_variational_bayesian_mixture,bishop_2006_ML_pattern_recognition} from the {\small scikit-learn} library, using the default parameters and setting the maximum number of components to $N_\mathrm{massive}$. This Gaussian Mixture method decomposes the cluster into several Gaussian density distributions, which are, by our definition, subclusters (see Figure \ref{fig:MSO_subcluster} for a cartoon illustration and Figure \ref{fig:cluster_zoom_massive} for later examples from the simulations). Unlike the DBSCAN algorithm we use to identify the clusters themselves, Gaussian Mixture Models in general require no specific length scale and the specific variational method can infer the appropriate number of Gaussian components, i.e. subclusters. Note that we do not use the Variational Bayesian Gaussian Mixture model to identify the clusters themselves in the simulations, because this method suffers from the same assignment persistence issues as other clustering algorithms without spatial scales (see \S\ref{sec:cluster_identification}).

%SO You use this the entire time not just for the early stages.
In order to account for cluster substructure, we introduce a second metric for mass segregation, the \emph{mass segregation offset (MSO)}:
\begin{equation}
\MSO \equiv \left.\left\langle \frac{ d_\mathrm{subcl} }{R_\mathrm{subcl}} \right\rangle_\mathrm{all}\middle/\left\langle \frac{ d_\mathrm{subcl} }{R_\mathrm{subcl}} \right\rangle_\mathrm{massive}\right., %\CM{Center to match equation formatting on prev. page}\DG{What do you mean?}
\label{eq:mso}
\end{equation}
where $d_\mathrm{subcl}$ is the distance from a star to the center of the nearest subcluster. For simplicity we disregard whether the star is a member of this nearest subcluster. $R_\mathrm{subcl}$ is the size scale of the subcluster (defined following Eq. \ref{eq:cluster_R}), while the $\langle...\rangle_\mathrm{massive}$ operation denotes averaging over all massive stars in the cluster (see Figure \ref{fig:MSO_subcluster}). Note that in this definition we introduce the concept of \myquote{subcluster}, which makes the definition of Equation \ref{eq:mso} in theory different from similar offset measures in the literature \citep{Kirk_2014_small_cluster_formation}, although it gives the same answer for small or highly centralized clusters.

Finally we define the mass segregation time scale from for a star of mass $M$ as \citep{Spitzer_69_cluster_nbody_mass_segregation,Binney_Tremaine_galactic_dynamics_book_1987}:
\be
t_\mathrm{seg}(M) = \frac{\langle m\rangle}{M}\frac{N}{8 \ln N}\frac{R}{\sigma_\mathrm{sys}}, %\CM{Center to match equation formatting on prev. page}
\label{eq:t_segregation}
\ee
where $\langle m\rangle$ is the average stellar mass in the cluster, while $N$ is the number of its members, R is the cluster size (Eq. \ref{eq:cluster_R}) and $\sigma_\mathrm{sys}$ is its velocity dispersion. Note that using the system velocity dispersion changes the results by a factor of 2. 
 
 % %%%%%%%%%%%%%%%%%%%%%%%%%%%%%%%%%%%%%%%%%%%%%%%%%% 
  \section{Cluster formation and evolution in the fiducial M2e4 run}\label{sec:fiducial_results}

In this section we detail the formation and evolution of clusters within our fiducial run (\textbf{M2e4}, Sphere) that are identified using the methodology described in \S\ref{sec:cluster_identification}.

\subsection{General behavior}

\begin{figure*}
\begin {center}
\includegraphics[width=0.32\linewidth]{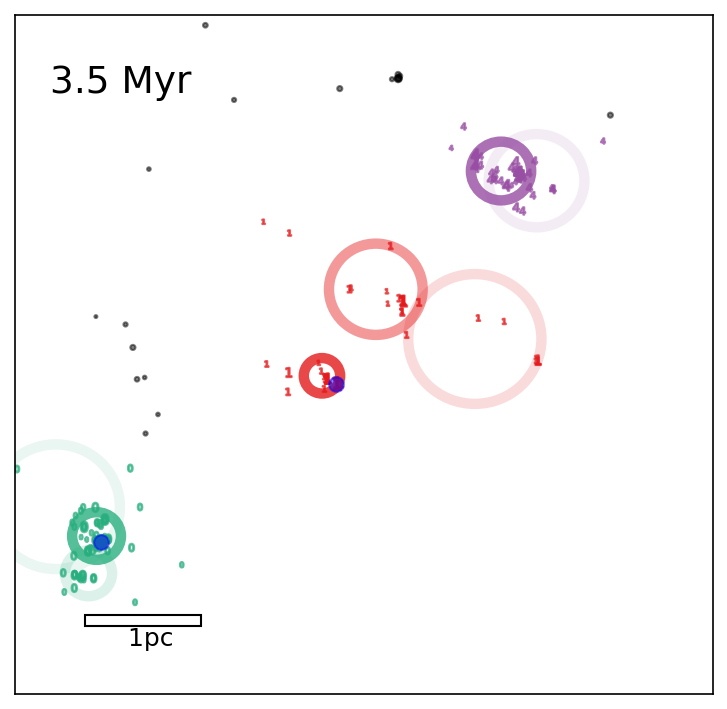}
\includegraphics[width=0.32\linewidth]{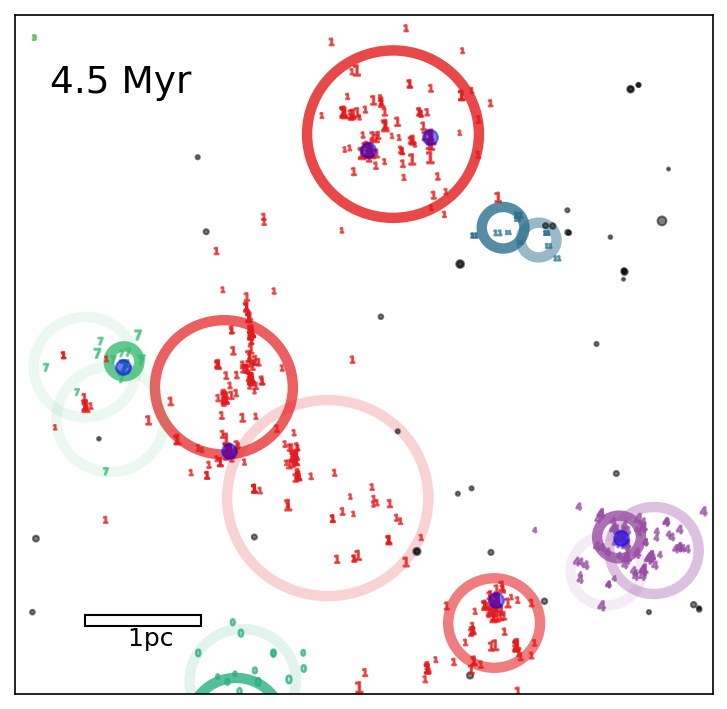}
\includegraphics[width=0.32\linewidth]{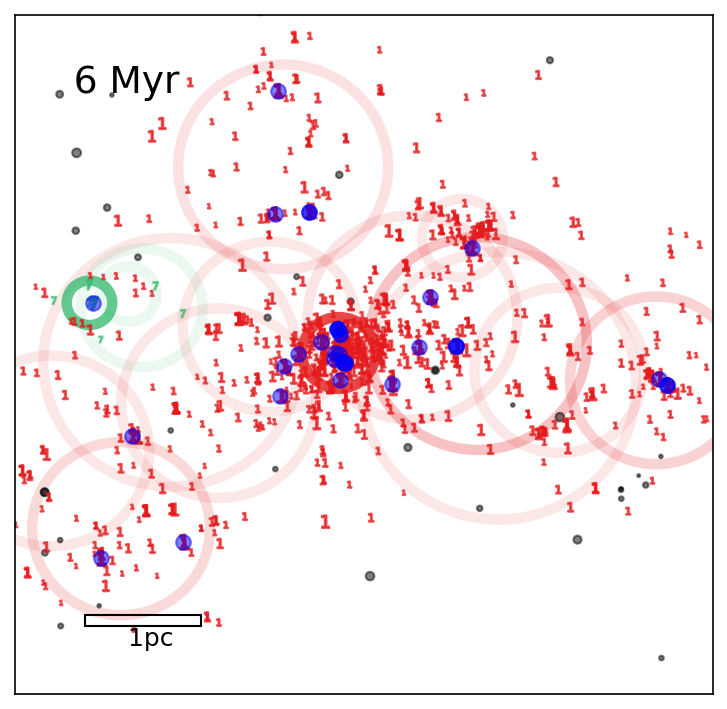}\\
\includegraphics[width=0.32\linewidth]{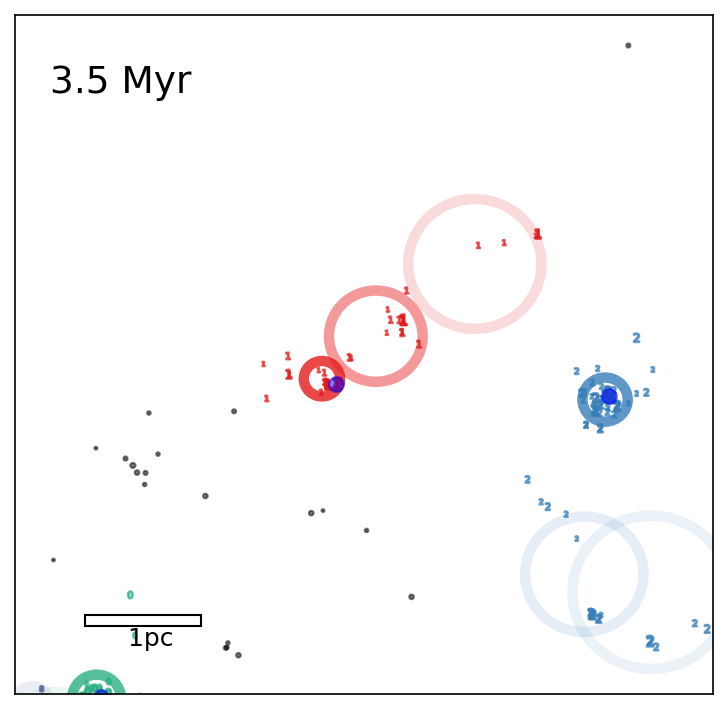}
\includegraphics[width=0.32\linewidth]{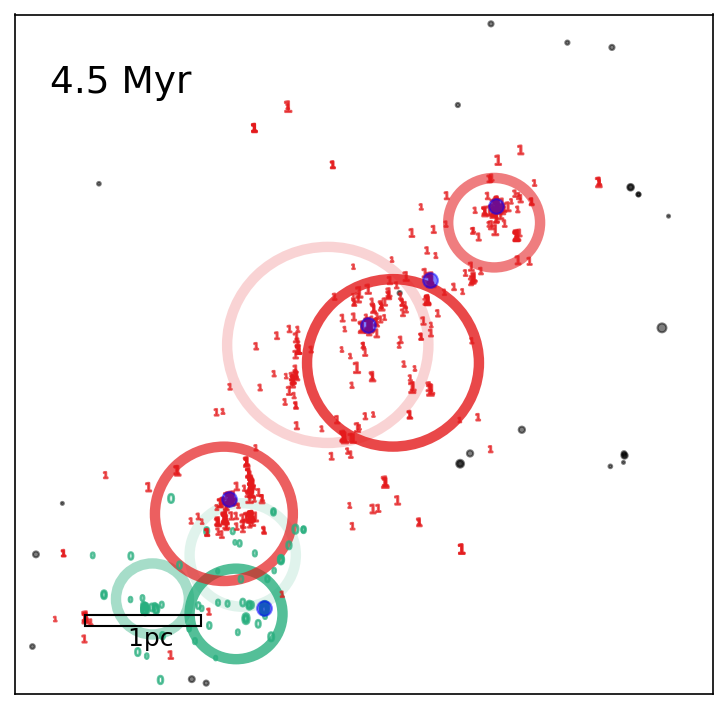}
\includegraphics[width=0.32\linewidth]{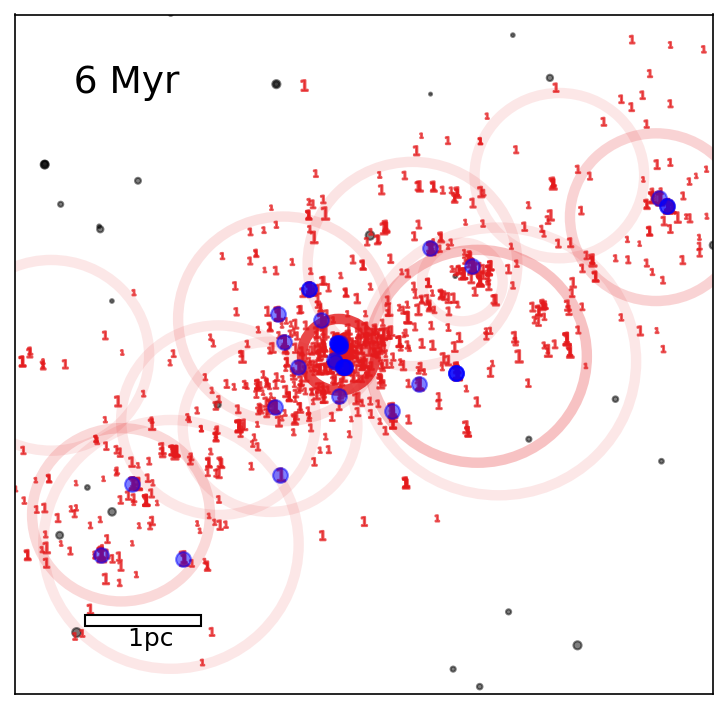}\\
\vspace{-0.4cm}
\caption{Cluster assignment of stars in the fiducial run (colored according to Figure \ref{fig:merger_history}), projected into the X-Y (top row) and X-Z (bottom row) planes. In addition, the position of massive stars ($>5\,\msun$) are marked with semi-transparent blue circles (note that some of them are in tight binaries, which makes them appear as a single opaque circle). Subclusters identified by the Variational Bayesian Gaussian Mixture model (see \S\ref{sec:cluster_mass-segregation} and Figure \ref{fig:MSO_subcluster}) are highlighted by rings (opacity increasing with mass contained). Each star formation site only has a few massive stars (usually a single or a binary) that are in the center. As these merge to form ever larger clusters, the centralized substructure remains until gas expulsion after which N-body dynamics relax the cluster into a centralized configuration. %\alr{I wonder if it would be useful to plot the newly fomred massive stars (which increase in \# across panels differently, might be useful because orignally you have $\sim$1 MS per subcluster, but then these are eventually merging leading to clusters containing multiple stars. Maybe a blue X? or different color circle?}
}
\label{fig:cluster_zoom_massive}
\vspace{-0.5cm}
\end {center}
\end{figure*}

We find that star formation begins at several locations in the cloud, which we refer to as star formation sites. These sites produce a few massive stars (often just a single one) as well as many lower mass stars, forming a small cluster (see Figure \ref{fig:cluster_zoom_massive}). These small clusters are still gas-dominated, actively accreting and star-forming when they encounter similar nearby clusters and merge with them, forming larger clusters, with these structures becoming subclusters in them. This behavior is similar to previous results claiming hierarchical cluster assembly from similar initial conditions (e.g., \citealt{bonnell:2003.hierarchical, grudic_2017}). The newly formed clusters continue accreting gas and forming new stars, as well as merging with other structures until feedback from massive stars terminates star formation and expels the remaining gas (see Figure \ref{fig:M2e4_series}). %We find that clusters gain most of their mass through directly accreting gas, but mergers also contribute a significant fraction of the total mass growth of clusters (see Figure \ref{fig:mass_source}).
%\alr{I recommend pointing to each panel in this figure as you describe it, also you jump right to figure 3 since figure 2 seems to be part of section 2, can't find where you mention it though. If figure 2 is referred to at first, since it describes some of the definitions in Section 2 then you should switch figs 1 and 2 and refer to them in order of appearance. I also recommend to add more panels to figure 1, say at 1 Myr intervals because the cluster structure/dynamics is dramatically changing. The second panel shows the subclusters that form at intersections of filaments but then the last panel shows that you many have one large cluster that forms due to gas ejection and stellar dynamical interactions. Showing more panels will show the evolution of the dispersal/mixing of the sub-clusters and support section 2.2.1 better.}

\begin{figure}
\begin {center}
\includegraphics[width=\linewidth]{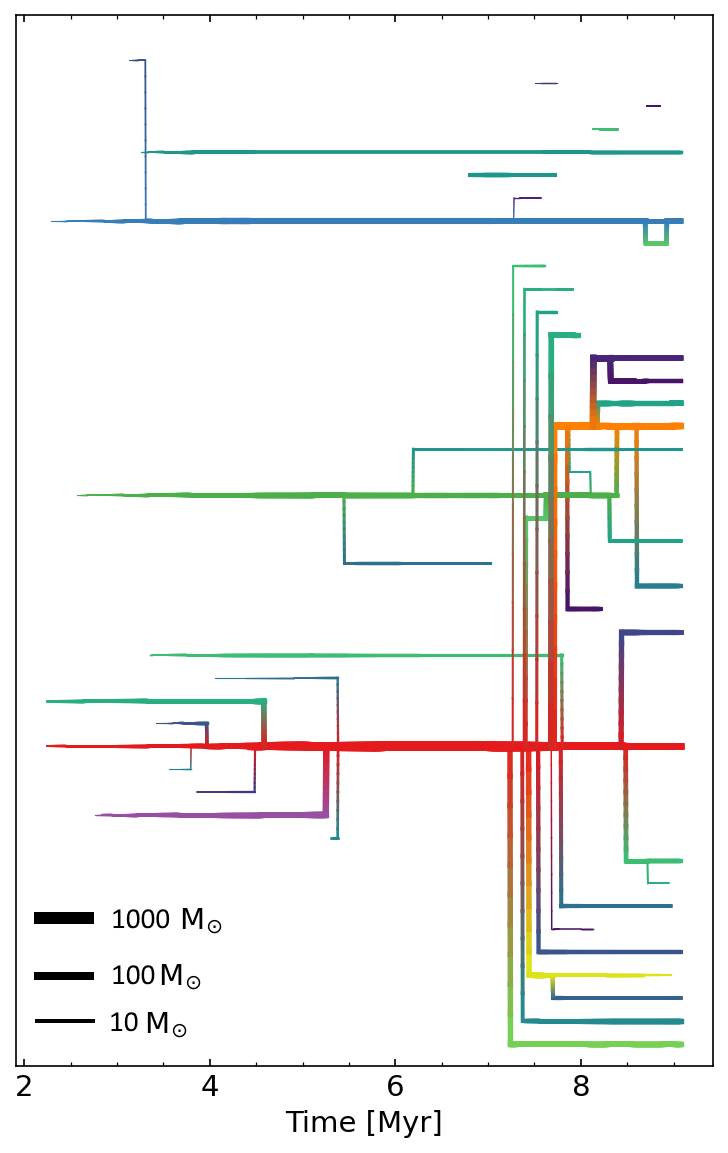}
\vspace{-0.4cm}
\caption{Merger history of clusters in the fiducial \textbf{M2e4} run. Each line represents a cluster (using its assigned color) with a width that logarithmically increases with mass. Mergers and splits are denoted by connecting the lines at the time of the event. The line representing the cluster ends if the cluster dissolves. The initial behavior is hierarchical assembly where clusters merge to form ever greater structures. This continues until feedback expels gas from the cluster, so it becomes unbound and begins to break into smaller clusters.}
\label{fig:merger_history}
\vspace{-0.5cm}
\end {center}
\end{figure}

Figure \ref{fig:merger_history} shows the formation and merger history of newly formed clusters depicting the hierarchical build-up of larger structures via mergers of smaller clusters. This leads to the formation of a \myquote{dominant} cluster that ultimately encompasses most of the stellar mass in the simulation. Once stellar feedback expels the gas from a cluster, the remaining stars are not gravitationally bound and the cluster starts breaking into smaller structures. This mainly affects the %\cut{dominant}
%SO you put dominant in quotes in the next section
largest cluster, which becomes unbound and expands, breaking up into many smaller clusters.
%SO rearranged
%as it breaks up into many smaller clusters as the original cluster becomes unbound and expands.

\subsection{Cluster properties}\label{sec:cluster_property_evol}

\begin{figure*}
\begin {center}
\includegraphics[width=0.33\linewidth]{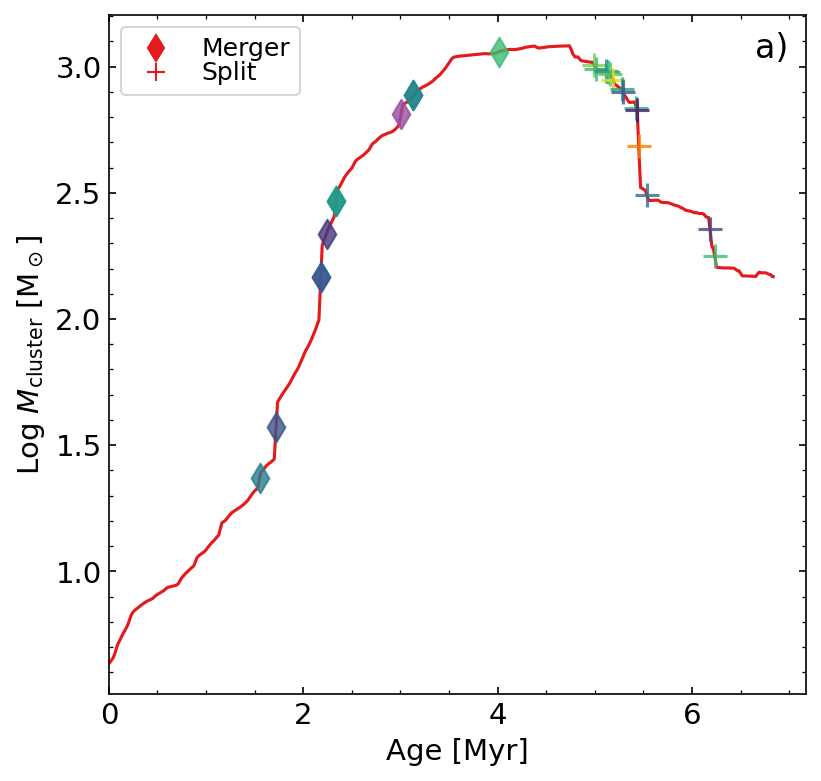}
\includegraphics[width=0.33\linewidth]{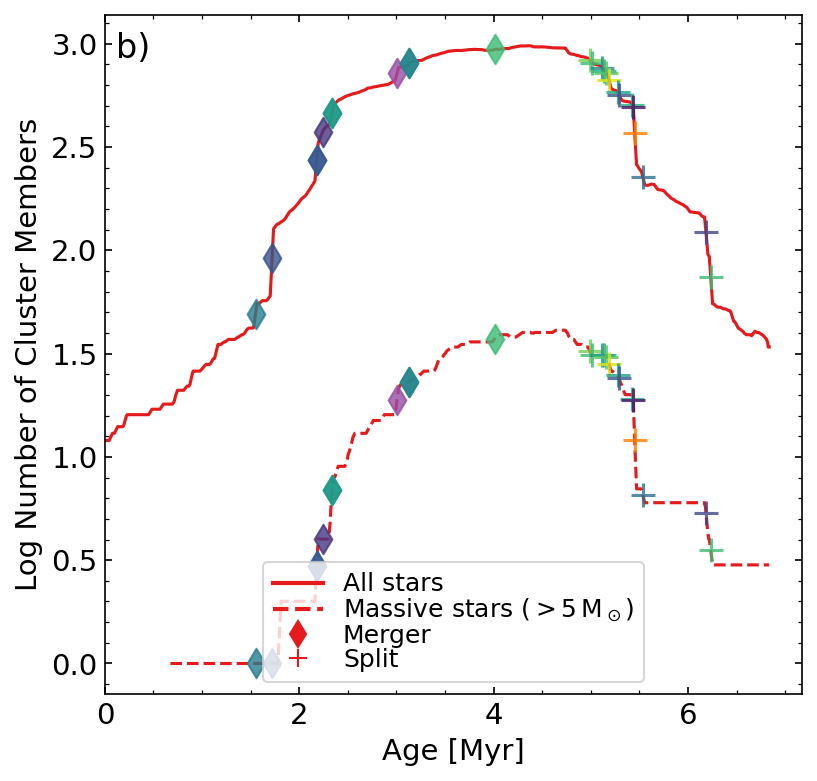}
\includegraphics[width=0.33\linewidth]{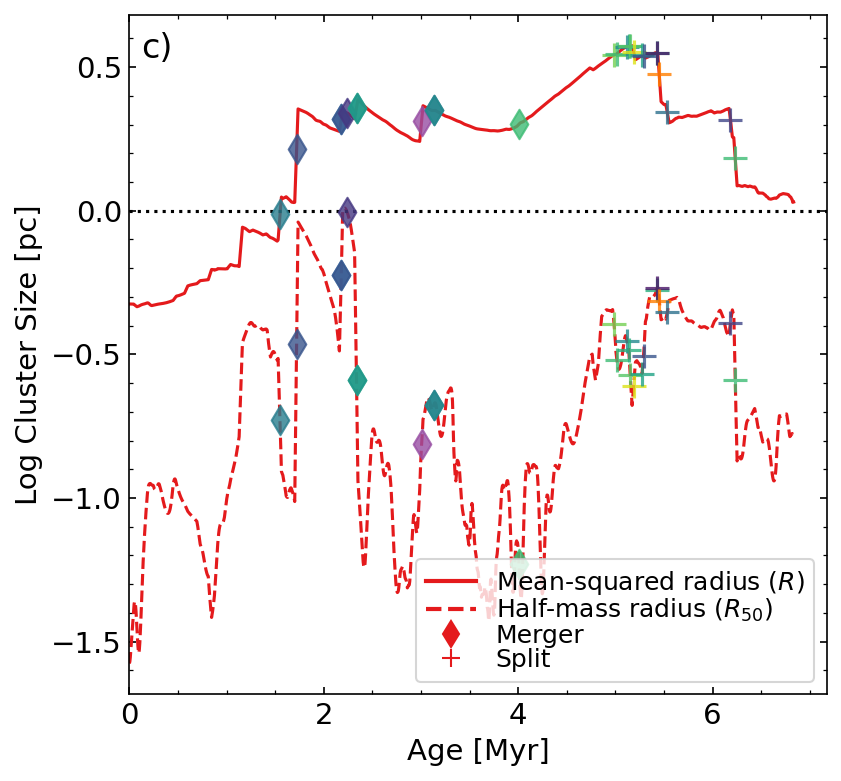}\\
\includegraphics[width=0.32\linewidth]{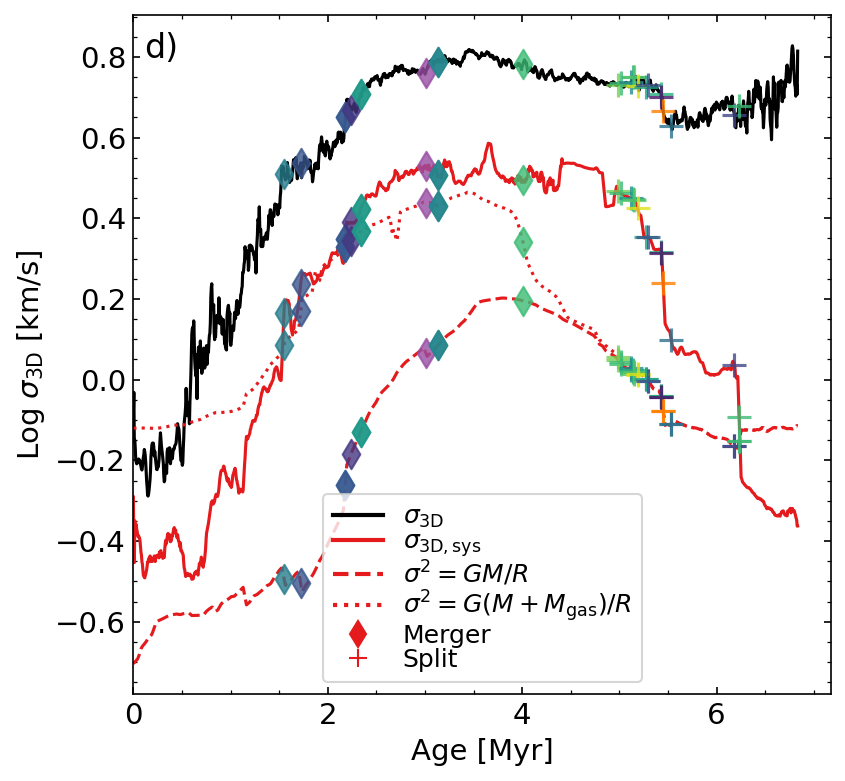}
\includegraphics[width=0.35\linewidth]{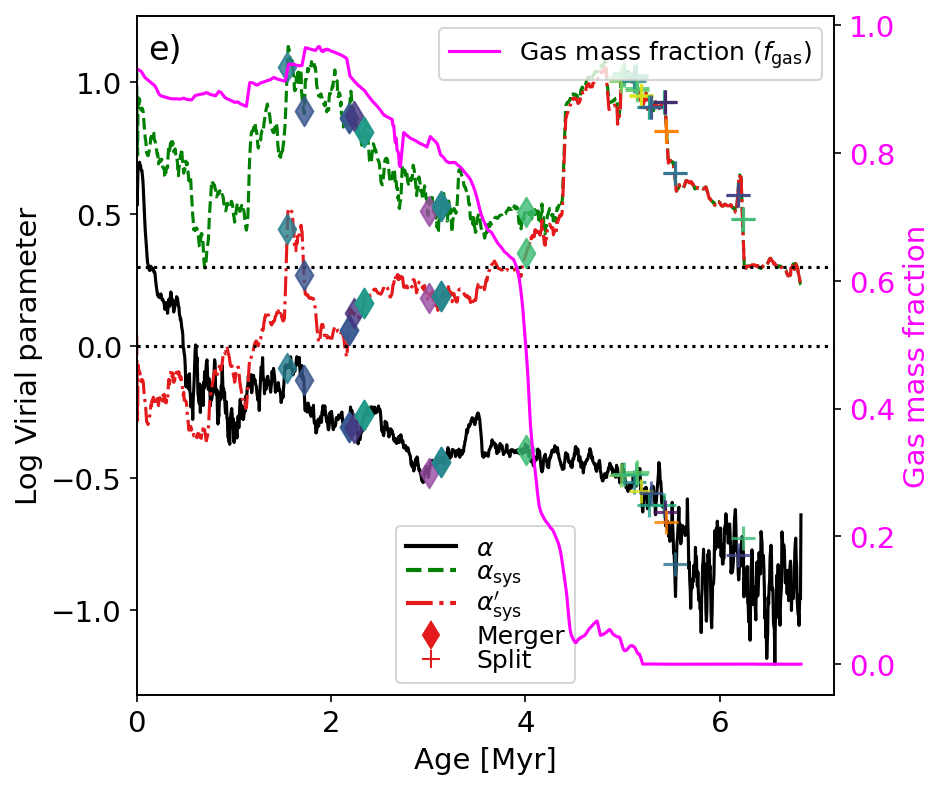}
\includegraphics[width=0.32\linewidth]{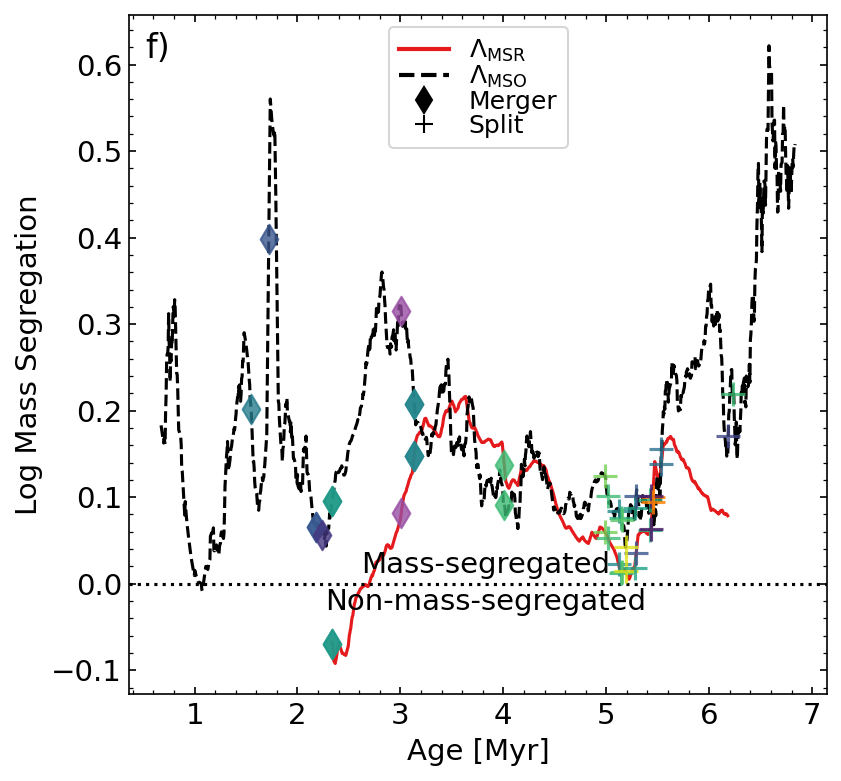}
\vspace{-0.4cm}
\caption{Evolution of the dominant cluster in the fiducial run (\textbf{M2e4}, Sphere), highlighting major events, i.e., mergers with another cluster (diamonds) and the splitting of the cluster (crosses), colored according to Figure \ref{fig:merger_history}. Dotted horizontal lines mark the characteristic scale of our cluster size definition (1 pc), the virial equilibrium and marginal boundedness ($\alpha=1$ and 2) and the boundary between regular and inverse mass segregation.  The \emph{top  row} shows the total stellar mass in the cluster (top left), number of cluster members (top middle) and the cluster size scales (top right, Eq. \ref{eq:cluster_R}) respectively. 
The \emph{bottom  row} shows the cluster velocity dispersion (left, Eq. \ref{eq:cluster_sigma}), virialization state (middle, Eqs. \ref{eq:alpha}-\ref{eq:alphag_sys}) and mass segregation (right, Eqs. \ref{eq:lambda_msr}-\ref{eq:mso}). Note that all values are smoothed with a 30 kyr averaging window to make the plots easier to read. For an analysis of the main trends see \S\ref{sec:cluster_property_evol}.
}
\label{fig:cluster_basic_evol_sphere}
\vspace{-0.5cm}
\end {center}
\end{figure*}

To illustrate the evolution of cluster properties in our simulations we focus on the \myquote{dominant} cluster that eventually encompasses the majority of stars at the end of the simulation. Figure \ref{fig:cluster_basic_evol_sphere} shows the cluster properties defined in \S\ref{sec:cluster_properties_defs}. The dominant cluster reaches about 1000 members and attains roughly $1000\,\msun$ by the time the cloud disrupts (see panels a-b), containing the majority of the total stellar mass. This run endes with $\SFE=M_{\star}/M_{0}\sim 7\%$, corresponding to $M_{\star}\sim 1400\,\msun$). 
Clusters form around individual star formation sites, and these structures merge to form larger objects, leading to \myquote{jumps} in the cluster size. 
%As one of these substructures often contains a significant portion of the mass, this leads to a growing discrepancy between the cluster $R$ mean-squared radius and the $R_\mathrm{50}$ half-mass radius. 
Although gravitational attraction between the various substructures and stellar interactions should shrink the cluster over time and increase the central stellar density \citep{krause2020}, the continuous formation of new stars from infalling gas and  mergers with other clusters causes the cluster to maintain its size until gas expulsion occurs (Figure \ref{fig:cluster_basic_evol_sphere} panel c). 

As the star cluster grows rapidly in both mass and size the velocity dispersion also increases (Figure \ref{fig:cluster_basic_evol_sphere}, panel d). Note that the  stellar velocity dispersion, $\sigma_\mathrm{3D}$, is super-virial due to the effect of close binaries. Meanwhile, the velocity dispersion for systems, $\sigma_\mathrm{3D,sys}$,  (Eq. \ref{eq:cluster_sigma_sys}) is fairly close to the virial value (if the gas potential is also taken into account). The stellar velocity dispersion peaks as the cluster reaches its maximum mass, just as gas expulsion starts. It declines subsequently as the cluster breaks apart. Thus, the statistics follow a shrinking fraction of the original cluster.

We expect that most close binaries are unresolved in observed clusters \citep[e.g.,][]{Foster_2015,Kerr_2021_clustering_hdbscan}, so the observationally inferred velocity dispersion should be close to $\sigma_\mathrm{3D,sys}$. Thus, simple estimates using the global cluster mass and size scales (after including the enclosed gas mass) would conclude these clusters are virialized during star formation (as in \citealt{Foster_2015}) and highly supervirial during the breakup phase after star formation ceases and/or a significant fraction of the gas mass has been expelled. %\SO{This sounds like a bad thing. E.g., I don't think Orion is found to be very super-virial.} \mike{1) most ($\sim 75\%$) of young SF regions studied so far with good kinematics are expanding and supervirial (Kuhn 2019), unlike the ONC which is probably bound and the exception to the rule. The Anvil is probably more comparable to M17 than ONC. 2) the virial parameter you would ``observe" depends sensitively on how it's measured (linewidth vs. proper + radial motion with parallax-limited resolution of  binaries), so be specific here} 
Meanwhile, direct calculation of the virial parameter using Equation \ref{eq:alpha} indicates the cluster is highly sub-virial with $\alpha\sim 1/2-1$ (Figure \ref{fig:cluster_basic_evol_sphere}, panel e). We find that the gravitational potential energy is dominated by hard binaries, leading to low $\alpha$ values. After merging these systems, i.e., using the definition from Eq. \ref{eq:alpha_sys}), $\alphasys$ is consistently above the boundedness limit of $\alpha=2$. However, initially the clusters are strongly gas-dominated, so after correcting for the gas potential (Eq. \ref{eq:alphag_sys}) we find the clusters are initially strongly bound ($\alphag<1$, similar to the results of \citealt{Offner_2009_young_cluster_kinetamtics}) and then become unbound after gas expulsion ($\alphag\sim 10$). The resulting unbound cluster immediately expands and breaks into smaller structures. Since the simulation stops shortly after gas expulsion, it is unclear what fraction of the original cluster will remain bound. This will be investigated in a future STARFORGE project.

\subsection{Evolution of Mass Segregation}\label{sec:cluster_mass-segregation}

Figure \ref{fig:cluster_basic_evol_sphere} shows that the dominant cluster develops mass segregation (panel f) according to both the $\MSR$ and $\MSO$ metrics (see Equations \ref{eq:lambda_msr}-\ref{eq:mso}). However, these metrics differ on the initial degree of mass-segregation, with $\MSO$ identifying segregation from the time the first massive stars form, while $\MSR$ only detecting it at much later times. In the early stage of cluster evolution, each cluster is composed of the stars formed in one to a few star formation site and hosts 
%SO I'm a little confused by what you are saying here -- a cluster is made up stars from a few SF sites? Or each defined cluster is the stars from a single SF site (which is what you implied earlier). It depends what you mean by 'early' here.
only a few massive stars in the center (see Figure \ref{fig:cluster_zoom_massive}), leading to a $\MSO>1$ (at this stage they have <5 massive stars so $\MSR$ is not defined). These sites continue to accrete, form more stars, and merge with others, thereby forming ever-larger clusters. %\alr{Is this due to turbulent fragmentation and feedback dropping the accretion rate so that you preferentially form more low-mass stars instead of massive ones? e.g., see Stella's 2009 paper that talks about turbulent fragmentation, could also be due to turbulence taking time to set up these over-densities to become Jeans unstable> I think it might be useful to provide a sentence on why SF proceeds this way. E.g., Pillai+2019 had a paper showing that pre-stellar clumps have both low- and high mass SF}. 
The resulting merged clusters inherit the centrally condensed substructures, thus maintaining $\MSO>1$. Note that it is not the case for $\MSR$, which drops at the start of mergers due to the initial distance between the subclusters dominating the MST edge lengths, Once the merger is underway the subclusters interact and sink towards the center, increasing $\MSR$. Eventually the cluster relaxes to a centrally condensed, \myquote{classical} star cluster with $\MSO>1$ and $\MSR>1$. This redistribution occurs on a timescale of $t_\mathrm{seg}\sim 2\,\mathrm{Myr}$ for massive stars (see Eq. \ref{eq:t_segregation}). Before the cluster can fully dynamically relax, stellar feedback expels the remaining gas and unbinds the cluster. Still, as gas expulsion begins, massive stars are already preferentially located near the center of the dominant cluster, leading to high $\MSR$ and $\MSO$ values. During the gas dispersal process massive stars that formerly reside near the center move outwards %SO This language implies a more violent dynamical interaction rather than (I think) the more gentle gravitational pushing/expansion caused by the feedback/changing potential.
with the rest of the cluster, causing both mass segregation metrics to drop. Note that at later times $\MSO$ may increase, but it is due to the cluster identification algorithm splitting the dominant cluster into smaller clusters, which tend to have a few massive stars at their centers.

\subsection{Mass distribution of stars inside and outside clusters}\label{sec:cluster_IMF}

\begin{figure}
\begin {center}
\includegraphics[width=\linewidth]{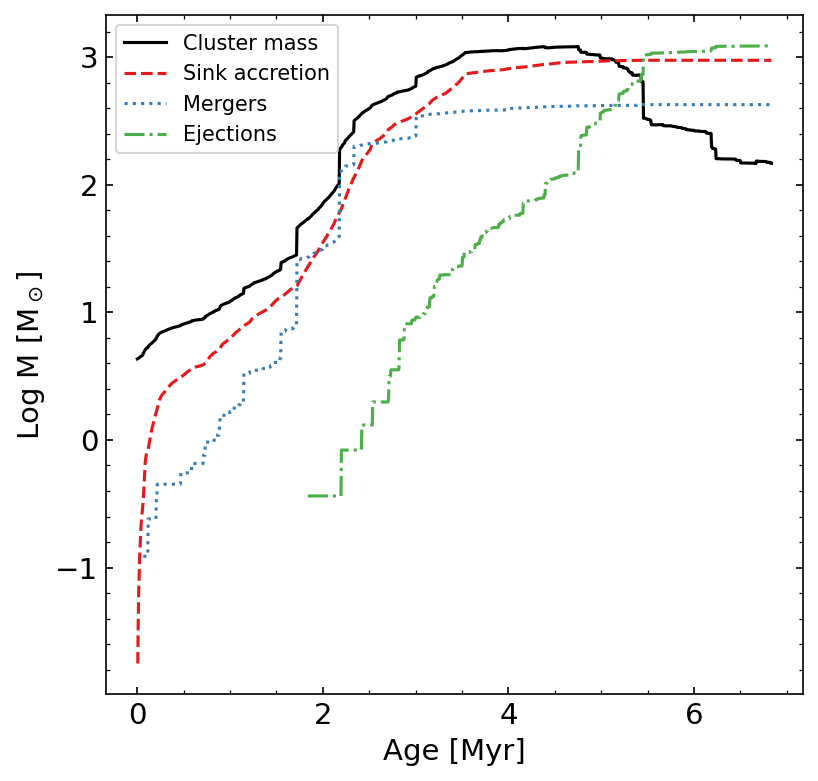}
\vspace{-0.4cm}
\caption{Mass history of the dominant cluster in the fiducial run (\textbf{M2e4}, Sphere). Lines show the current cluster mass (solid), cumulative mass the cluster acquired by stellar members, by accreting gas (dashed) and through cluster mergers (dotted). The total mass of stars ejected from the cluster is shown by the dash-dotted line. Over its lifetime the cluster gains most of its mass by accreting gas, while mergers and accretion contribute roughly equally to the cluster growth while there are clusters to merge with. Before gas expulsion (around 4.5 Myr, see Figure \ref{fig:cluster_basic_evol_sphere}) the cluster ejects roughly 10\% of its stellar mass.}
\label{fig:mass_source}
\vspace{-0.5cm}
\end {center}
\end{figure}

We find that the majority of stars that form in our simulations end up in clusters, although a significant fraction ($~\sim 10\%$) are ejected before gas expulsion (see Table \ref{tab:final_properties} and Figure \ref{fig:mass_source}). During its lifetime the dominant cluster in the fiducial simulation gains stellar mass from two sources: 1) mergers with other clusters and 2) gas accretion by its stellar members or newly formed stars within the cluster. We find these two mechanisms to have roughly similar weight during most of the cluster lifetime, with accretion becoming more important after most stars have merged with the dominant cluster, leaving no other clusters to merge with. After gas expulsion the cluster becomes unbound and loses an order unity of its mass.

\begin{figure*}
\begin {center}
\includegraphics[width=0.45\linewidth]{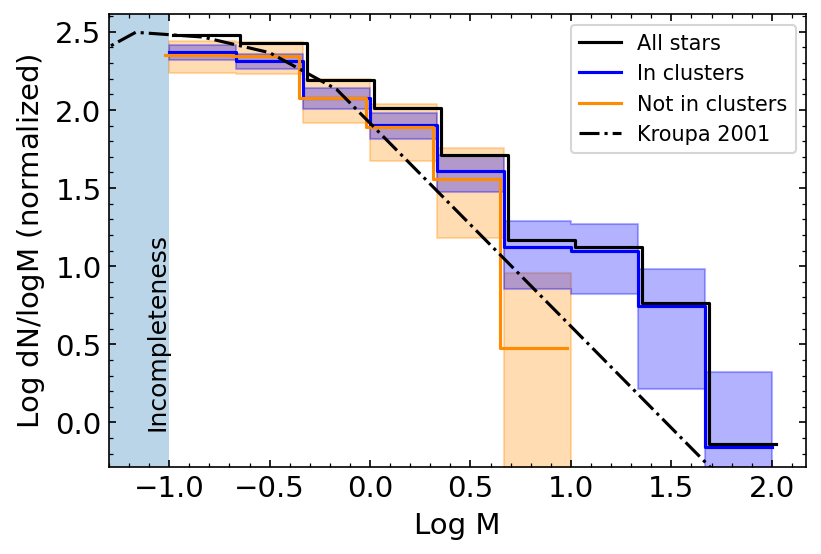}
\includegraphics[width=0.45\linewidth]{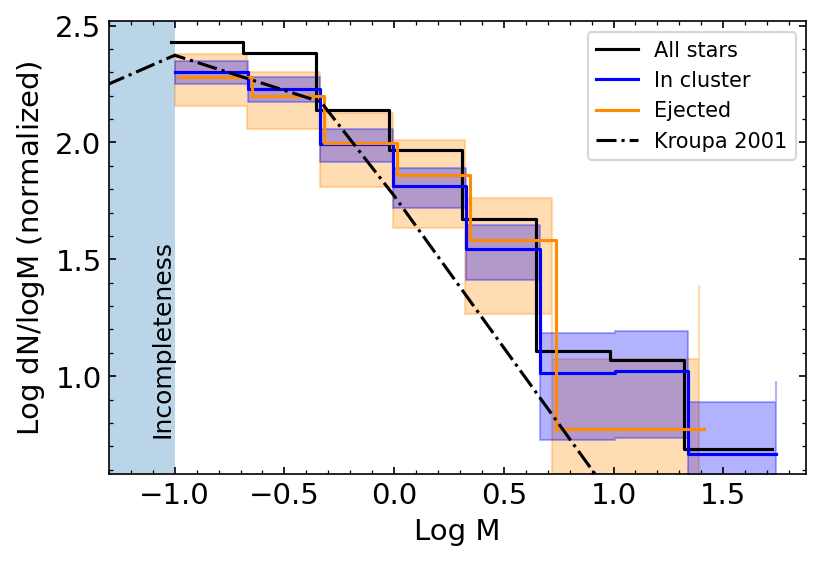}
\vspace{-0.4cm}
\caption{\textit{Left}: Mass distribution of sink particles (stars), comparing three populations:  those assigned to clusters (blue), those that are not assigned to clusters (orange) and the full stellar population (black). The spectrum is taken at the time when the dominant cluster reaches its maximum mass (i.e., just as gas expulsion starts). Shaded regions denote the two sigma Poisson uncertainties. We also show  the \citet{kroupa_imf} canonical fitting function for the MW IMF from the literature and mark the completeness limit of the simulation. \textit{Right}: Similar to the left panel, but concentrating on the dominant cluster only. We compare three populations: those assigned to the dominant cluster (blue), those that were ejected from it (orange) and the full stellar population (black).}
\label{fig:ejected_vs_clusters}
\vspace{-0.5cm}
\end {center}
\end{figure*}

Figure \ref{fig:ejected_vs_clusters} shows the mass distribution of various stellar populations in the fiducial simulation just as gas expulsion starts due to strong radiative feedback (around 4.5 Myr into the simulation). We find that about 80\% of stars are cluster members where more than 90\% of these belong to the dominant cluster. The mass distributions of the clustered and non-clustered stars are similar up to $10\,\solarmass$, but we can not rule out that the distribution deviate at the high mass end. The distributions are statistically consistent with the \citet{kroupa_imf} IMF fitting function except at very high masses, because the overall simulation IMF is slightly top-heavy (to be described in more detail in an upcoming paper). Figure \ref{fig:ejected_vs_clusters} further shows that the mass distribution for the dominant cluster is similar to that of the ejected stars (with a significance of $p=0.96$ obtained from a two-sample Kolmogorov-Smirnov test). There is no indication of preferential ejection of more massive stars before gas expulsion. Unsurprisingly the stellar mass distributions of the dominant cluster and the full simulation are also similar.

% %%%%%%%%%%%%%%%%%%%%%%%%%%%%%%%%%%%%%%%%%%%%%%%%%% 
\section{Effects of initial condition variations on clusters}\label{sec:variation_results}

In this section we investigate the effects of turbulent driving (\S\ref{sec:variations_results_box}), the initial level of turbulence (\S\ref{sec:variations_results_alpha}) and surface density (\S\ref{sec:variations_results_sigma}) on the properties of the dominant cluster. These properties have a significant effect on the star formation history of the system, and here we examine the impact of these properties on clustering. %\cut{ In general we find that the turbulent virial parameter and surface density have little qualitative effect on cluster evolution, unlike turbulent driving and the Box boundary condition (see Table \ref{tab:final_properties} and \S\ref{sec:variations_results_box}-\S\ref{sec:variations_results_sigma}).} \SO{These are results, which I don't think belong in a section intro, before the analysis is discussed. }

\begin{table*}
    \setlength\tabcolsep{2.0pt} %compress table
	\centering
	\begin{tabular}{|c|c|c|c|}
	    \hline
		Label & SFE ($M_\mathrm{*,final}/M_\mathrm{cloud}$) & $\mathrm{max}(M_\mathrm{in\,clusters})/M_\mathrm{*,final}$ & $\mathrm{max}(M_\mathrm{dominant\,cluster})/\mathrm{max}(M_\mathrm{in\,clusters})$  \\
		\hline
		\textbf{M2e4} & 8\% & 83\% & 94\% \\
		\hline
		\textbf{M2e4} Box, with driving & 5\% & 45\% & 64\% \\
		\hline
		\textbf{M2e4} Box, no driving & 9\% & 73\% & 46\% \\
		\hline
		\textbf{M2e4\_a1} & 11\% & 80\% & 94\% \\
		\hline
		\textbf{M2e4\_a4} & 4\% & 67\% & 79\% \\
		\hline
		\textbf{M2e4\_R3} & 14\% & 87\% & 99\% \\
		\hline
	\end{tabular}
        \vspace{-0.1cm}
 \caption{General cluster and sink properties of various runs (see Table \ref{tab:IC_phys}), including the final star formation efficiency ($\SFE=M_\mathrm{*,final}/M_\mathrm{cloud}$), relative mass of stars in clusters vs outside clusters (calculated as $\mathrm{max}(M_\mathrm{in\,clusters})/M_\mathrm{*,final}$) and the relative weight of the dominant cluster ($\mathrm{max}(M_\mathrm{dominant\,cluster})/\mathrm{max}(M_\mathrm{in\,clusters})$). %\alr{I suggest adding a column for the final snapshot times since the SFE quoted does not correspond to the same final times.}\DG{True but it should not matter and the stopping point is somewhat arbitrary so it doe snot really say much. In the IMF paper I will report cloud disruption times.}
 }
 \label{tab:final_properties}\vspace{-0.5cm}
\end{table*}

\subsection{Cloud setup and turbulent driving (Box vs. Sphere)}\label{sec:variations_results_box}
 
As noted in \S\ref{sec:initial_conditions}, the Sphere vs Box configurations have two important differences, which may lead to different clustering properties. First, the periodic boundary conditions of the Box setup leads to both an order-of-magnitude shallower gravitational potential \citep{federrath_sim_2012} and prevents the escape of radiation and gas. Second, the Box setup starts from a self-consistent, pre-stirred state and this external driving is continuous throughout the run, providing energy for turbulent modes on the box scale that cascade down to smaller scales. To disentangle the effects of these two factors, we compare three \textbf{M2e4} runs (Table \ref{tab:IC_phys}): 1) our fiducial Sphere run, 2) a Box run with continuous external driving and 3) a Box run where we turn off the driving after the initial \myquote{stirring} phase. %\alr{which is turned off once we switch on self-gravity so that collapse and star formation can occur}.\SO{Isn't this detail given in methods already?}

We find that both periodic boundary conditions and turbulent driving significantly affect cluster properties. As mentioned in \S\ref{sec:initial_conditions} periodic boundaries prevent both material and radiation from escaping the cloud, %\alr{(I find this misleading because material and radiation can escape from one side of the box but is returned to the other side due to the periodic boundary conditions - I suggest rephrasing this to include that effect, some readers may be less familiar with this method. Maybe just include this effect in more detail in \S2 when you describe the box setup)}, 
thus these runs never experience gas ejection that may lead to cloud disruption. 

\begin{figure*}
\setlength\tabcolsep{0.0pt} %compress table
\begin {center}
\begin{tabular}{cccc}
\multicolumn{2}{c}{\large \bf Fiducial run (M2e4 Sphere)} & \multicolumn{2}{c}{\large \bf Box run with driven turbulence (M2e4 Box)}\\ 
\includegraphics[width=0.23\linewidth]{figures/M2e4_C_M_J_RT_W/SurfaceDensity_0810.0.png} &
\includegraphics[width=0.24\linewidth]{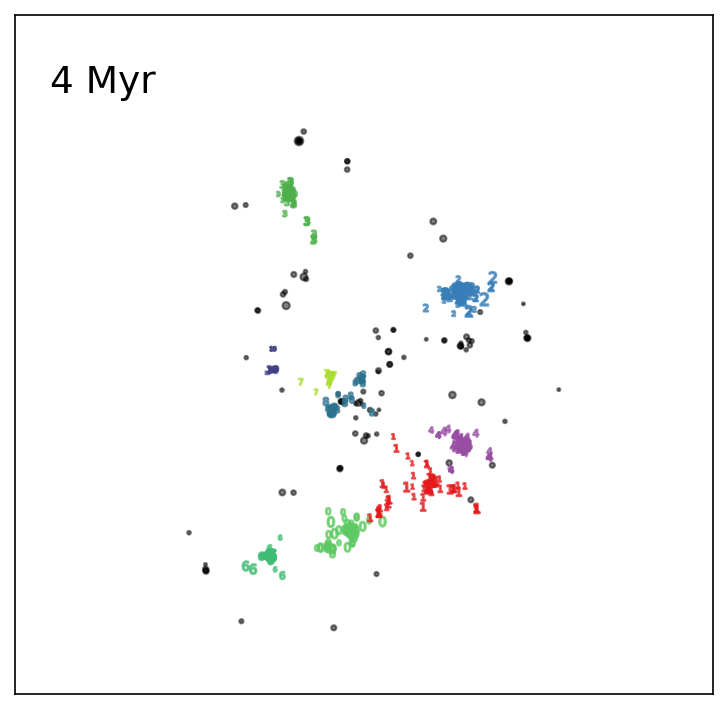} &
\includegraphics[width=0.23\linewidth]{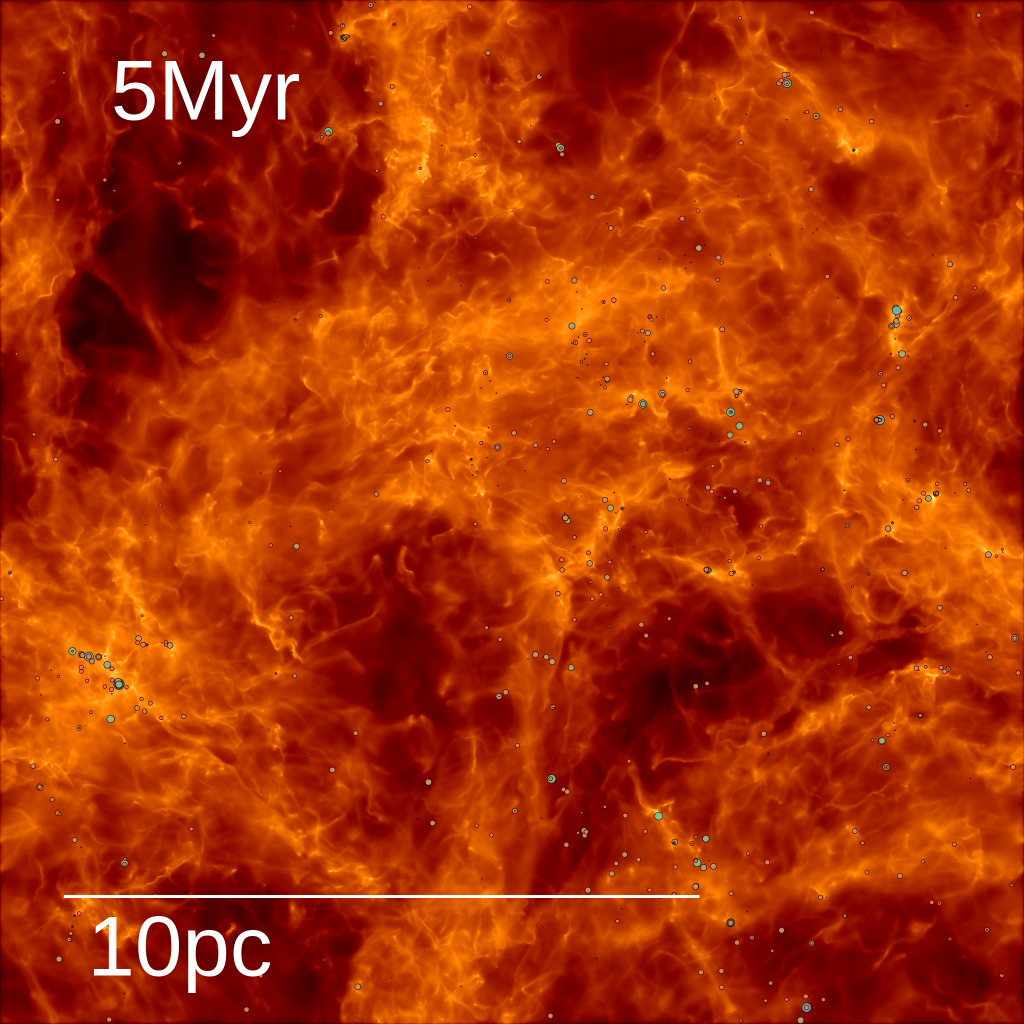} &
\includegraphics[width=0.24\linewidth]{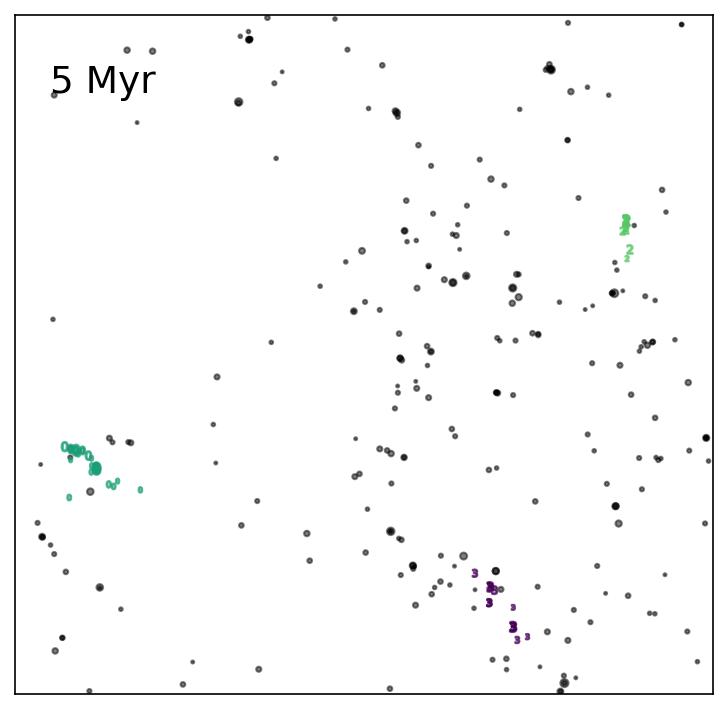} \\

\includegraphics[width=0.23\linewidth]{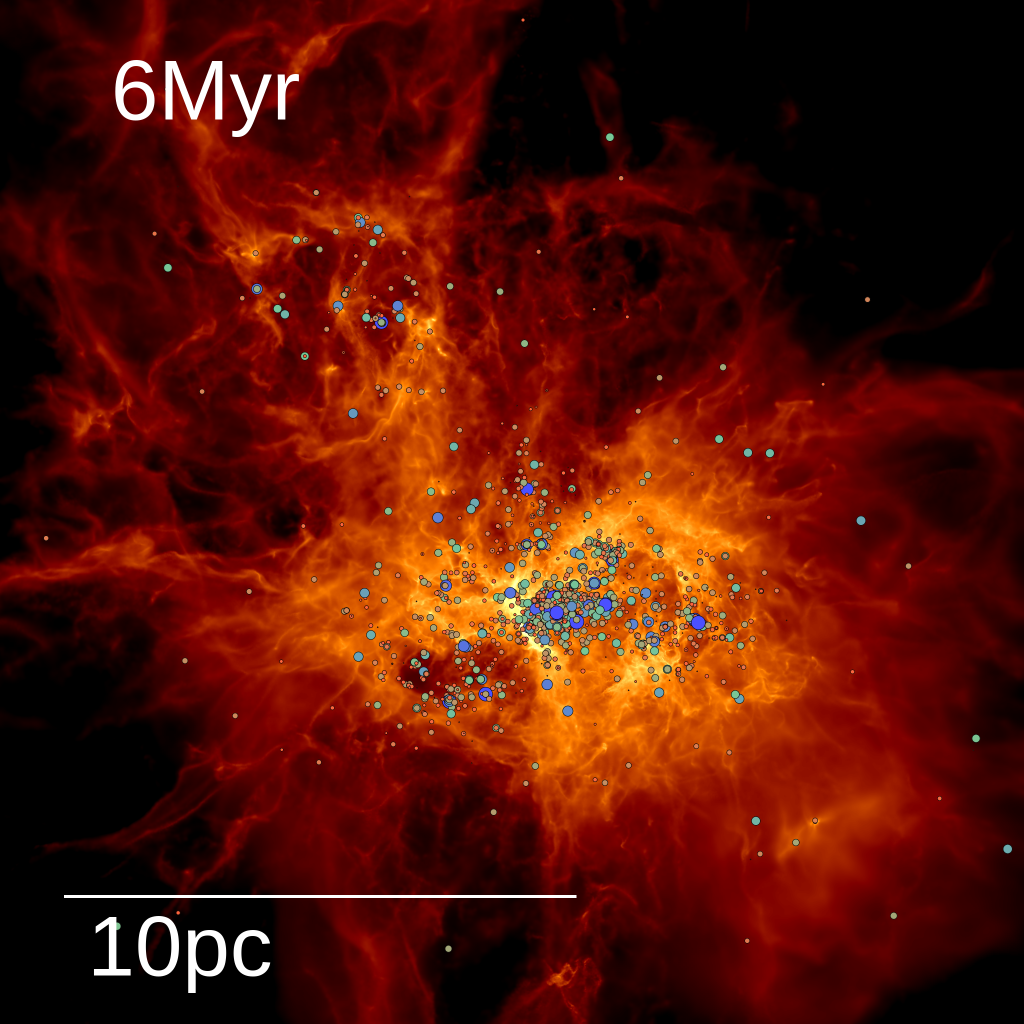} &
\includegraphics[width=0.24\linewidth]{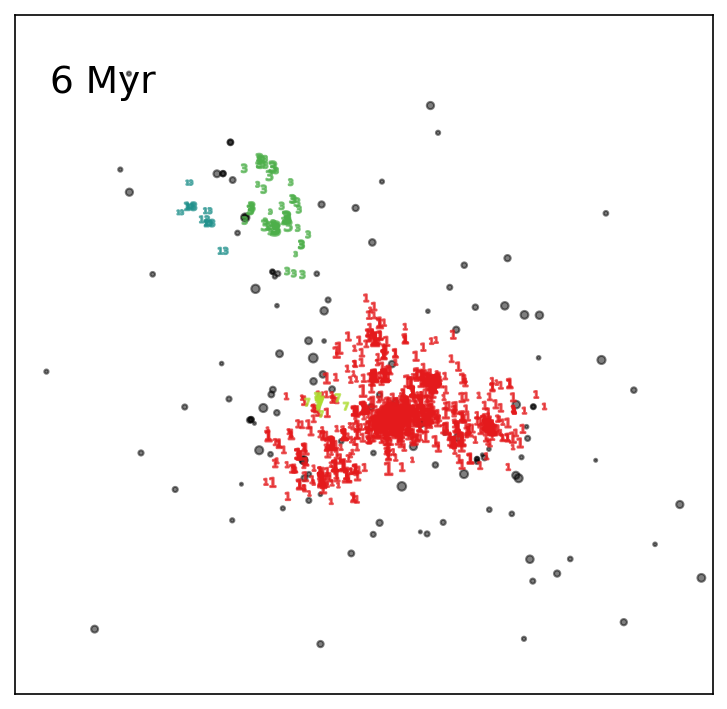} &
\includegraphics[width=0.23\linewidth]{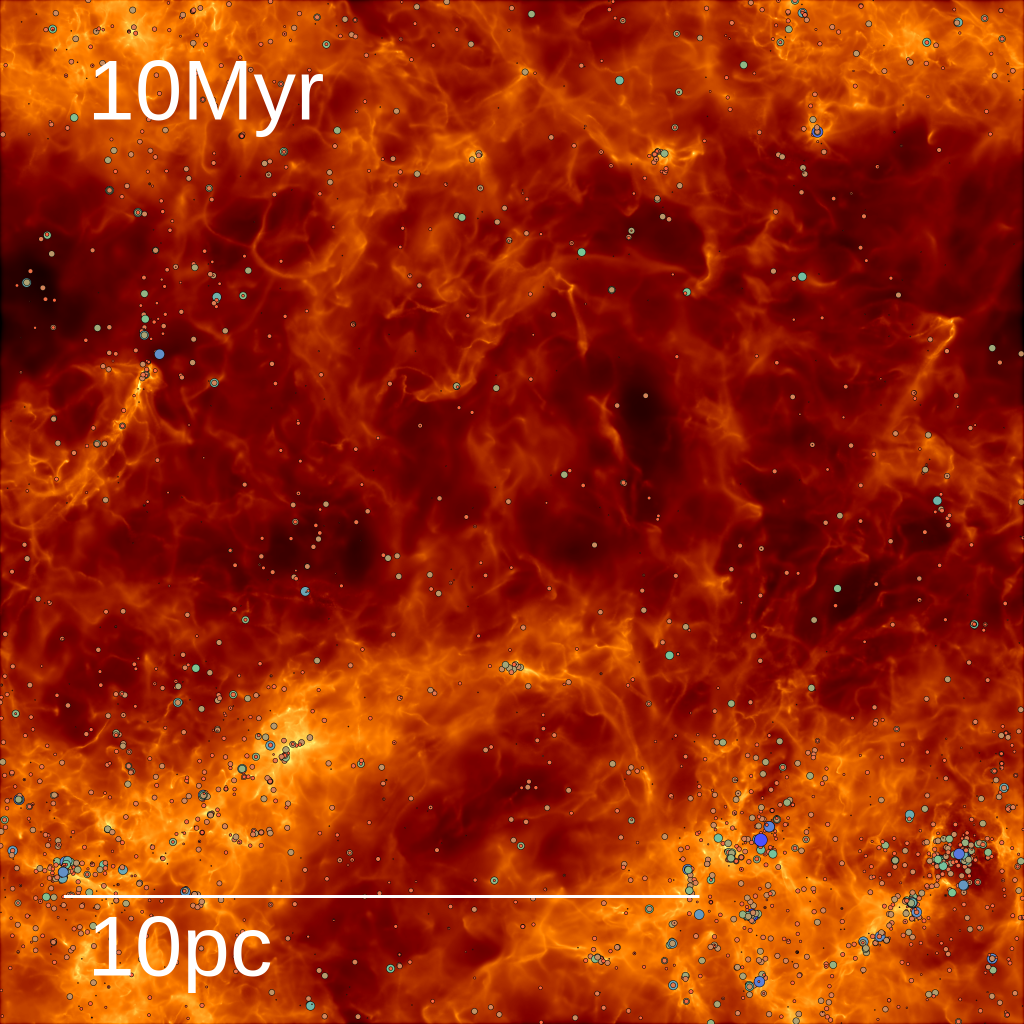} &
\includegraphics[width=0.24\linewidth]{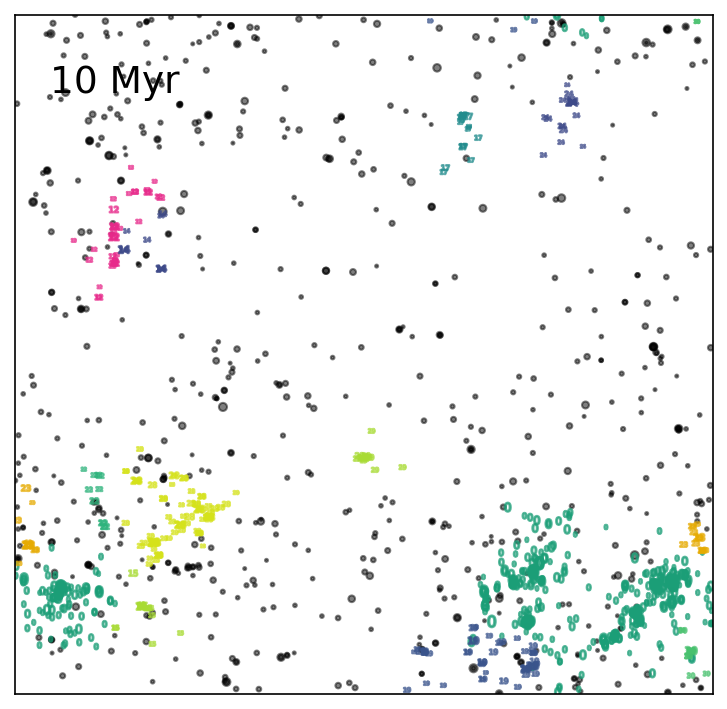} \\

\includegraphics[width=0.23\linewidth]{figures/M2e4_C_M_J_RT_W/SurfaceDensity_1630.0.png} &
\includegraphics[width=0.24\linewidth]{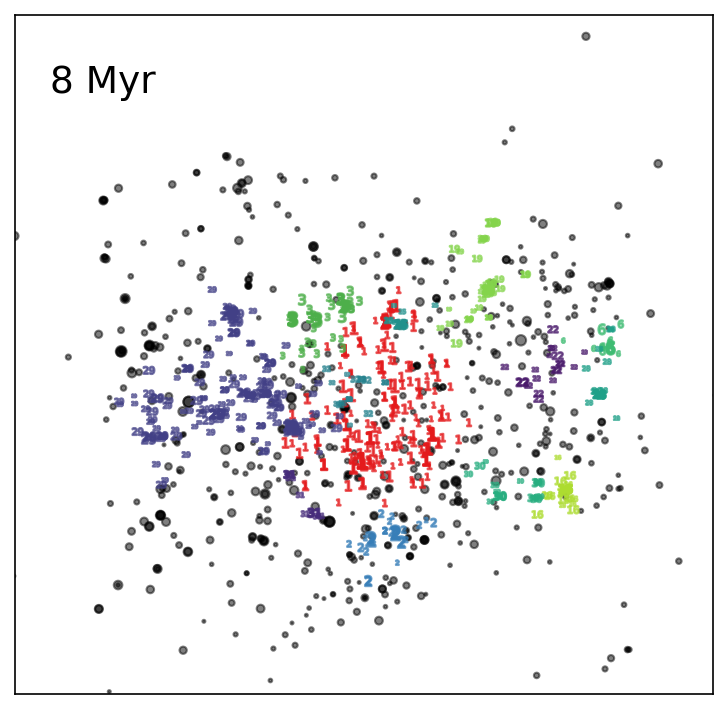} &
\includegraphics[width=0.23\linewidth]{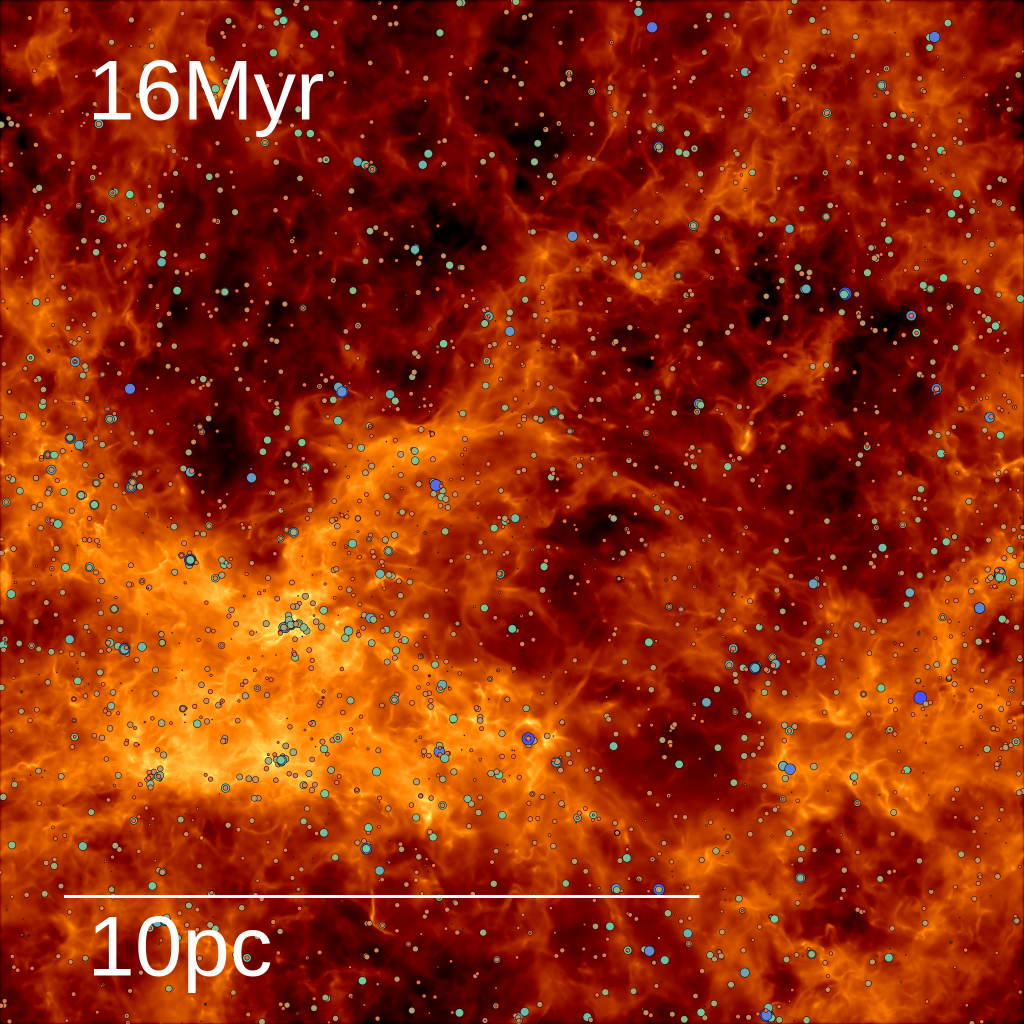} &
\includegraphics[width=0.24\linewidth]{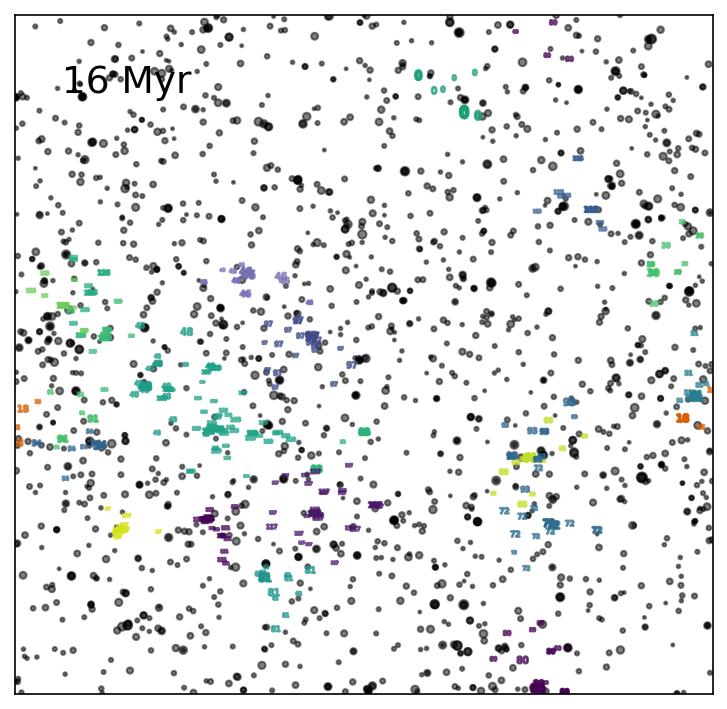} \\
\end{tabular}
%\vspace{-0.4cm}
\caption{The surface density maps and cluster assignments of the \textbf{M2e4} runs with both \myquote{Sphere} and \myquote{Box} ICs. The surface density maps use the same conventions as Figure \ref{fig:M2e4_series}. On the cluster assignment maps each star is represented with its cluster ID and colored with the color assigned to the cluster, while stars not assigned to clusters are marked with black circles.}
\label{fig:cluster_assignment_compare_box}
\vspace{-0.5cm}
\end {center}
\end{figure*}

\begin{figure*}
\setlength\tabcolsep{0.0pt} %compress table
\begin {center}
\begin{tabular}{cccc}
\multicolumn{2}{c}{\large \bf Box run with driven turbulence} & \multicolumn{2}{c}{\large \bf Box run with decaying turbulence}\\ 
\includegraphics[width=0.23\linewidth]{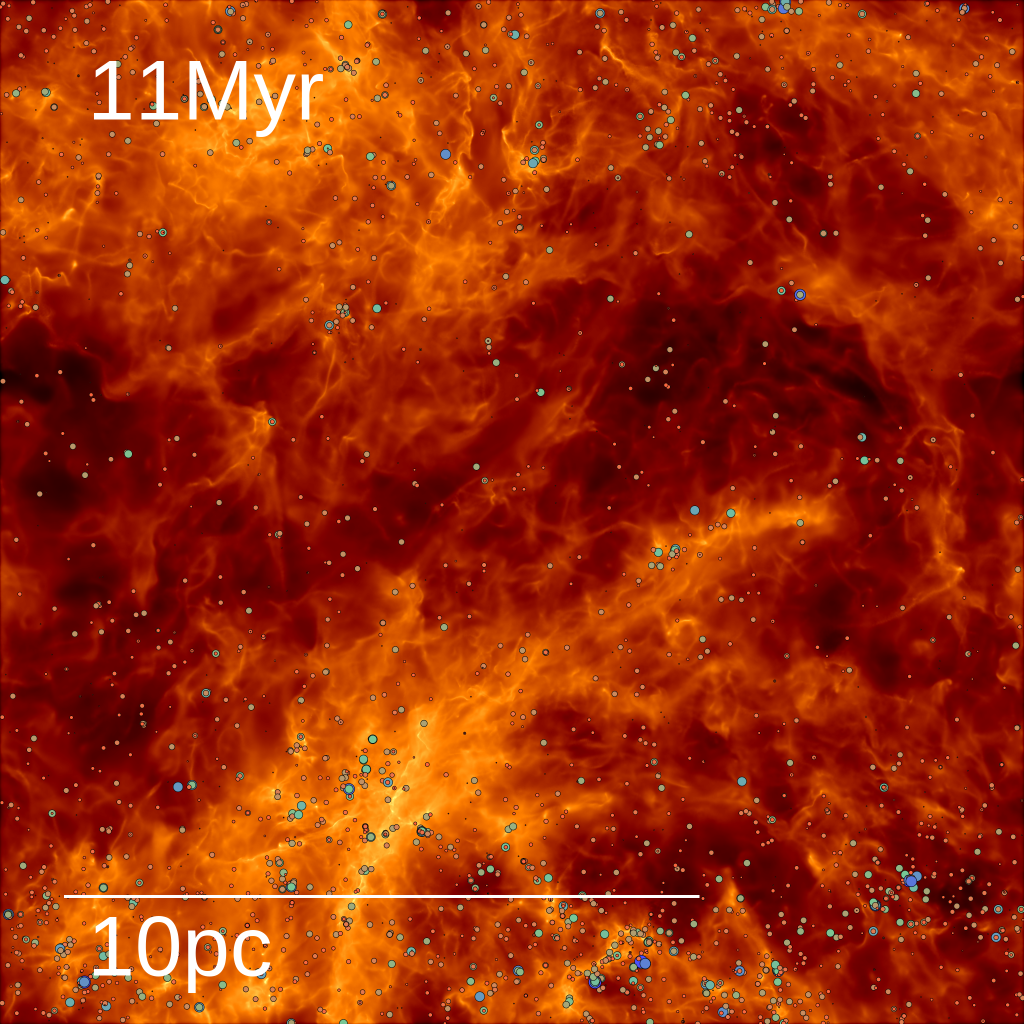} &
\includegraphics[width=0.24\linewidth]{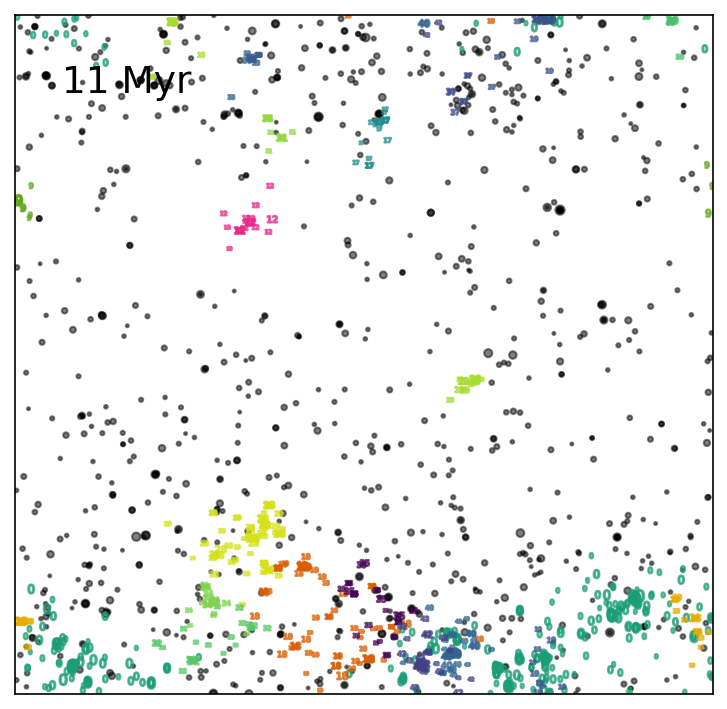} &
\includegraphics[width=0.23\linewidth]{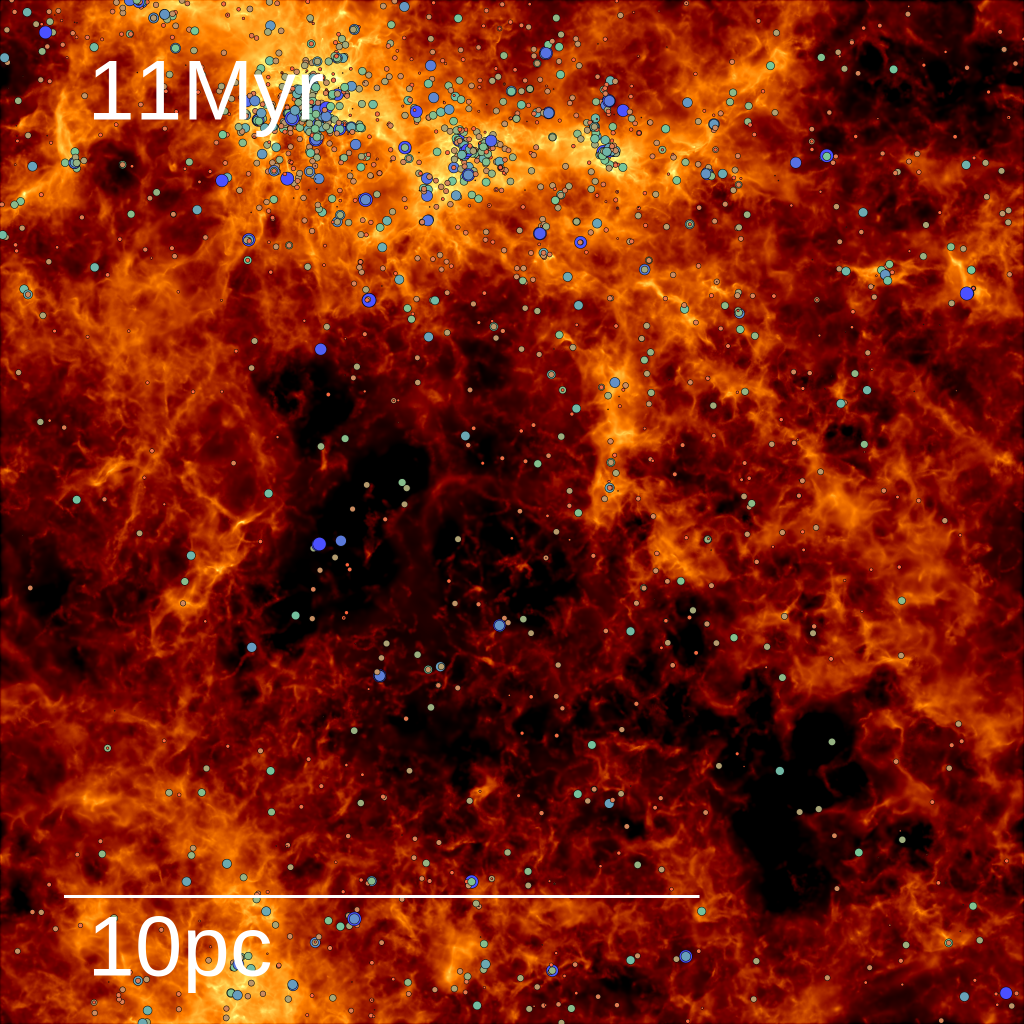} &
\includegraphics[width=0.24\linewidth]{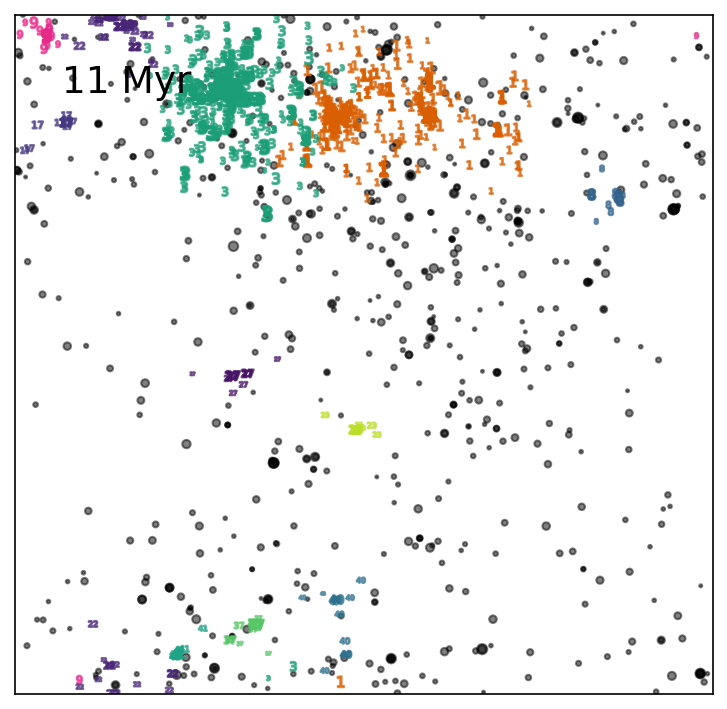} \\
\end{tabular}
%\vspace{-0.4cm}
\caption{Same as Figure \ref{fig:cluster_assignment_compare_box} for \textbf{M2e4} Box runs with and without turbulent driving.}
\label{fig:cluster_assignment_compare_driving}
\vspace{-0.5cm}
\end {center}
\end{figure*}

\begin{figure}
\begin {center}
\includegraphics[width=\linewidth]{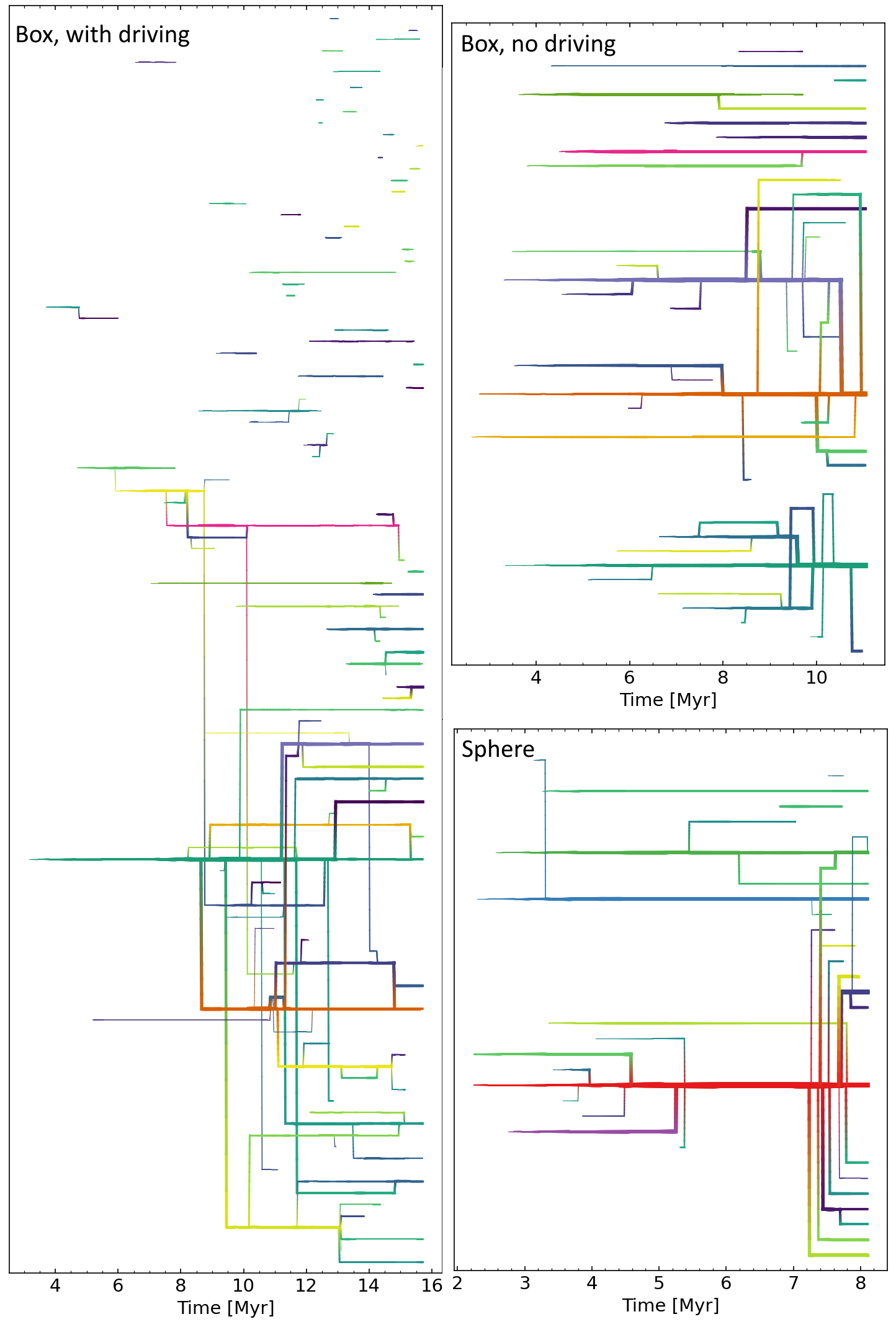}
\vspace{-0.4cm}
\caption{Merger histories for the \textbf{M2e4} Box runs with and without turbulent driving and the fiducial Sphere run, same as Figure \ref{fig:merger_history}. Driving prevents the hierarchical merging of smaller clusters and leads to the formation of many transient clusters. The periodic boundary conditions of the Box run also reduce gravitational focusing relative to the Sphere run, suppressing mergers.}
\label{fig:box_merger_history_compare}
\vspace{-0.5cm}
\end {center}
\end{figure}

%SO new paragraph
Figure \ref{fig:cluster_assignment_compare_box} shows that while the evolution of the Sphere run is well described as the hierarchical assembly of one dominant cluster, this is not the case in the driven Box run. Turning off turbulent driving restores this behavior, as the gas undergoes global gravitational collapse once the initial turbulence decays (see Figure \ref{fig:cluster_assignment_compare_driving}). Note that even without turbulent driving the shallower gravitational potential of the Box run relative to the Sphere run leads to weaker gravitational focusing, delaying mergers (see Figure \ref{fig:box_merger_history_compare}). We also find that continued turbulent driving leads to the formation of a significant number of transient clusters that survive for a few 100 kyr before dissolving.

\begin{figure*}
\begin {center}
\includegraphics[width=0.33\linewidth]{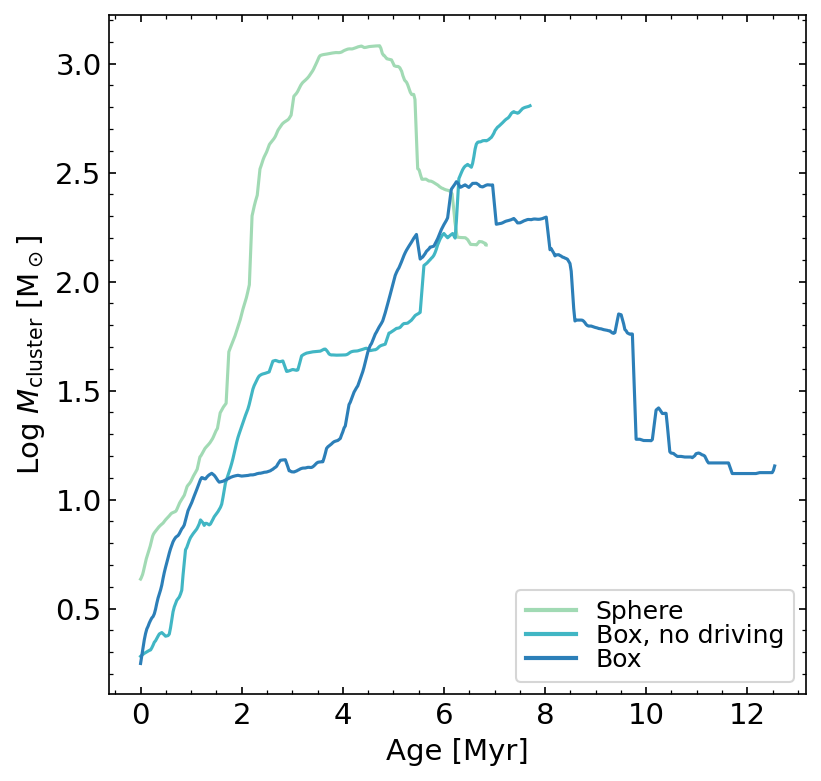}
\includegraphics[width=0.33\linewidth]{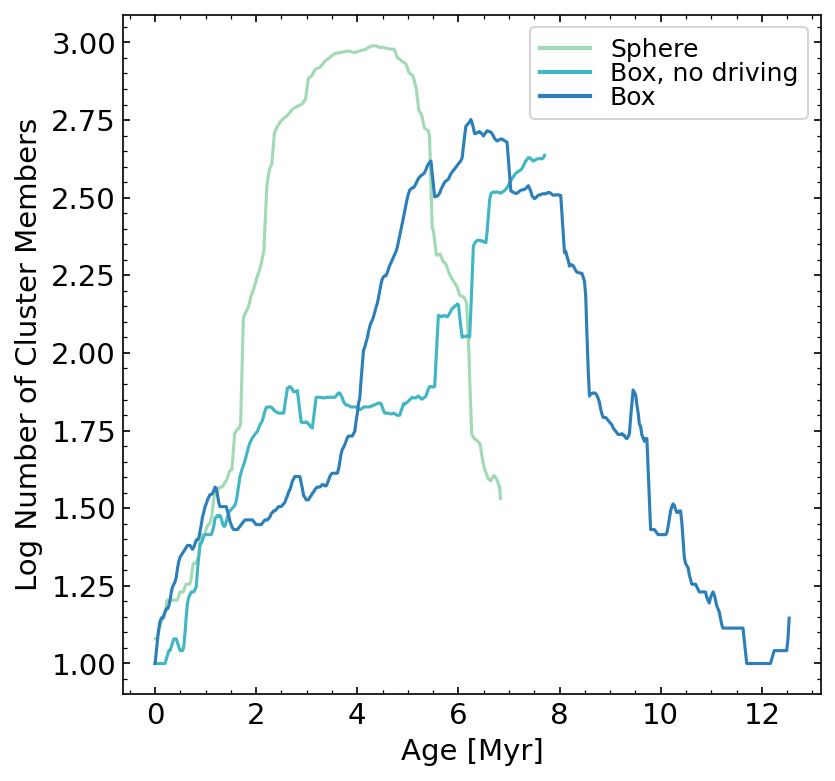}
\includegraphics[width=0.33\linewidth]{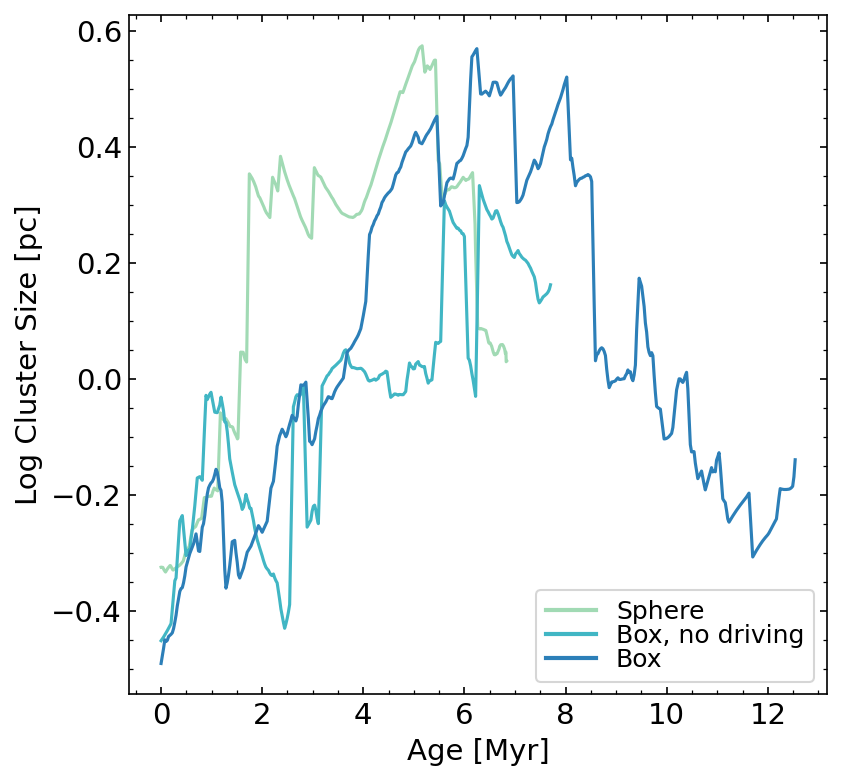}\\
\includegraphics[width=0.33\linewidth]{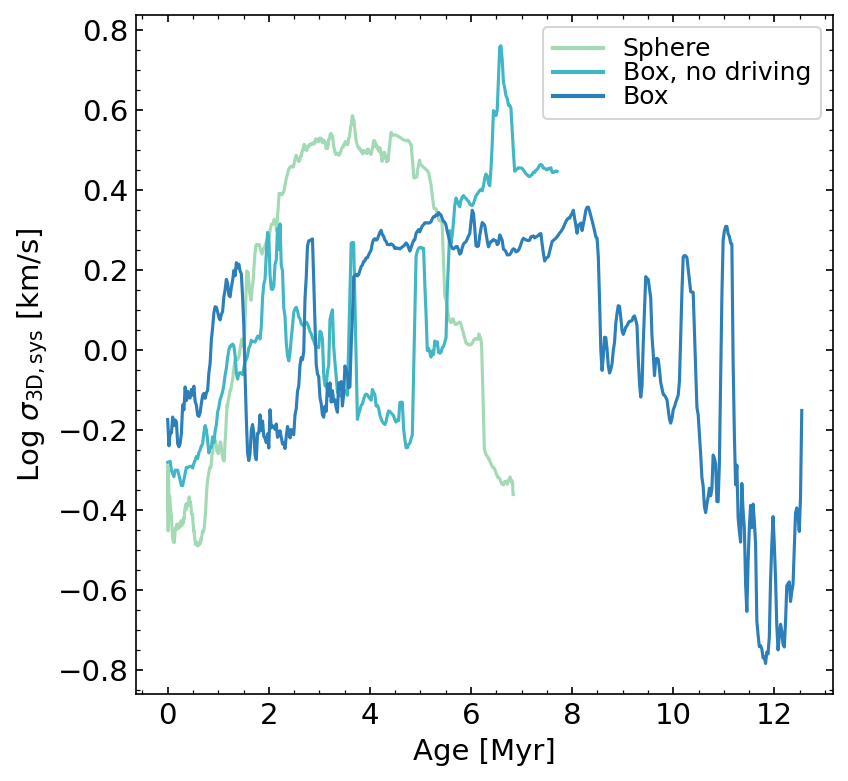}
\includegraphics[width=0.33\linewidth]{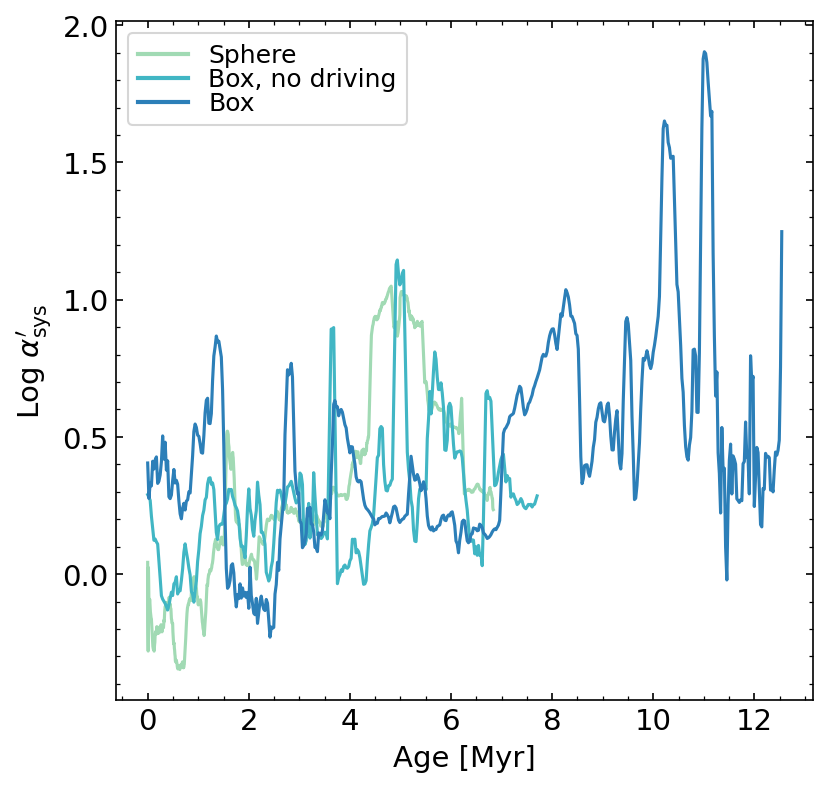}
\includegraphics[width=0.33\linewidth]{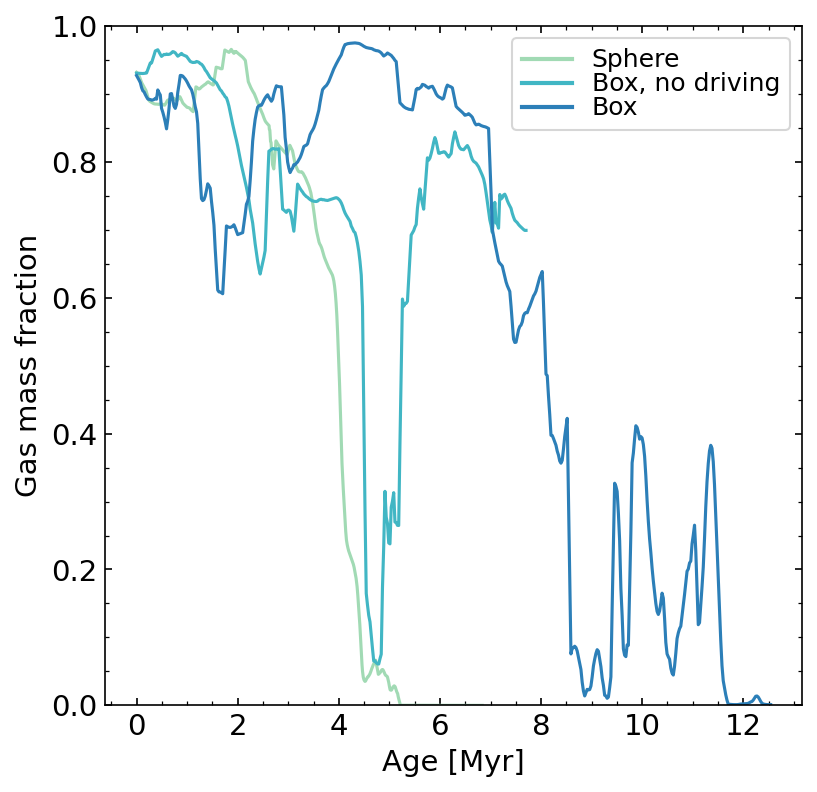}\\
\includegraphics[width=0.33\linewidth]{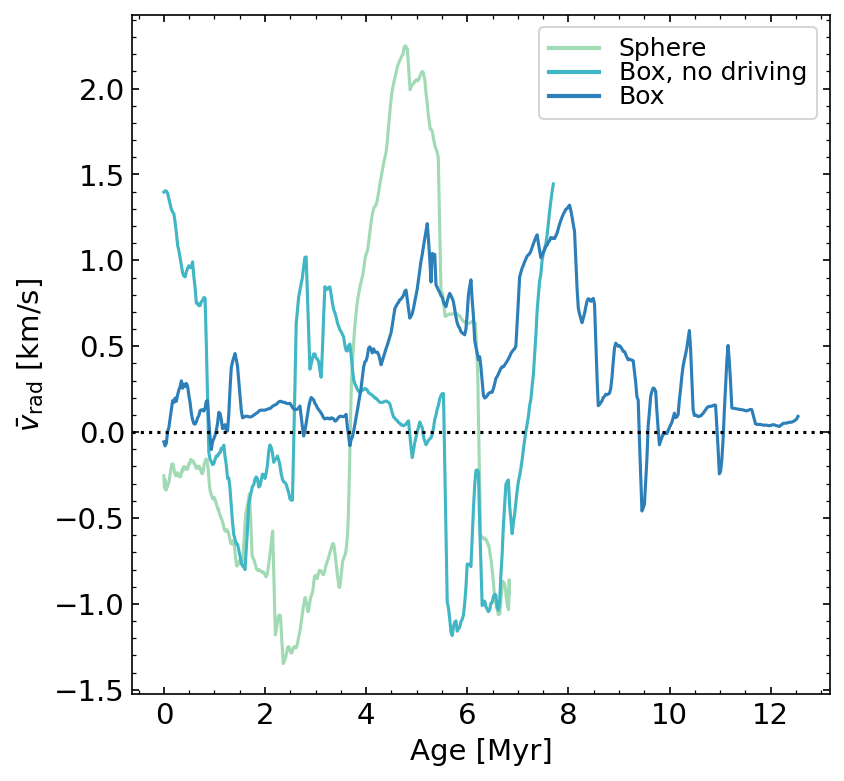}
\includegraphics[width=0.33\linewidth]{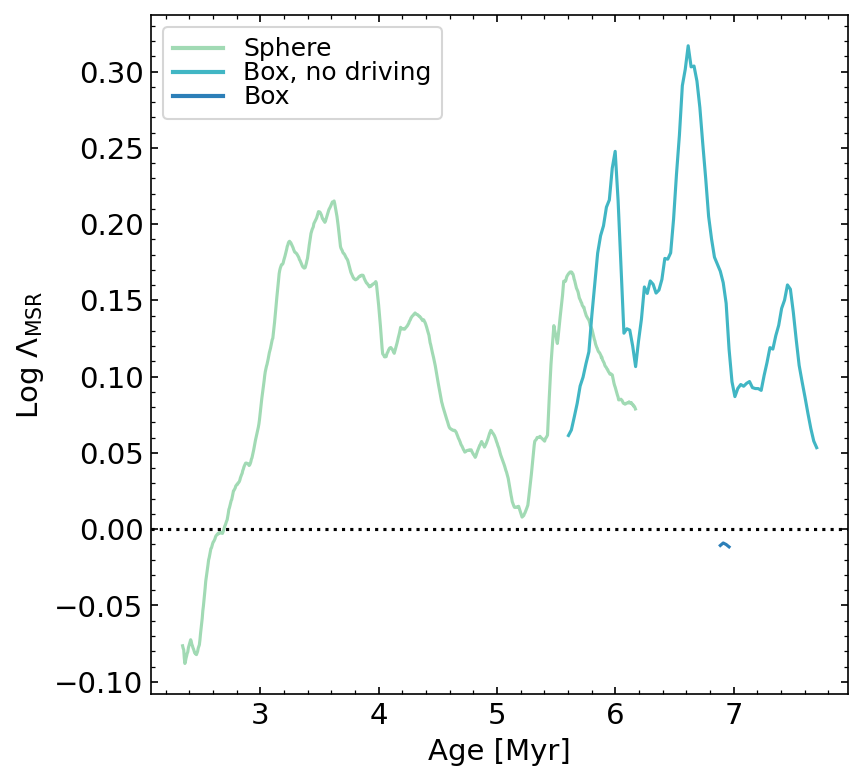}
\includegraphics[width=0.33\linewidth]{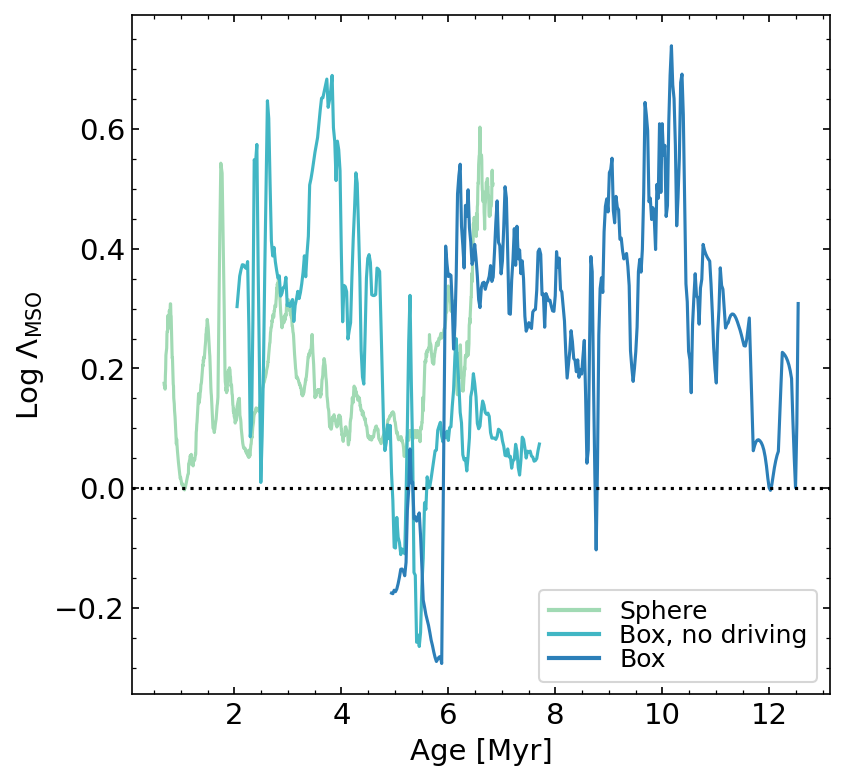}
\vspace{-0.5cm}
\caption{Evolution of the dominant cluster in a set of runs using both Sphere and Box initial conditions for \textbf{M2e4} (see Table \ref{tab:IC_phys}). The \emph{top  row} shows the total stellar mass in the cluster (left), number of cluster members (middle) and the cluster size scales (right, Eq. \ref{eq:cluster_R})\add{,} respectively. 
The \emph{middle  row} shows the cluster velocity dispersion (left, Eq. \ref{eq:cluster_sigma}), virialization state (middle, Eq. \ref{eq:alpha_sys}) and the gas mass fraction within the cluster radius (like Figure \ref{fig:cluster_basic_evol_sphere} panel e). The \emph{bottom row} shows the $\bar{v}_\mathrm{rad}$ mass-weighted mean radial velocity for the cluster (left, Eq. \ref{eq:radial_v}), as well as the mass segregation ratio (right, Eq. \ref{eq:lambda_msr}) and mass segregation offset (right, Eq. \ref{eq:mso}). Note defining $\MSR$ requires a minimum of 5 massive stars to be in the cluster, while $\MSO$ requires only a single massive star. All values are smoothed with a 50 kyr averaging window to make the plots easier to read. For an analysis of the main trends see \S\ref{sec:variations_results_box} in the main text.}
\label{fig:compare_clusters_box}
\vspace{-0.5cm}
\end {center}
\end{figure*}

%In \citetalias{guszejnov_starforge_imf} we show that 
Turbulent driving dramatically slows down star formation in the cloud ($\SFE\propto t^2$ vs $\SFE\propto t^3$), while in the non-driven case star formation is only suppressed until the initial turbulent velocity field decays. This is due to the weaker gravitational potential in the Box runs \citep{federrath_sim_2012}, which produces weaker gravitational focusing in addition to the external driving that prevents global gravitational collapse. This is apparent in Figure \ref{fig:compare_clusters_box} as the dominant cluster in both Box runs grow significantly slower than in the Sphere one. While the initial cluster masses are similar between the driven and decaying runs, the decaying run has (on average) more massive members due to the slightly less top heavy IMF in the Box runs. Unlike the Sphere run, the Box runs experience no cloud disruption, and stellar feedback is unable to permanently expel gas from the cluster before the gas in the simulation volume is heated to unphysical temperatures by radiation trapped by the periodic boundary condition. There is no permanent gas expulsion, the clusters themselves do not suddenly become unbound (like in the Sphere run). Their future evolution, however, is uncertain as we stop the simulation when it reaches the unphysical, radiation-filled regime.

\subsection{Initial level of turbulence}\label{sec:variations_results_alpha}

\begin{figure*}
\setlength\tabcolsep{0.0pt} %compress table
\begin {center}
\begin{tabular}{cccc}
\multicolumn{2}{c}{\large \bf $\alphaturb=1$ (M2e4\_a1)} & \multicolumn{2}{c}{\large \bf  $\alphaturb=4$ (M2e4\_a4)}\\ 
\includegraphics[width=0.23\linewidth]{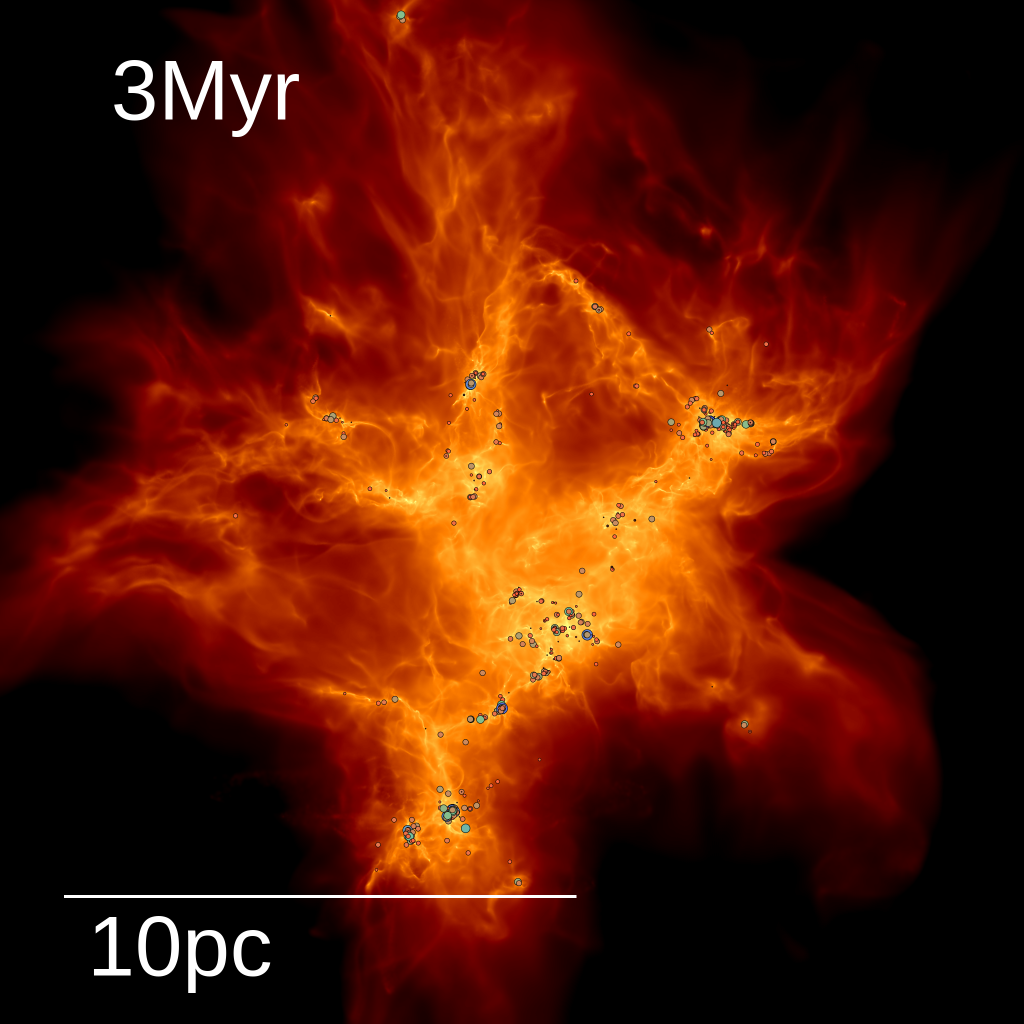} &
\includegraphics[width=0.24\linewidth]{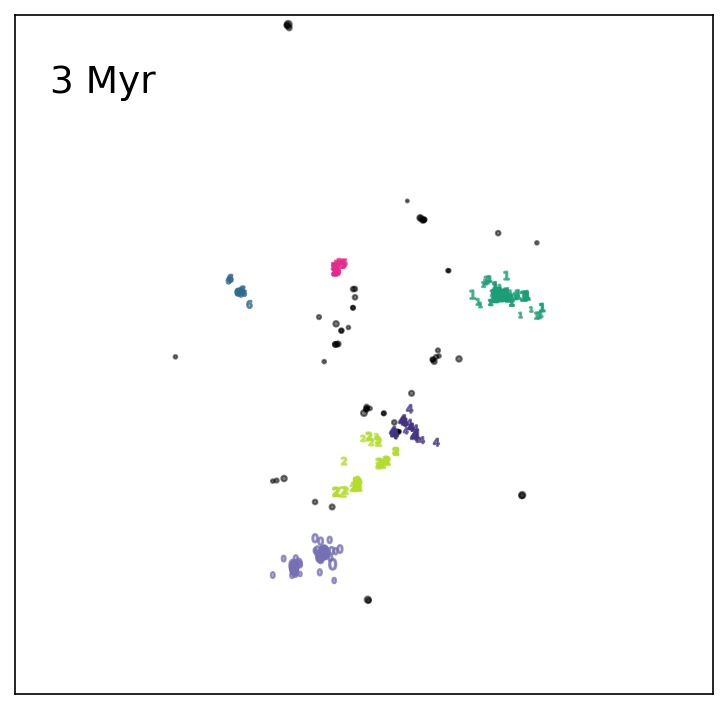} &
\includegraphics[width=0.23\linewidth]{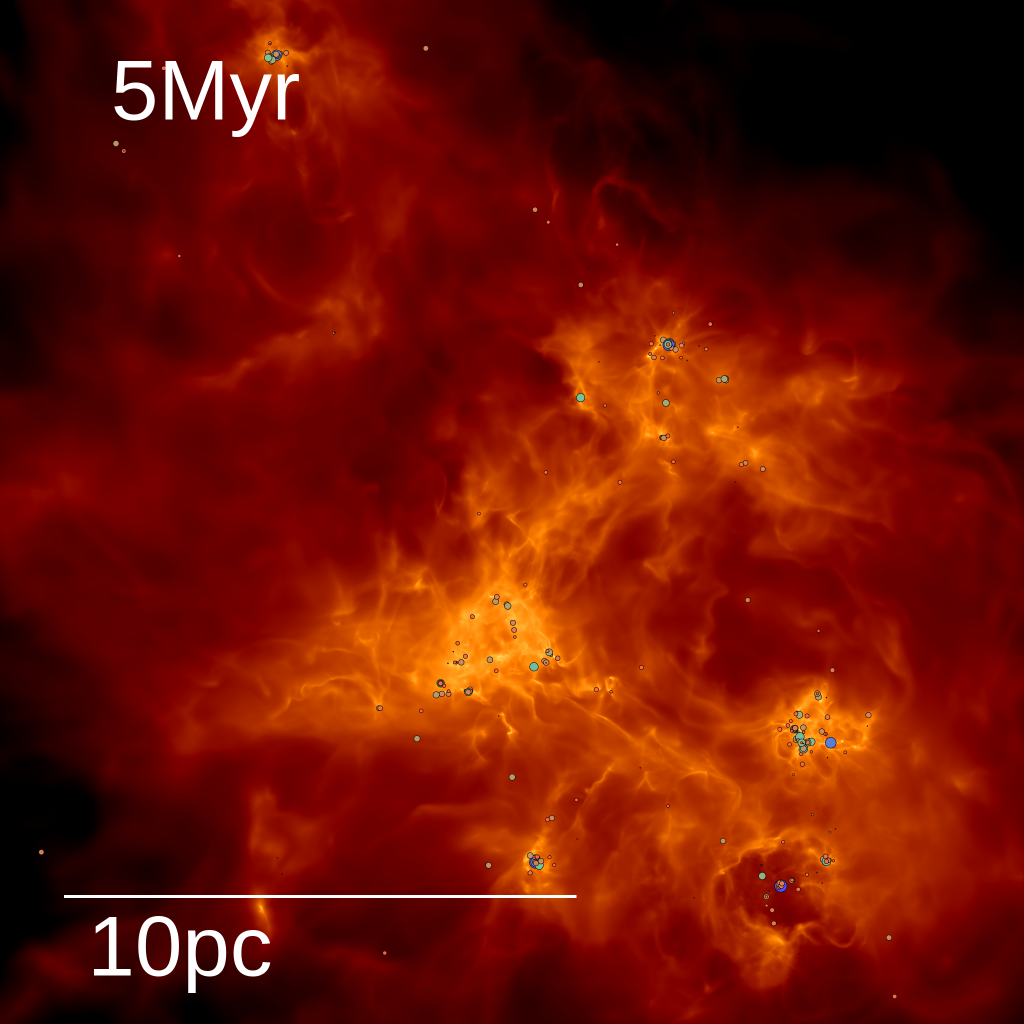} &
\includegraphics[width=0.24\linewidth]{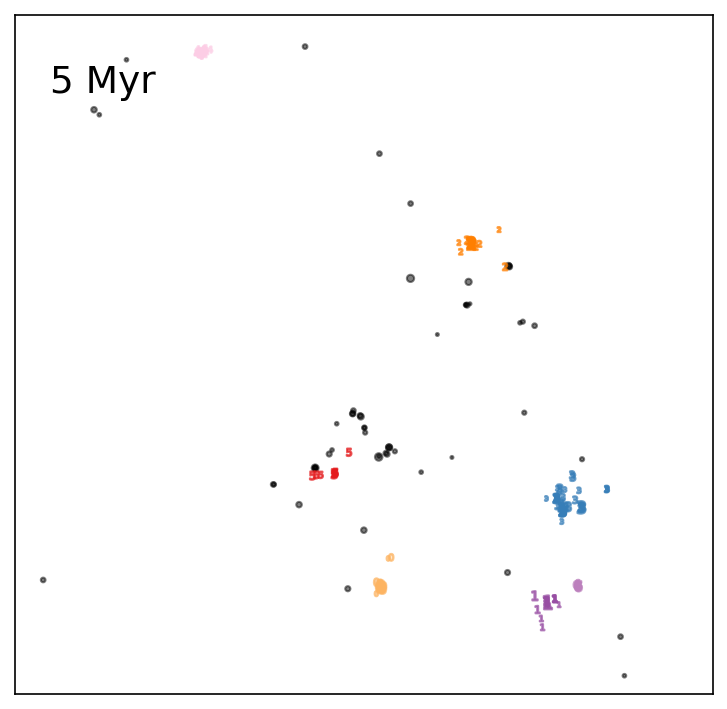} \\

\includegraphics[width=0.23\linewidth]{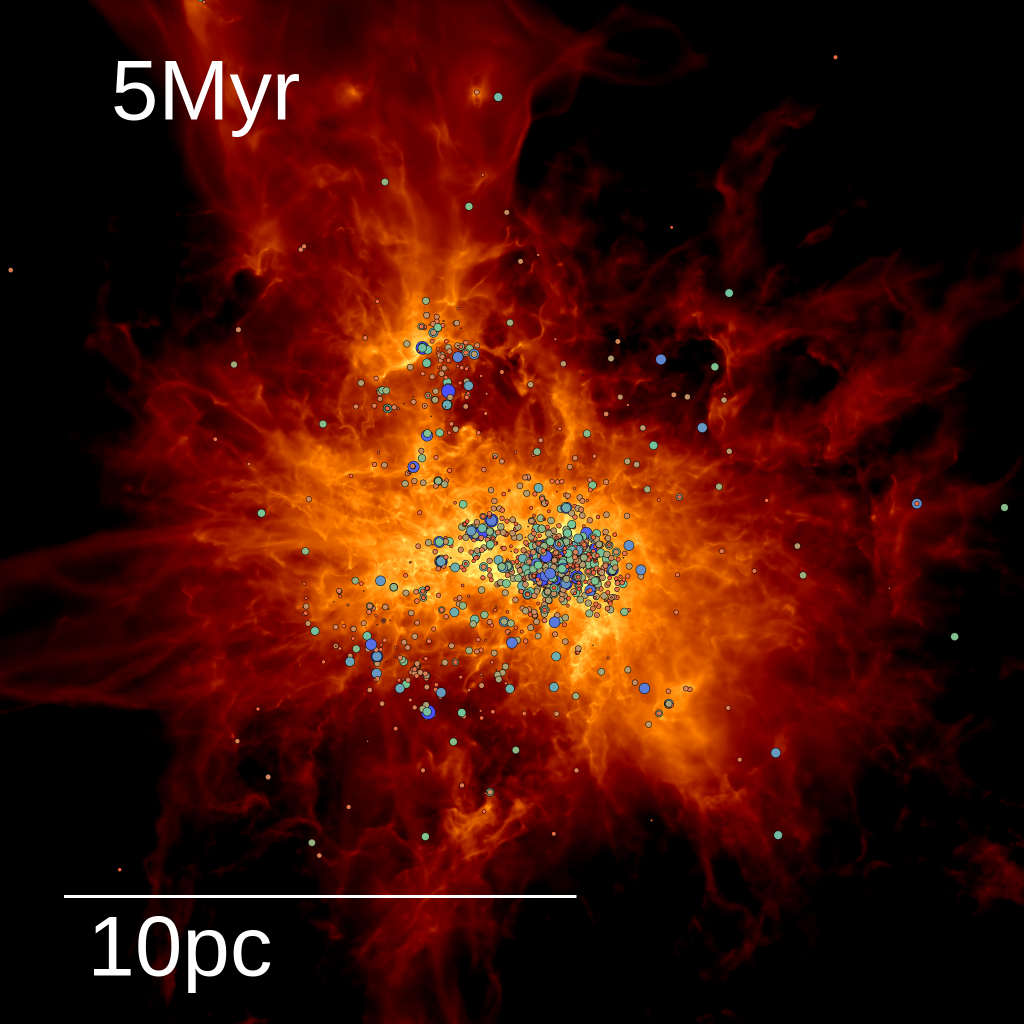} &
\includegraphics[width=0.24\linewidth]{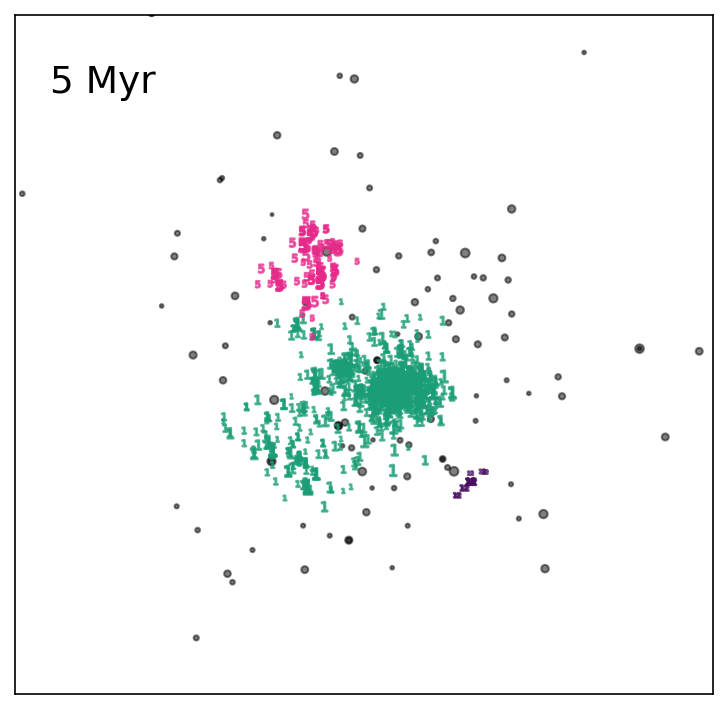} &
\includegraphics[width=0.23\linewidth]{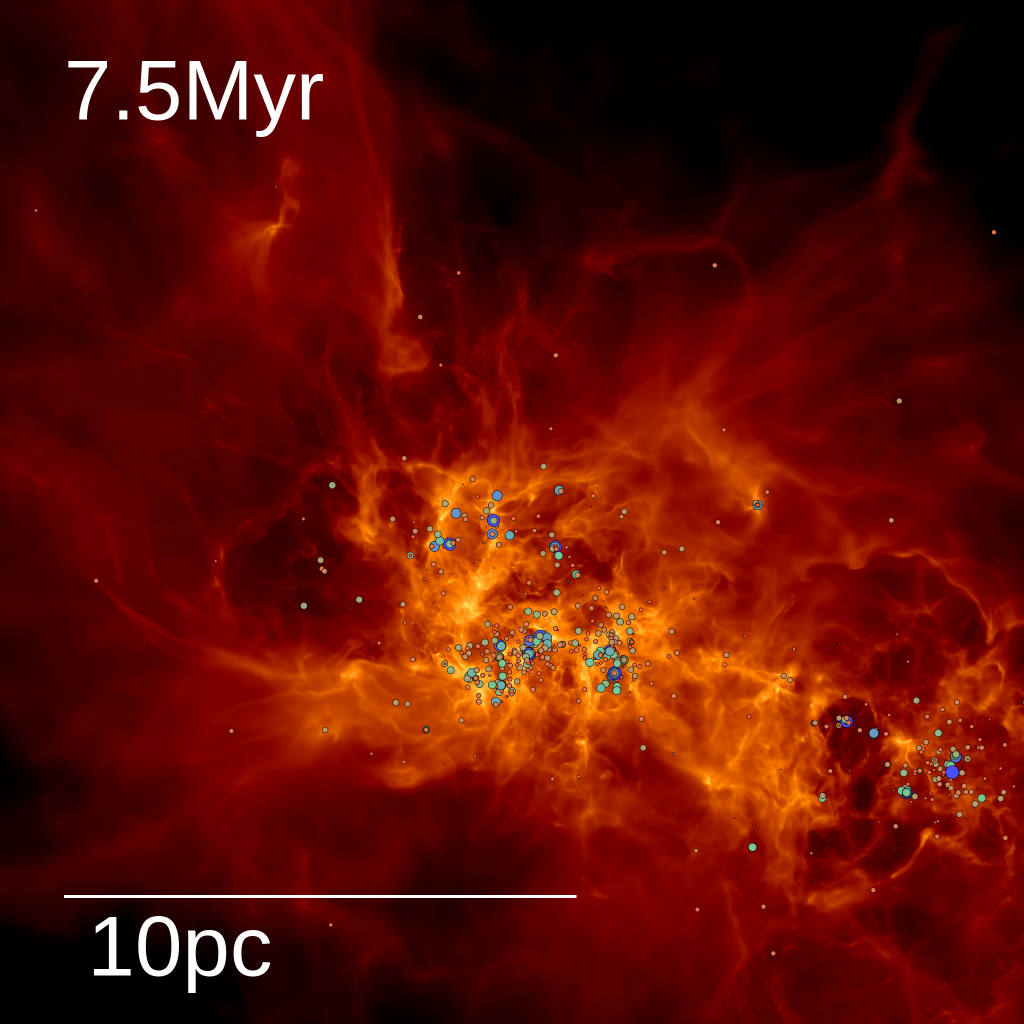} &
\includegraphics[width=0.24\linewidth]{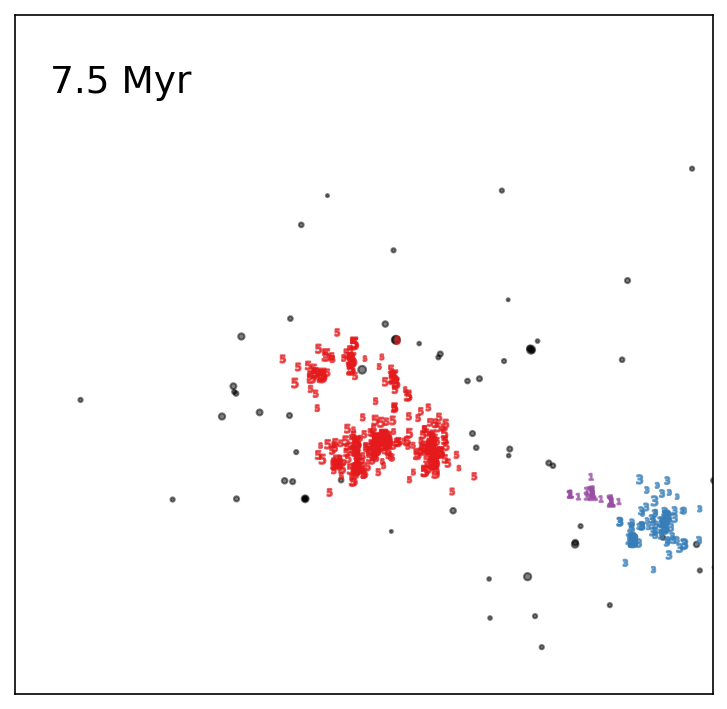} \\

\includegraphics[width=0.23\linewidth]{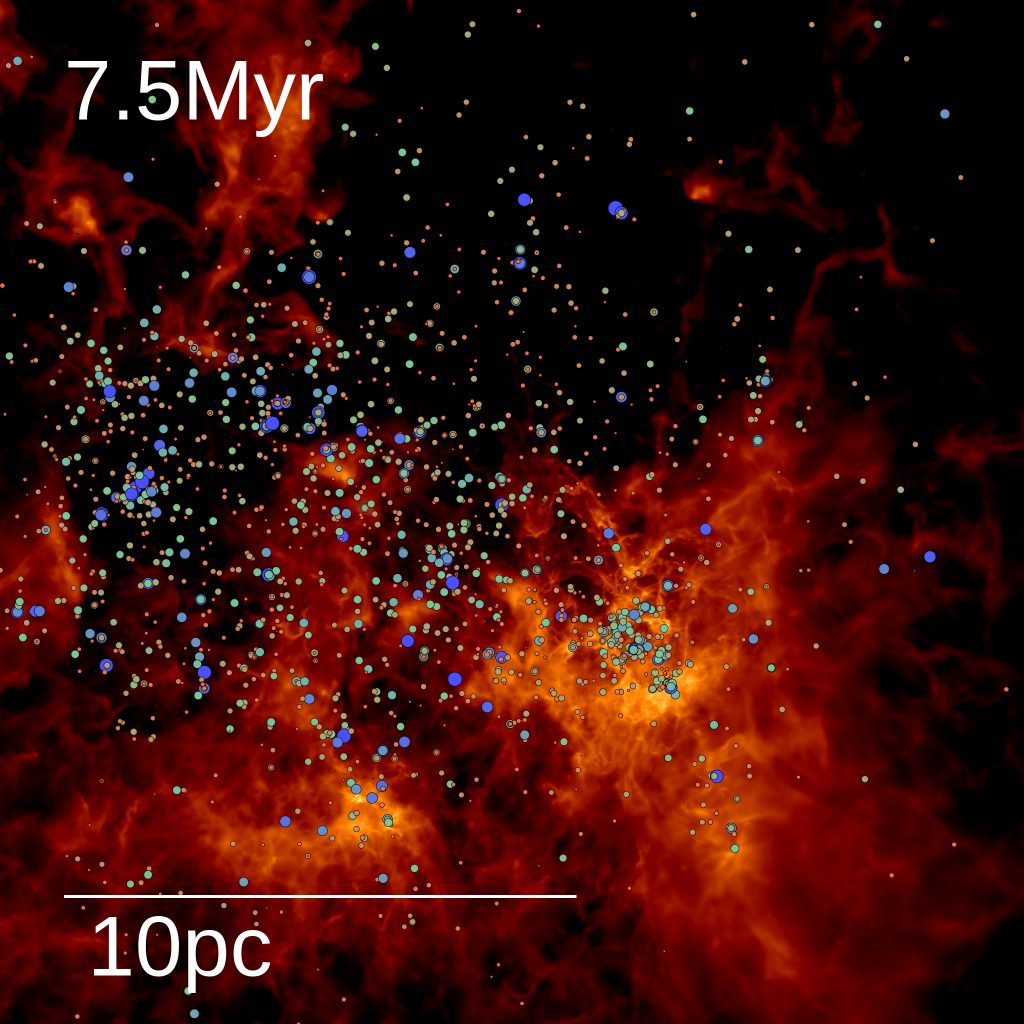} &
\includegraphics[width=0.24\linewidth]{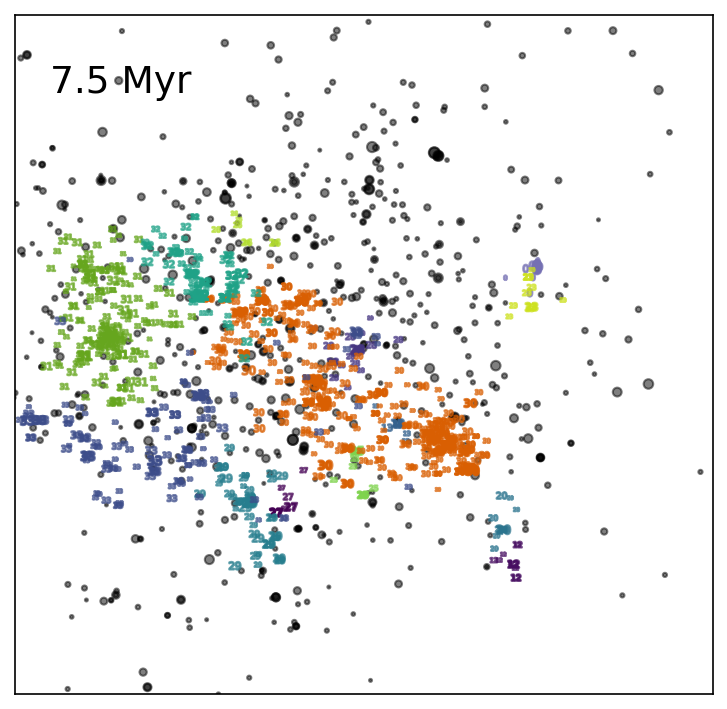} &
\includegraphics[width=0.23\linewidth]{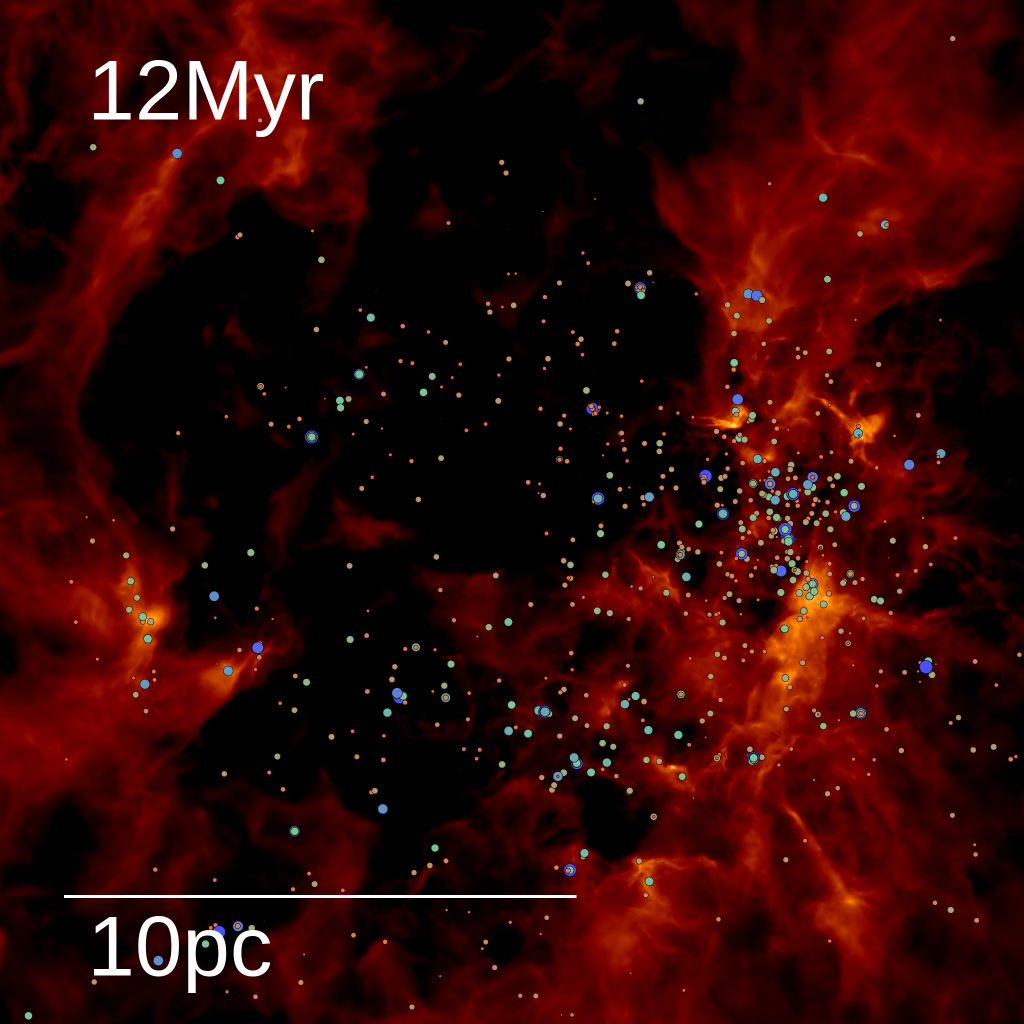} &
\includegraphics[width=0.24\linewidth]{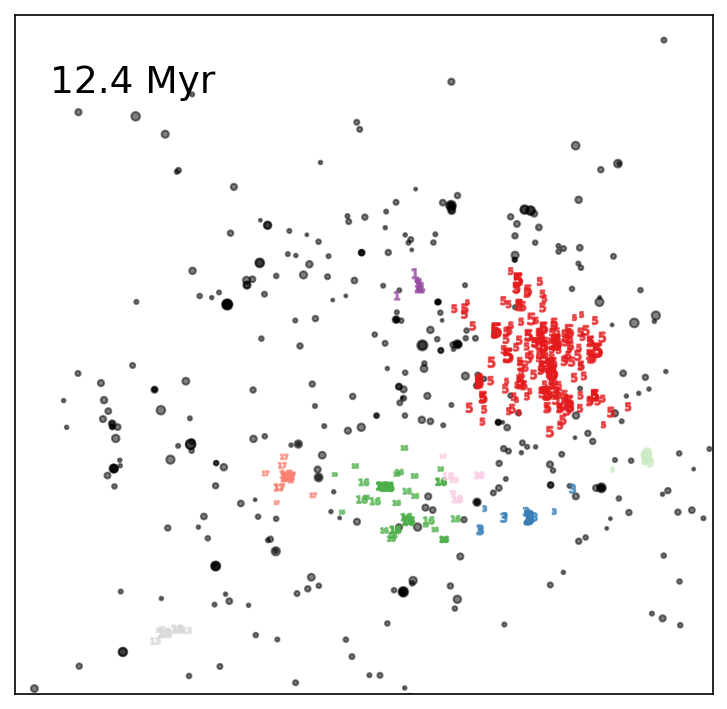} \\
\end{tabular}
%\vspace{-0.4cm}
\caption{Same as Figure \ref{fig:cluster_assignment_compare_box} but for runs with initial turbulent virial parameters of $\alphaturb=1$ and $\alphaturb=4$ respectively (\textbf{M2e4\_a1} and \textbf{M2e4\_a4}).}
\label{fig:cluster_assignment_compare_alpha}
\vspace{-0.5cm}
\end {center}
\end{figure*}

%SO This needs some kind of intro sentence. I rewrote this -- it was a franken-sentence.
%By analysing Sphere runs with different levels of initial turbulence by varying the initial gas velocity dispersion, by comparing runs with $\alphaturb$ values of 1,2 and 4, corresponding to nominally bound, marginally bound and unbound clouds. 
In this section we vary the initial velocity dispersion to determine the impact of the cloud turbulence on clustering. We compare Sphere runs with $\alphaturb$ values of 1,2 and 4, which correspond to bound, marginally bound and unbound clouds, respectively.
We find that the final star formation efficiency decreases with increasing $\alphaturb$ (see Table \ref{tab:final_properties}). Figure \ref{fig:cluster_assignment_compare_alpha} shows that higher $\alphaturb$ also leads to less concentrated star formation and weaker global gravitational collapse due to increased turbulent support. However, the hierarchical cluster formation picture that we find for the fiducial run ($\alphaturb=2$) still qualitatively applies (see Figures \ref{fig:cluster_assignment_compare_box} and \ref{fig:cluster_assignment_compare_alpha}). %\alr{(You can also add that the low alpha case does have clusters with a greater \# of stars and larger effective cluster sizes)}

\begin{figure*}
\begin {center}
\includegraphics[width=0.33\linewidth]{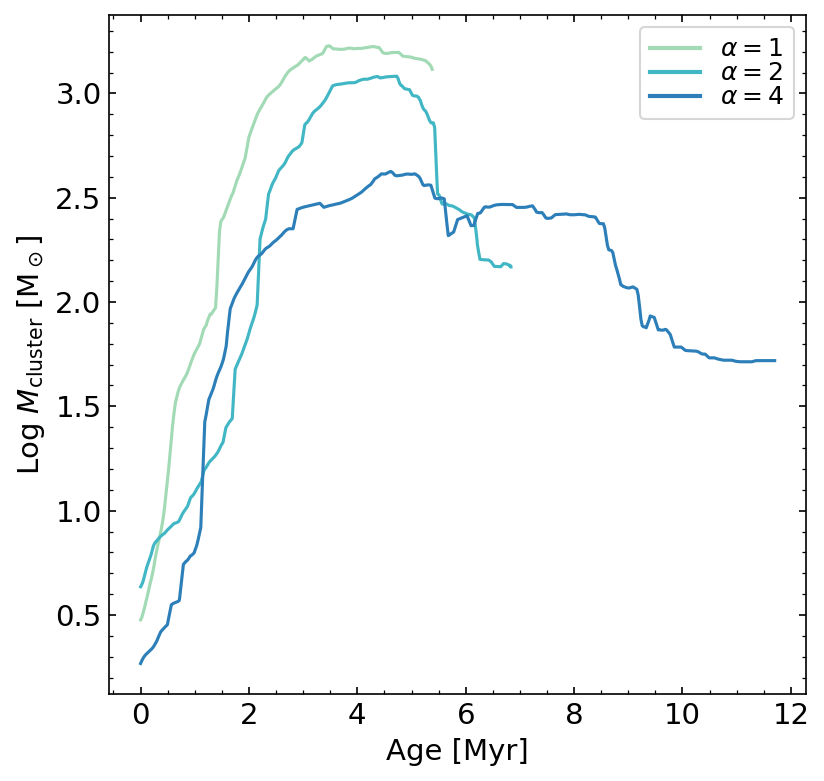}
\includegraphics[width=0.33\linewidth]{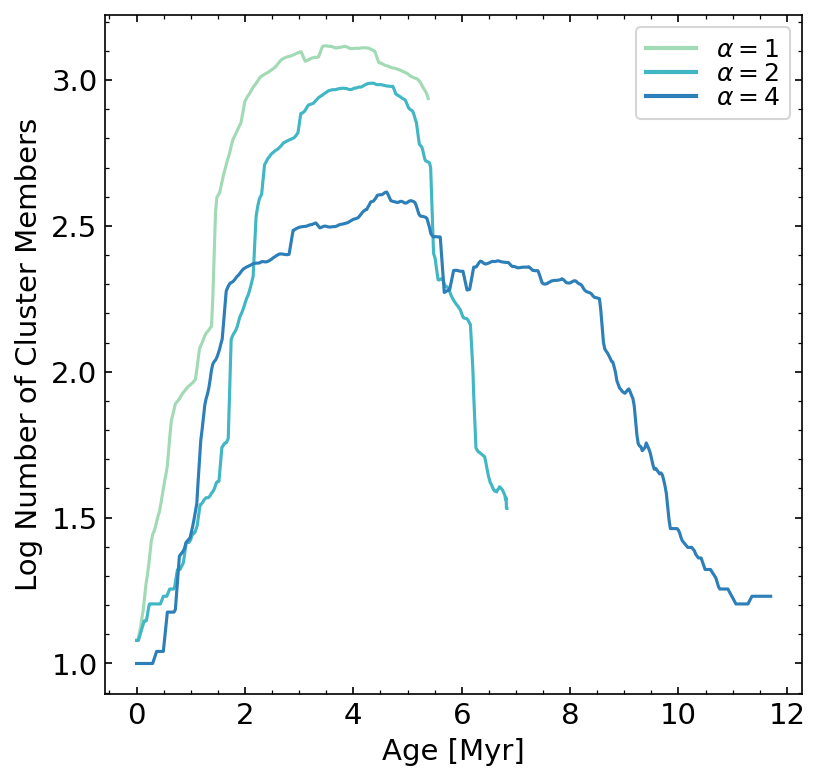}
\includegraphics[width=0.33\linewidth]{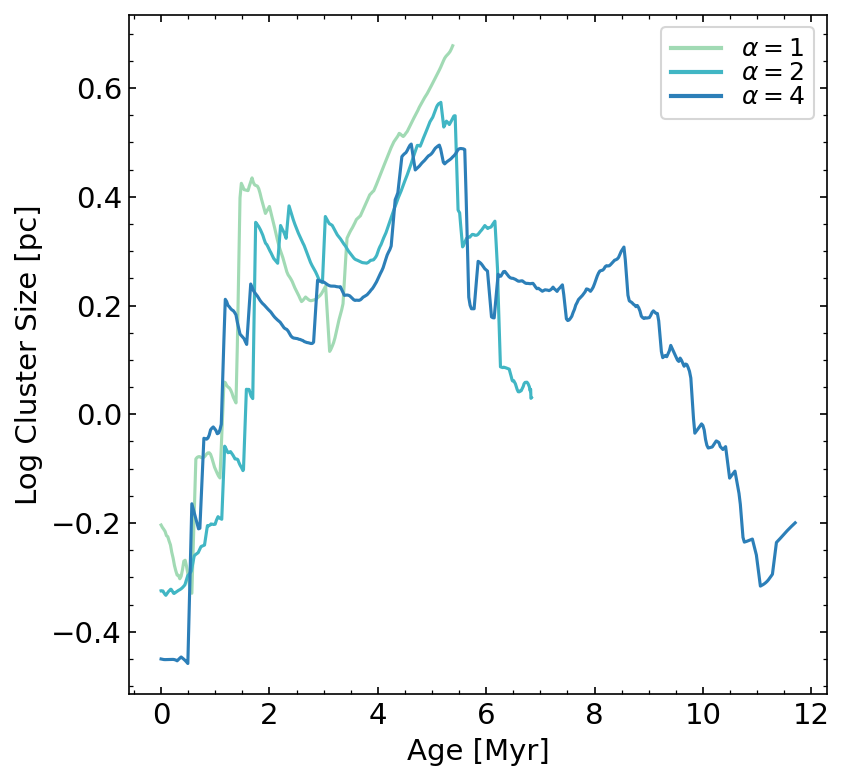}\\
\includegraphics[width=0.33\linewidth]{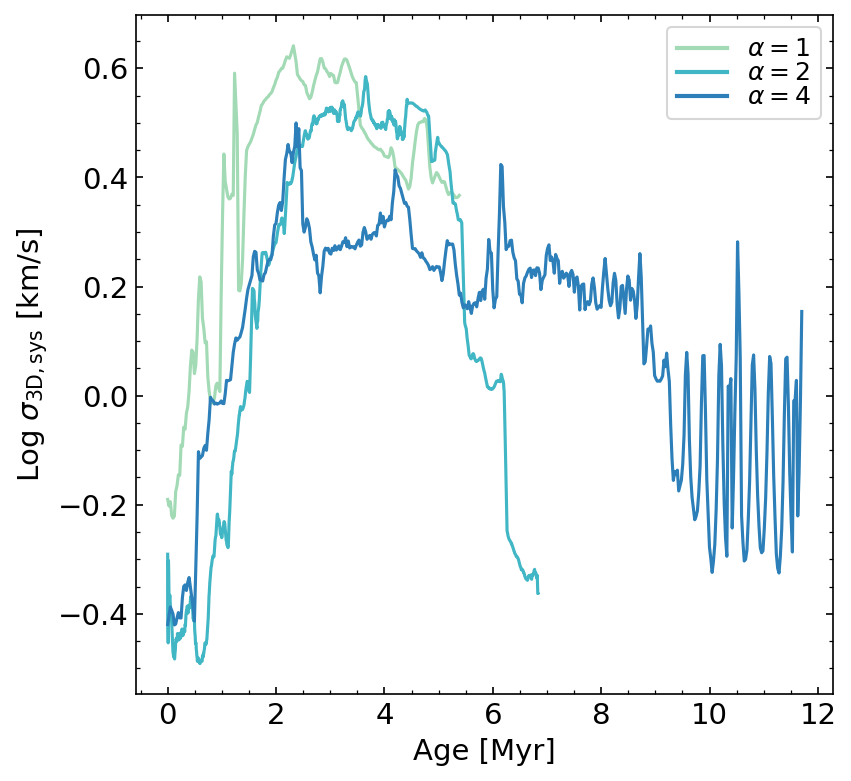}
\includegraphics[width=0.33\linewidth]{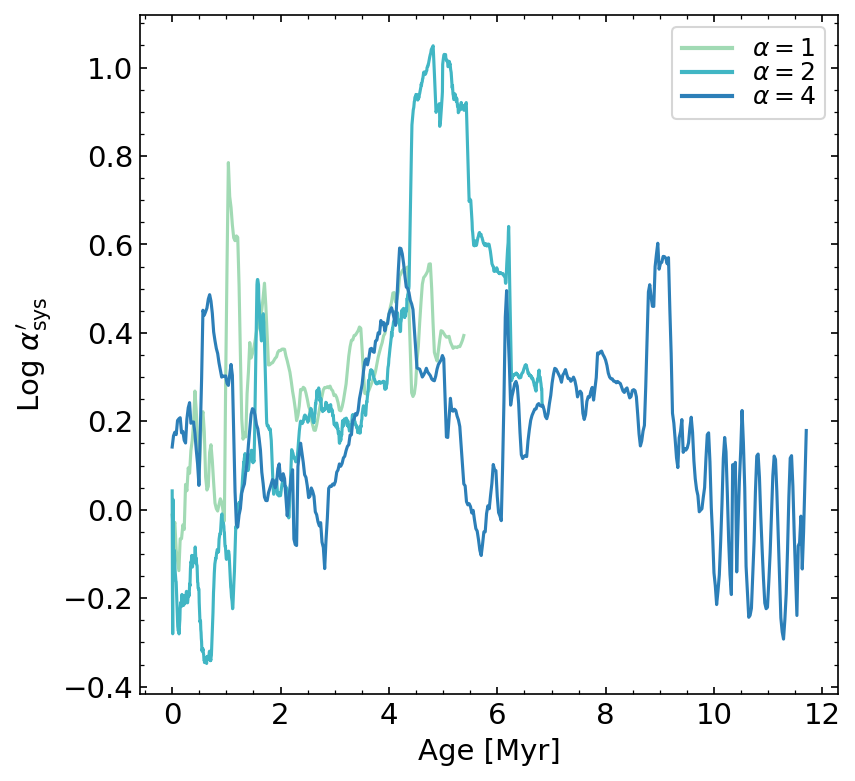}
\includegraphics[width=0.33\linewidth]{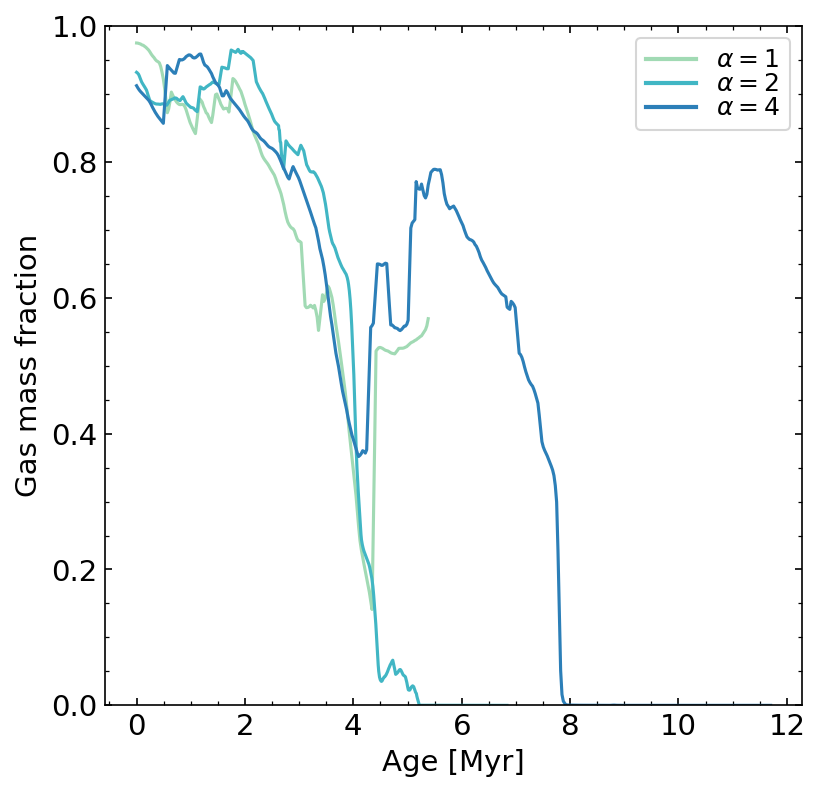}\\
\includegraphics[width=0.33\linewidth]{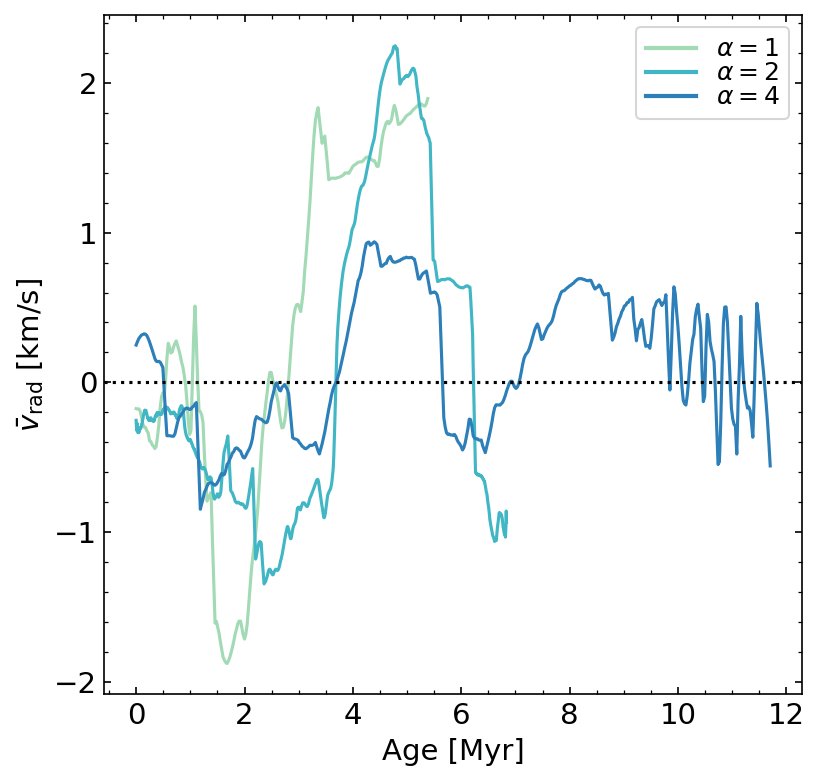}
\includegraphics[width=0.33\linewidth]{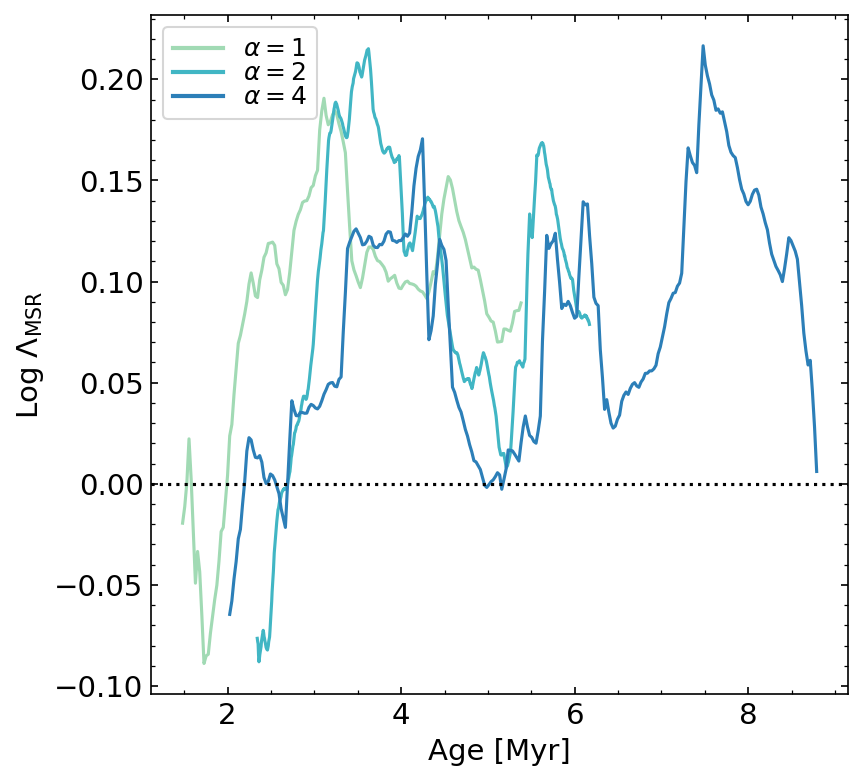}
\includegraphics[width=0.33\linewidth]{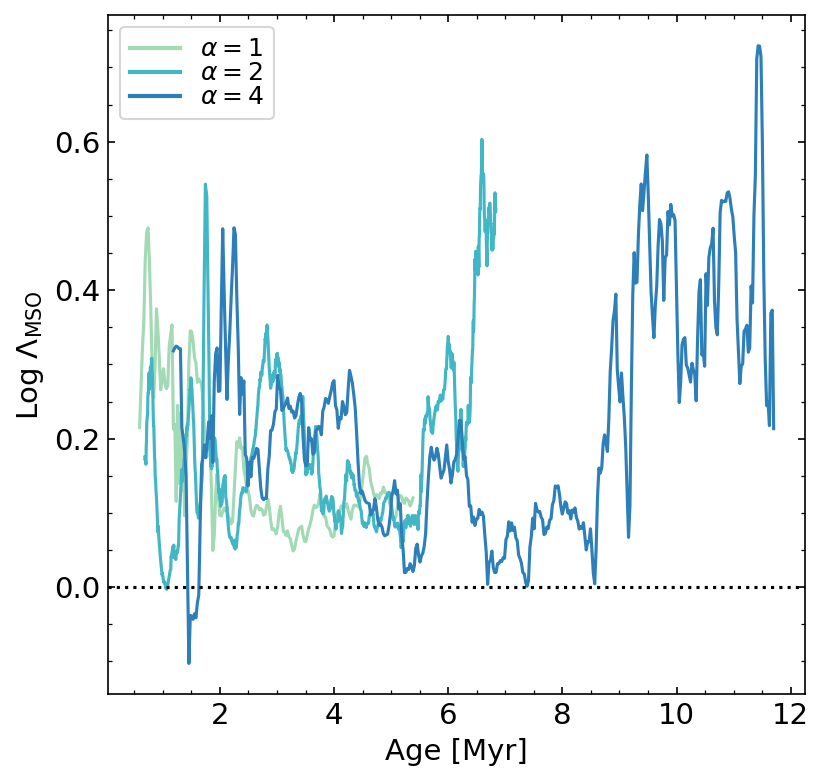}
\vspace{-0.4cm}
\caption{Evolution of the dominant cluster in a set of runs with different levels of initial turbulence (\textbf{M2e4\_a1}, \textbf{M2e4} and \textbf{M2e4\_a4}, see Table \ref{tab:IC_phys}), similar to Figure \ref{fig:compare_clusters_box}. The \emph{top  row} shows the total stellar mass in the cluster (left), number of cluster members (middle) and the cluster size scales (right, Eq. \ref{eq:cluster_R}) respectively. 
The \emph{middle  row} shows the cluster velocity dispersion (left, Eq. \ref{eq:cluster_sigma}), virialization state (middle, Eq. \ref{eq:alpha_sys}) and the gas mass fraction within the cluster radius (like Figure \ref{fig:cluster_basic_evol_sphere} panel e). The \emph{bottom row} shows the $\bar{v}_\mathrm{rad}$ mass-weighted mean radial velocity for the cluster (left, Eq. \ref{eq:radial_v}), as well as the mass segregation ratio (right, Eq. \ref{eq:lambda_msr}) and mass segregation offset (right, Eq. \ref{eq:mso}). For an analysis of the main trends see \S\ref{sec:variations_results_alpha} in the main text.}
\label{fig:compare_clusters_alpha}
\vspace{-0.5cm}
\end {center}
\end{figure*}

All runs produce a single dominant cluster. %SO Rearranged. Note this sentence was very similar to one in the above paragraph.
The final mass of this dominant cluster decreases with increased turbulence, mostly due to the lower final SFE values of the clouds (Table \ref{tab:final_properties}). Figure \ref{fig:compare_clusters_alpha} shows that the dominant cluster follows a qualitatively similar evolutionary trend in all runs, with higher masses and consequently higher velocity dispersions for runs with higher SFE (i.e., lower $\alphaturb$). 
The dominant cluster becomes unbound once stellar feedback expels the remaining gas, leading to its expansion and breakup into smaller clusters. In Figure \ref{fig:compare_clusters_alpha} it appears as if the $\alphaturb=4$ run had a much longer cluster lifetime. In fact, feedback causes the cluster to expel its gas content and become unbound in roughly the same time (4 Myr) as in the other cases (see gas mass fraction and $\alphag_\mathrm{sys}$ panels of Figure \ref{fig:compare_clusters_alpha}). At the same time the dominant cluster merges with two neighboring clusters, increasing the effective size of the resulting cluster and the relative gas mass content.

%SO New paragraph:
We find that all three runs are mass-segregated from early times, and we see no clear trend in either $\MSR$ or $\MSO$ as a function of $\alphaturb$. Therefore, we conclude that mass segregation is not very sensitive to modest changes in the initial cloud virial parameter.
%SO You already said this.
%\cut{, and therefore we find that all simulations broadly follow the same trends as the fiducial run.} %\alr{I'm a little confused by the gas mass fraction plot in Figure 13, what leads to the jump in the gas mass fraction? It increases for both the 1 and 4 runs, but not for the 2 run. Is this due to the volume you consider for measuring the gas mass fraction? If so, you should describe how you calculate this value.}\DG{They coincide with size jumps, so essentially the cluster merges with another that has gas in it.}

\subsection{Surface density}\label{sec:variations_results_sigma}

Cloud surface density is thought to be a key parameter of star formation \citep{krumholz08a, fkm2010, grudic_2020_cluster_formation} due to its influence on the dynamics of fragmentation and degree of stellar feedback. %\cut{the force balance of momentum feedback and gravity}. %SO I think this second part of the sentence is confusing (e.g., momentum feedback is not a force, gravity is covered by 'fragmentation')
Although we present only one run with a different surface density (Sphere run with a factor 10 times increase in $\Sigma$; \textbf{M2e4\_R3}), we also ran a calculation with 10 times lower surface density, but it had a final SFE value of only 1\% and produced no clusters with more than 20 stars, preventing a meaningful cluster analysis.% (see \citetalias{guszejnov_starforge_imf} for details on this run).

\begin{figure}
\setlength\tabcolsep{0.0pt} %compress table
\begin {center}
\begin{tabular}{cc}
\multicolumn{2}{c}{\large \bf $\Sigma=10\times\Sigma_\mathrm{MW}=630\msun/\pc^2$ (M2e4\_R3)} \\ 
\includegraphics[width=0.46\linewidth]{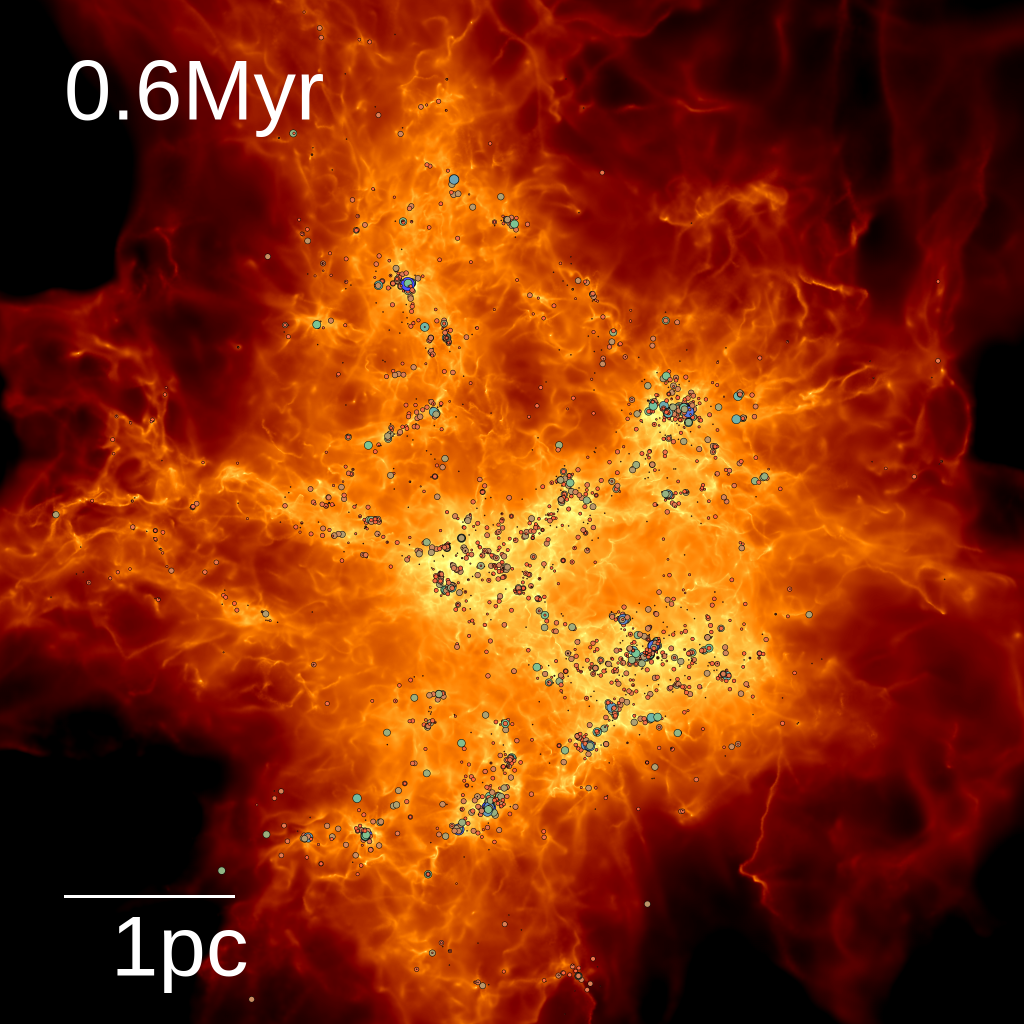} &
\includegraphics[width=0.48\linewidth]{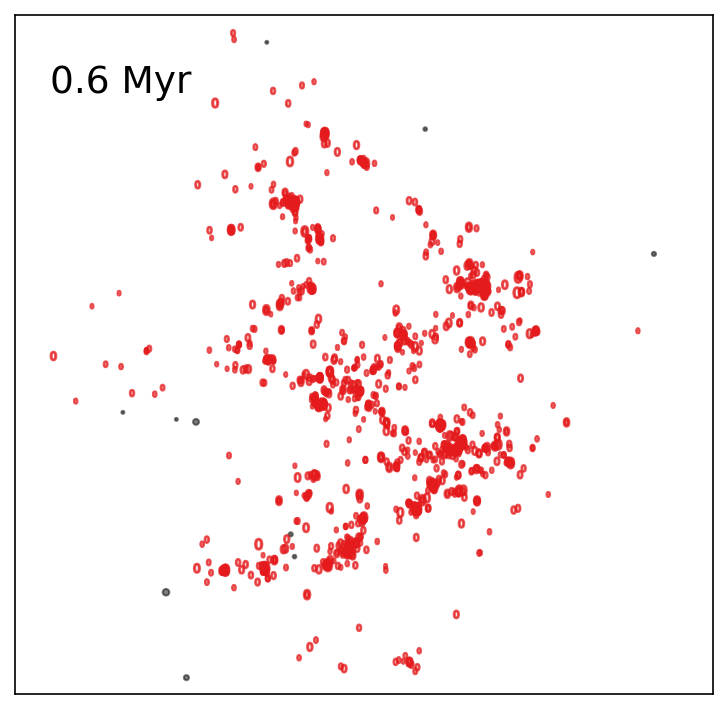} \\

\includegraphics[width=0.46\linewidth]{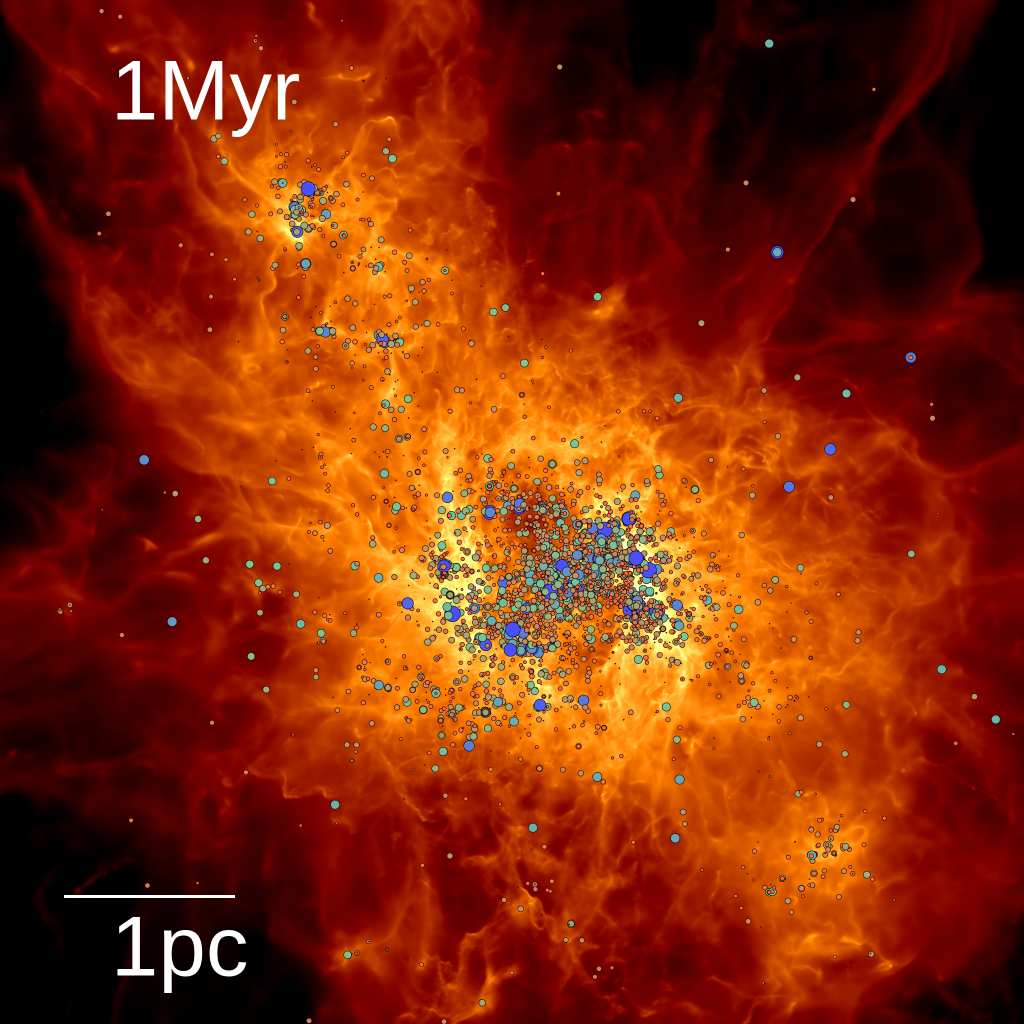} &
\includegraphics[width=0.48\linewidth]{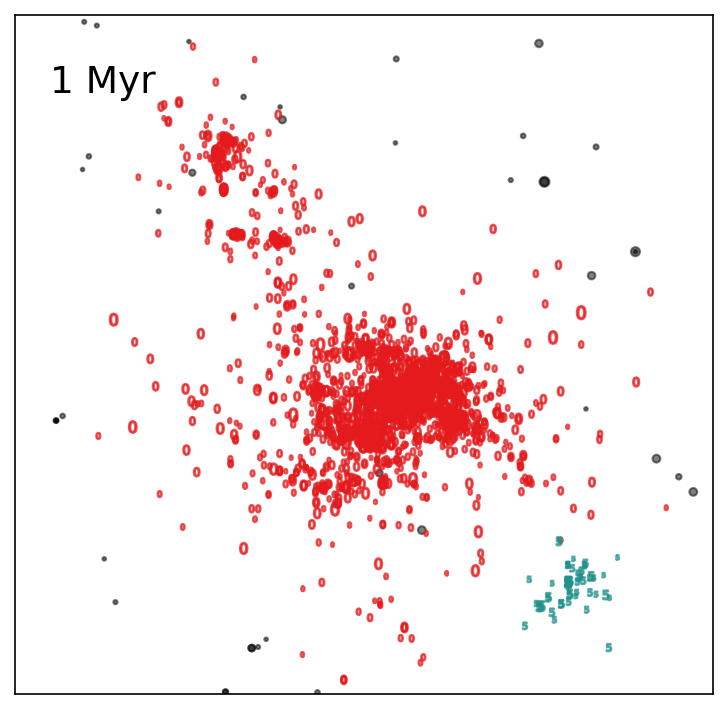} \\

\includegraphics[width=0.46\linewidth]{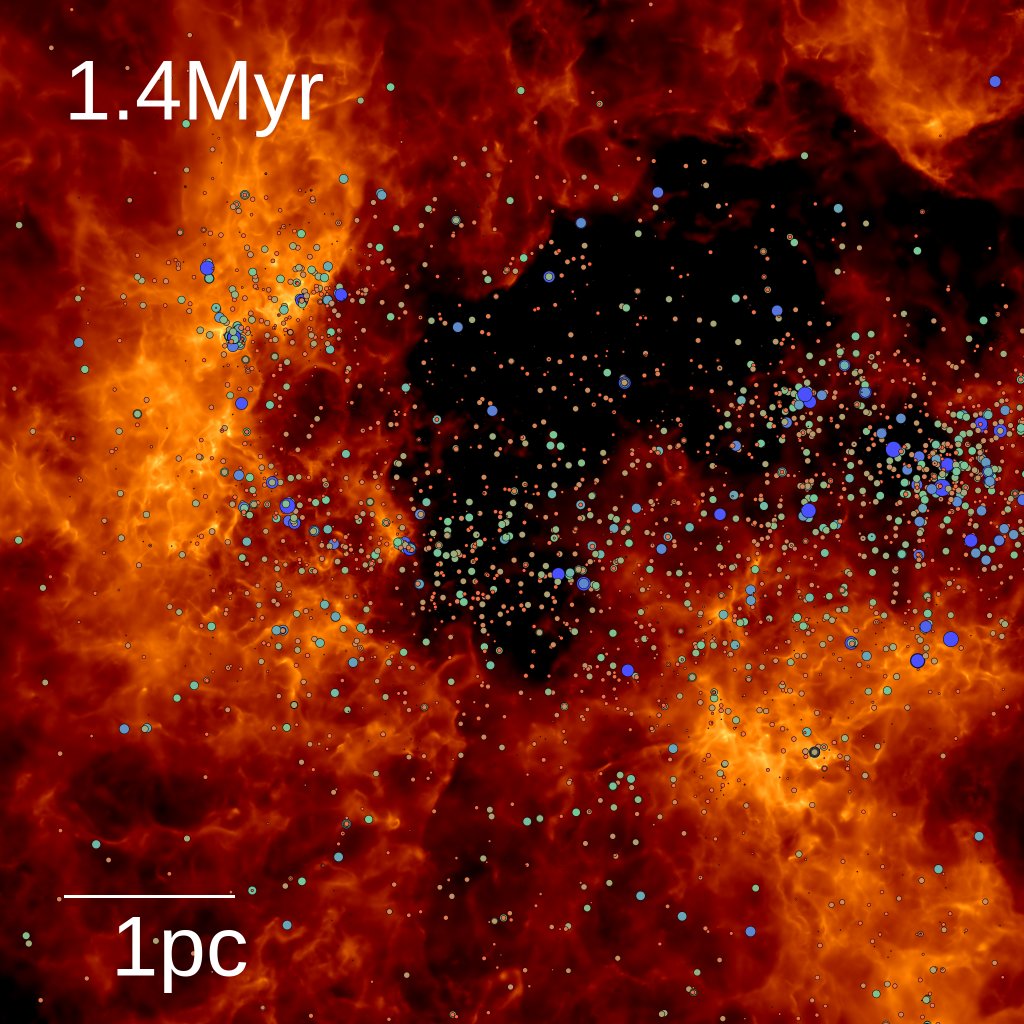} &
\includegraphics[width=0.48\linewidth]{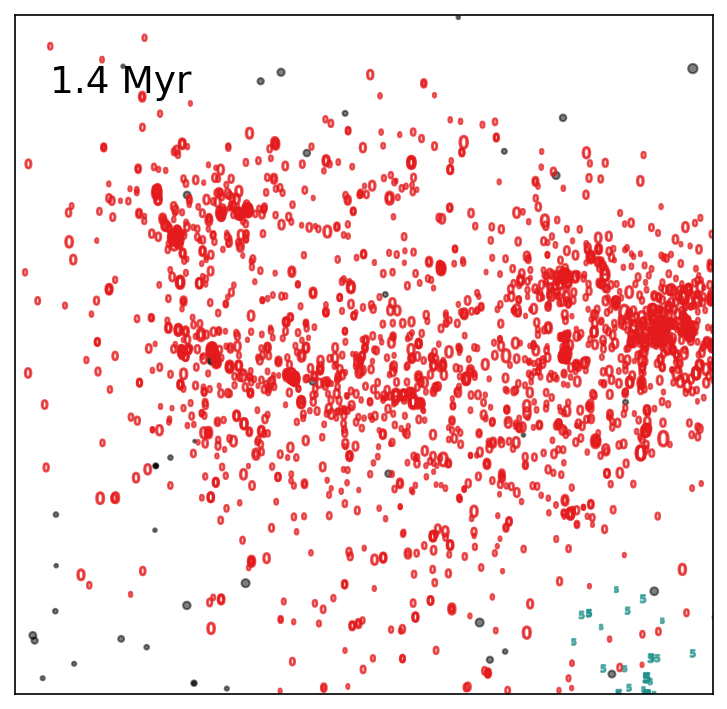} \\
\end{tabular}
%\vspace{-0.4cm}
\caption{Same as Figure \ref{fig:cluster_assignment_compare_box} but for a run with 10 times higher initial surface density (\textbf{M2e4\_R3}).}
\label{fig:cluster_assignment_compare_sigma}
\vspace{-0.5cm}
\end {center}
\end{figure}

\begin{figure*}
\begin {center}
\includegraphics[width=0.33\linewidth]{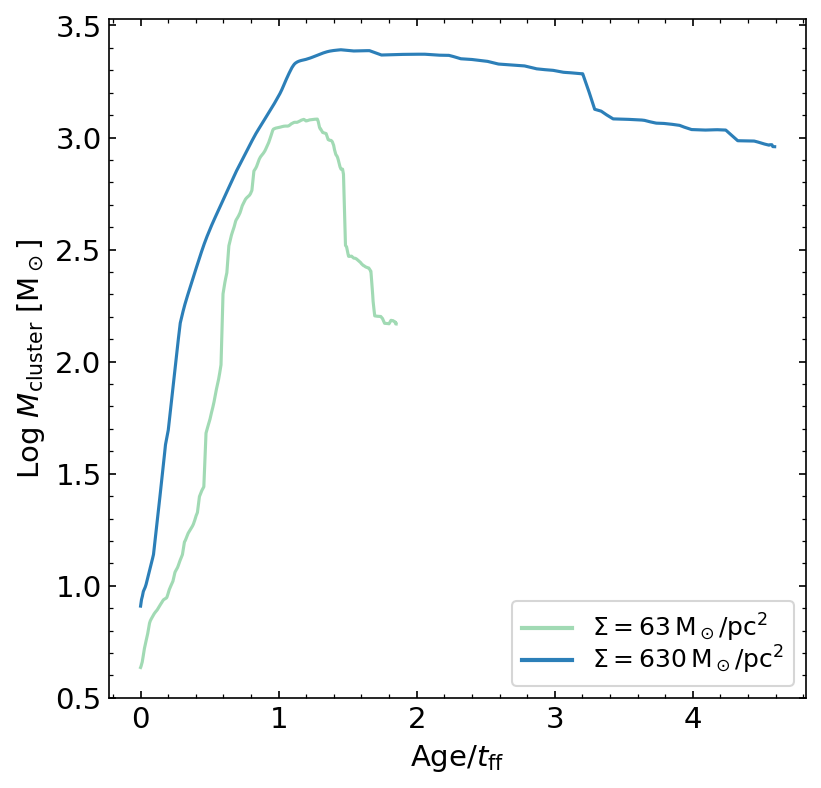}
\includegraphics[width=0.33\linewidth]{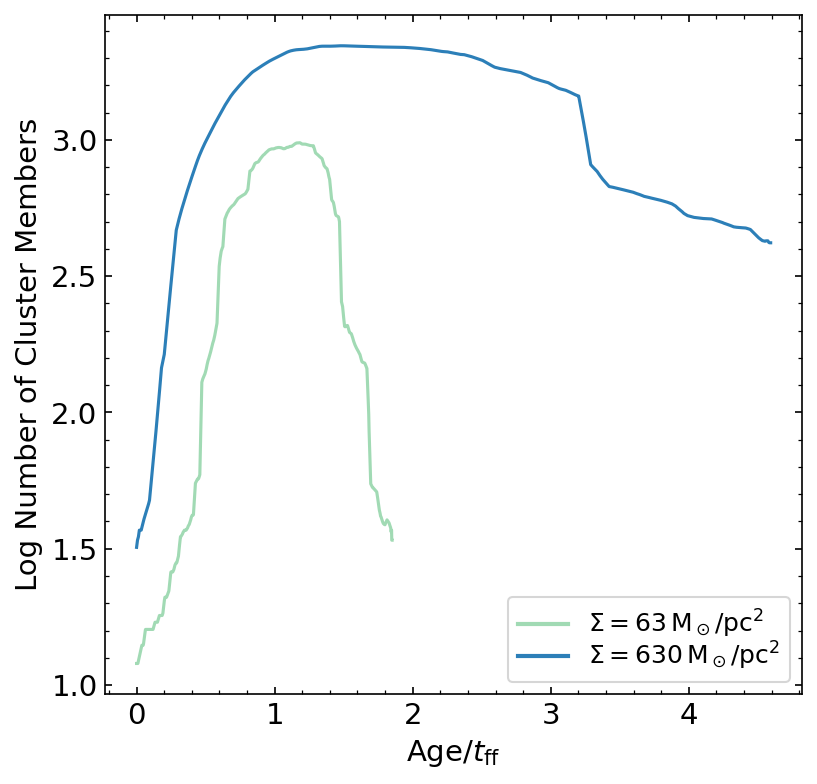}
\includegraphics[width=0.33\linewidth]{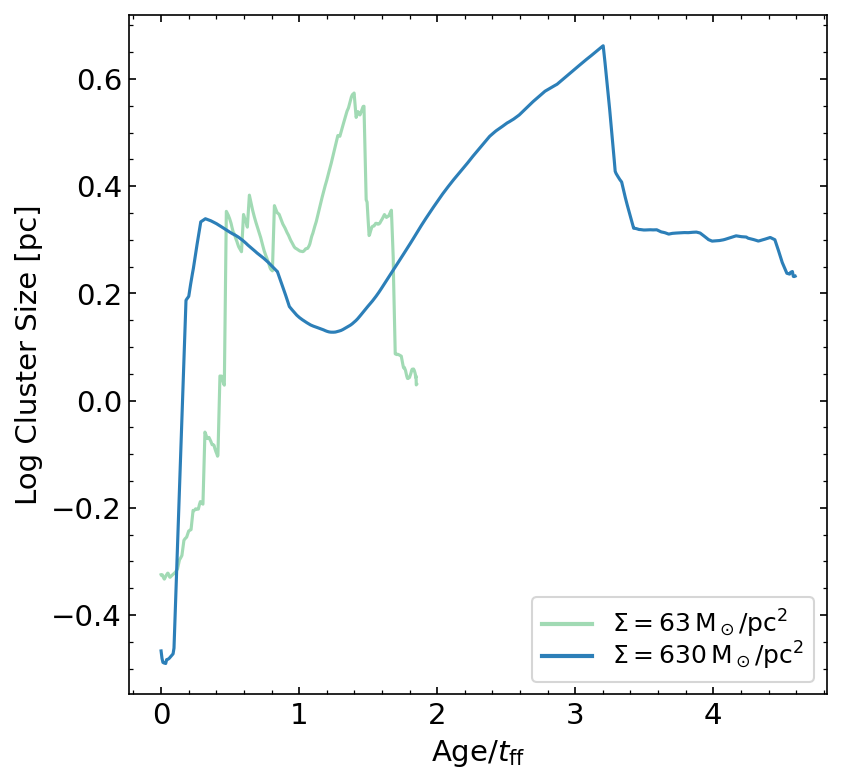}\\
\includegraphics[width=0.33\linewidth]{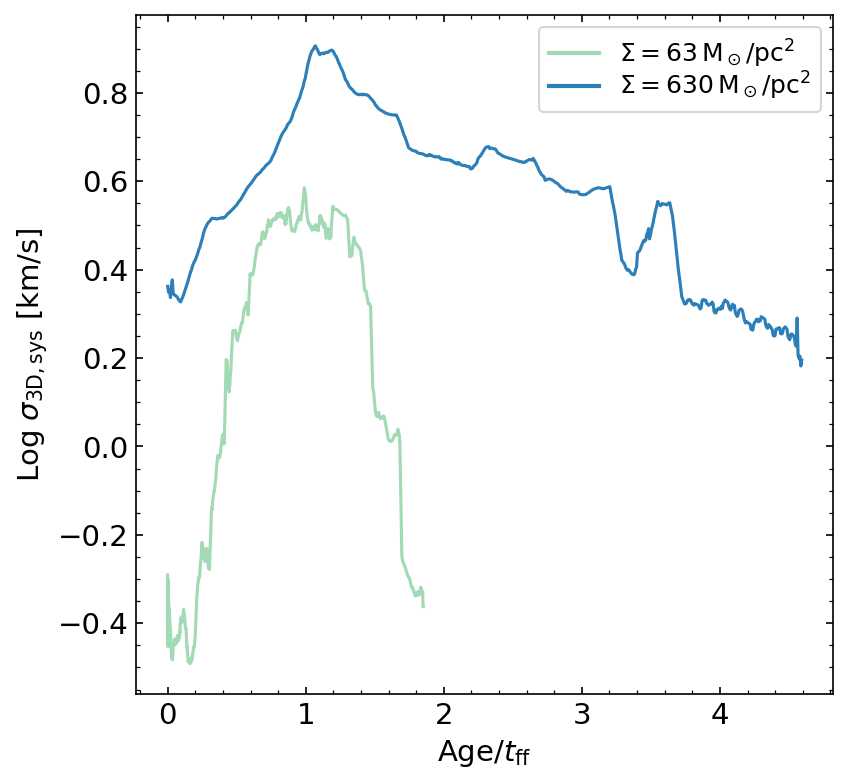}
\includegraphics[width=0.33\linewidth]{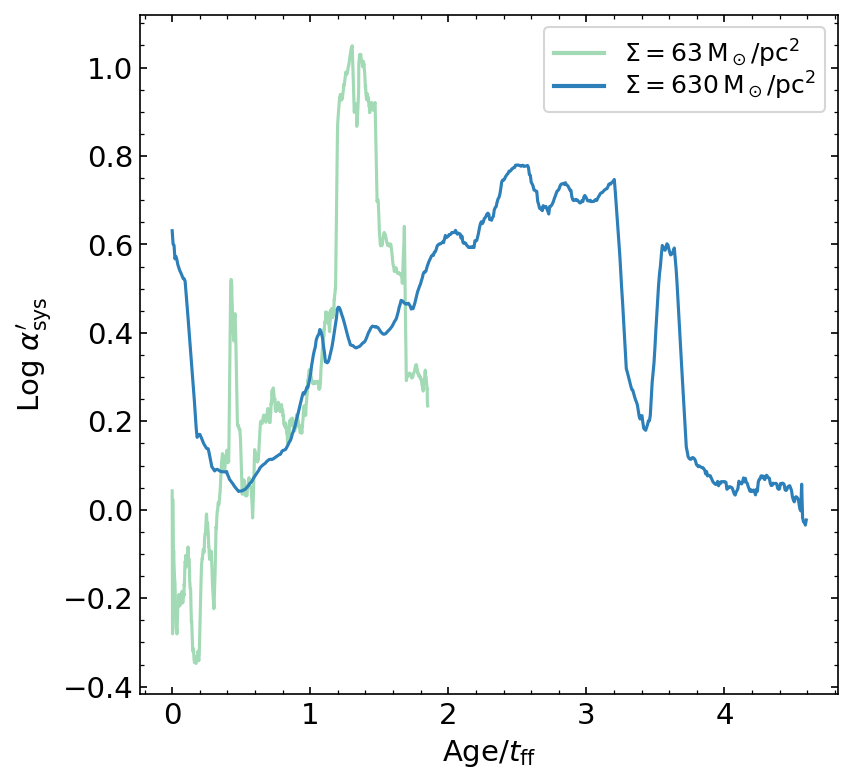}
\includegraphics[width=0.33\linewidth]{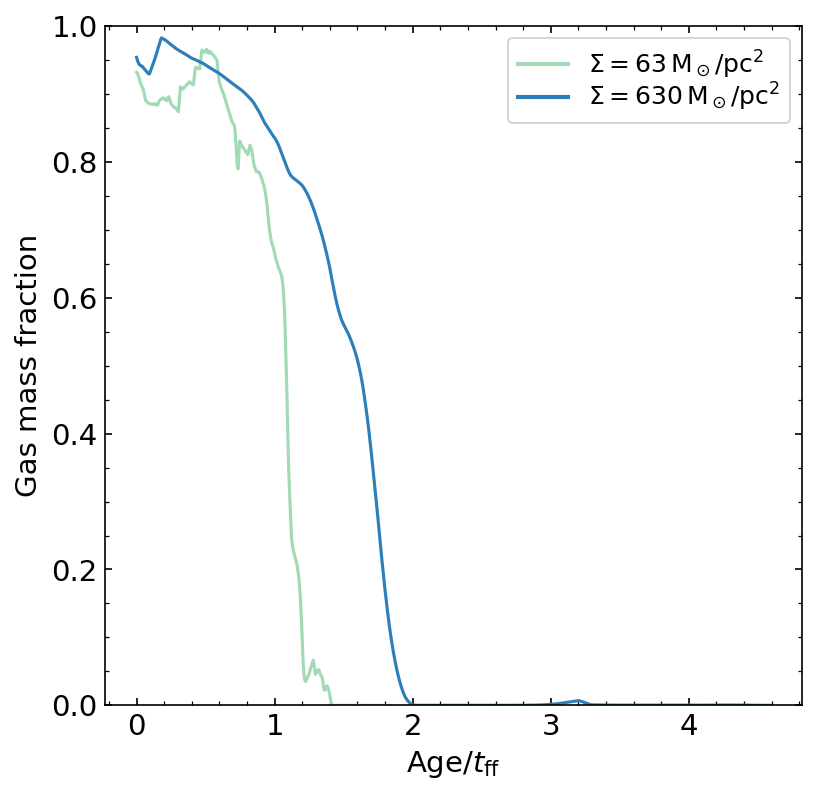}\\
\includegraphics[width=0.33\linewidth]{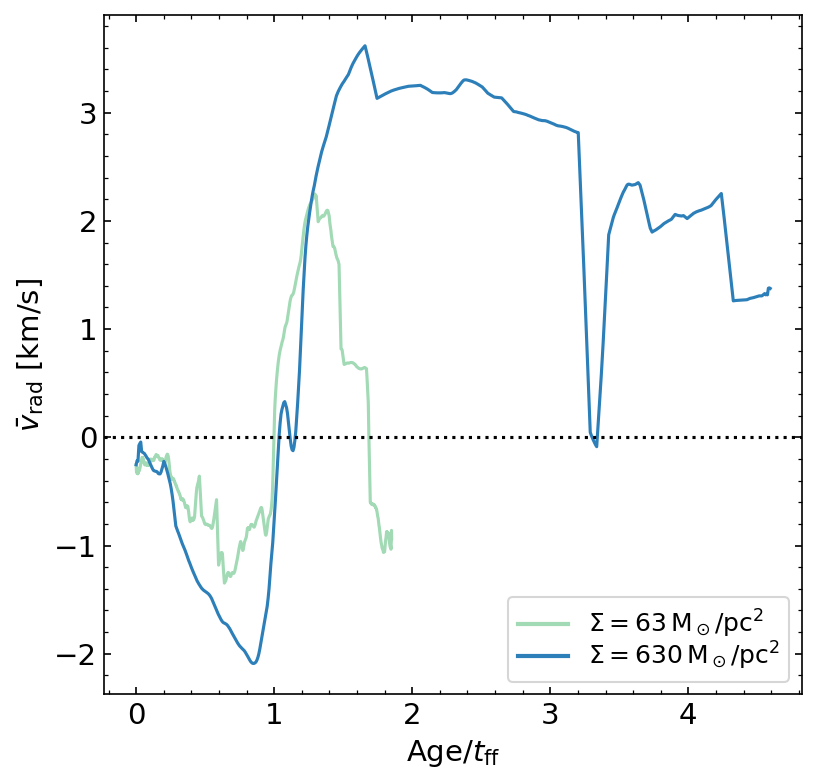}
\includegraphics[width=0.33\linewidth]{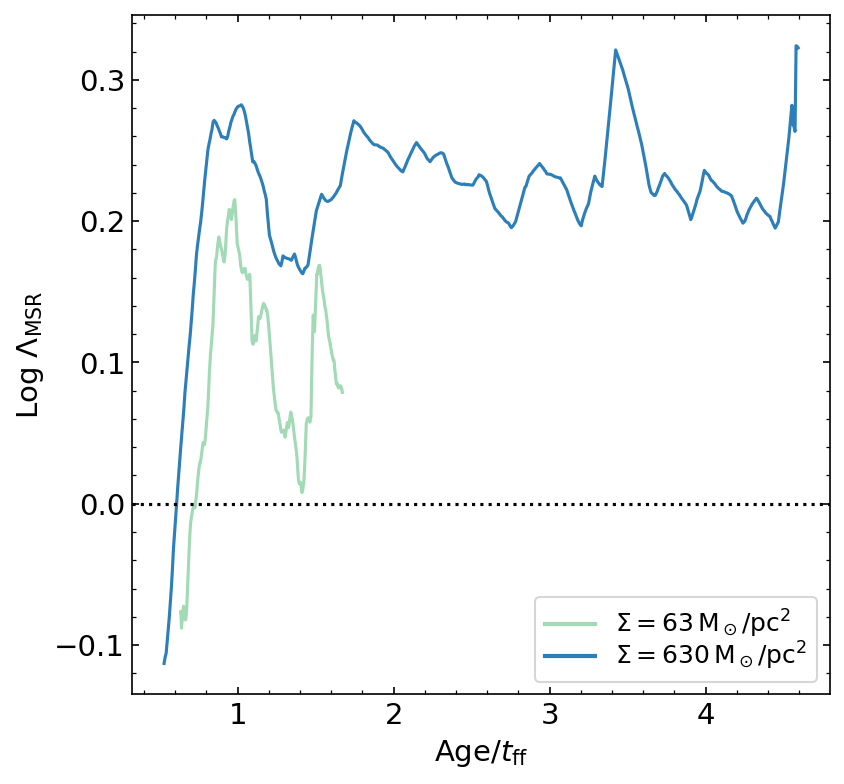}
\includegraphics[width=0.33\linewidth]{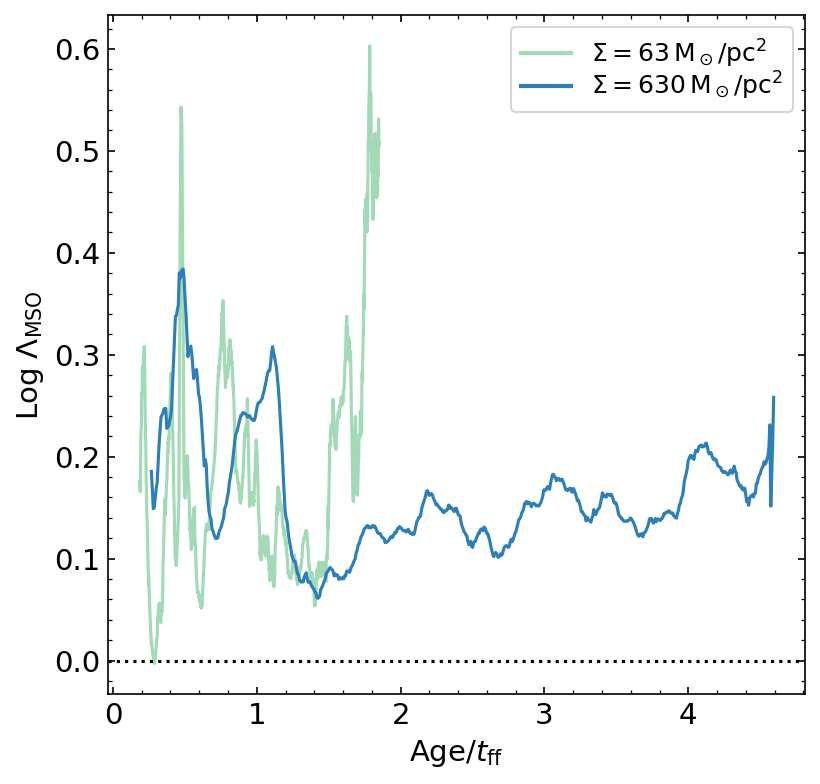}
\vspace{-0.4cm}
\caption{Evolution of the most dominant cluster in a set of runs with different initial surface densities (\textbf{M2e4} and \textbf{M2e4\_R3}, see Table \ref{tab:IC_phys}), similar to Figure \ref{fig:compare_clusters_box}. To make the runs easier to compare, we normalized the time evolutions with the freefall times of the respective initial clouds (3.7 and 0.6 Myr for\textbf{M2e4} and \textbf{M2e4\_R3} respectively). The \emph{top  row} shows the total stellar mass in the cluster (left), number of cluster members (middle) and the cluster size scales (top right, Eq. \ref{eq:cluster_R}) respectively. 
The \emph{middle  row} shows the cluster velocity dispersion (left, Eq. \ref{eq:cluster_sigma}), virialization state (middle, Eq. \ref{eq:alpha_sys}) and the gas mass fraction within the cluster radius (like Figure \ref{fig:cluster_basic_evol_sphere} panel e). The \emph{bottom row} shows the $\bar{v}_\mathrm{rad}$ mass-weighted mean radial velocity for the cluster (left, Eq. \ref{eq:radial_v}), as well as the mass segregation ratio (right, Eq. \ref{eq:lambda_msr}) and mass segregation offset (right, Eq. \ref{eq:mso}). For an analysis of the main trends see \S\ref{sec:variations_results_sigma} in the main text.}
\label{fig:compare_clusters_sigma}
\vspace{-0.5cm}
\end {center}
\end{figure*}

As expected, increasing the surface density leads to enhanced star formation and a higher final SFE (Table \ref{tab:final_properties}). Higher surface density also means that the cloud is smaller, making it easier for the clustering algorithm (see \S\ref{sec:cluster_identification}) to join star formation sites. Consequently, nearly all stars end up in one massive cluster (Figure \ref{fig:cluster_assignment_compare_sigma}). Similar to the fiducial run, the dominant cluster is gas-dominated and becomes unbound once stellar feedback expels the gas. The characteristic timescale of cloud evolution is the freefall time, which, due to the higher overall density, is significantly shorter than than that of our fiducial run (Figure \ref{fig:compare_clusters_sigma}). Note that the cluster assembly phase is mainly determined by this timescale, while the length of the following gas expulsion phase depends on both the freefall time and the timescales for stellar evolution. Apart from this non-trivial rescaling, the time evolution of the dominant cluster is similar in the fiducial and the high surface density runs.

% %%%%%%%%%%%%%%%%%%%%%%%%%%%%%%%%%%%%%%%%%%%%%%%%%% 
\section{Discussion}\label{sec:discussion}
 %\mike{At some point in this section should compare and contrast with the results of \cite{McMillan_2008_merged_mass_segregetaion} who found that an initially-mass-segregated cluster population will preserve mass segregation throughout the assembly process}\DG{I couldn't find where that paper says that}
 
%\mike{should also make sure to remind the reader that we are talking about mainly unbound systems that mostly lack any phase of gas-free dynamical evolution - this is {\bf not} the context in which mass seg has been traditionally studied in theory}
%\alr{I think this part should have a subsection heading like ``Cluster Assembly and Ejection of Cluster Members/Stars or Dynamical Effects" or something similar, looks odd to have all this text alone and then a subsection for caveats.}

\subsection{Cluster assembly and dynamical effects}

Clusters in our simulations form through the mergers of accreting subclusters, which in turn merge to form ever larger structures, corresponding to a hierarchical assembly of clusters. This is similar to the behavior found by recent simulations of larger ISM regions with lower mass resolution and a different subsets of the physics included in this work \citep{Rieder_2022_galaxy_cluster_sim,Dobbs_2022_massive_cluster_formation_sim}. By the time stellar feedback becomes important in the simulation most stars are concentrated in one or a few clusters. Stellar feedback eventually expels the remaining gas and the clusters become unbound, leading to their expansion and breakup \citep{Tutukov_1978_cluster_sim,Hills_1980_cluster_analytic,Mathieu_1983_cluster_SFE}.

This qualitative picture appears to be robust to changes in initial conditions, such as turbulence or surface density, even though these simulations have significantly different star formation histories. We find that using periodic boundary conditions (Box runs) significantly hinders hierarchical cluster assembly, as it significantly decreases the overall gravitational potential \citep{federrath_sim_2012}, weakening gravitational focusing. In the Box run with external driving, we find that continued driving effectively prevents hierarchical merging. In this case a significant fraction of newly formed clusters are transient and dissolve after $<$2~Myr (Figure \ref{fig:box_merger_history_compare}). We find that turbulent driving significantly increases the number of stars outside clusters (Table \ref{tab:final_properties}), making it one of the few parameters that can significantly affect clustering. %\alr{(Is this because you are sustaining turbulence to allow for more isolated SF throughout the cloud, whereas when driving is turned off you can collapse regions of the box to form more subclusters?) Maybe one thing to note with this IC is that the Box run is essentially modeling a patch of a cloud rather than an isolated GMC.} \DG{Sustaining turbulence in a Box run means that it is effectively equivalent to an alpha>>10 sphere run, so there is no glpbal collapse, so things have a harder time clustering/clumping as turbulence can shear structures apart.}

Recent observations found the high mass slope of the initial mass function in young clusters is steeper than the Milky Way average \citep{weisz_2015_survey}. One possible explanation for the deficit of high mass stars is dynamical interactions that preferentially eject massive members (see \citealt{Oh_Kroupa_2016A_massive_star_ejection}). This is supported by recent observations finding a significant number of \myquote{runaway} O and B stars ejected from young star clusters \citep{Lennon_2018_R136_massive_eject,Zeidler_2021_Westerlund_2_cluster}. In our simulations we find no evidence that high-mass stars are preferentially ejected. Stars within clusters, outside clusters and ejected from clusters all appear to be drawn from the same underlying mass distribution. It should, however, be noted that the clusters in our simulations are actively accreting and gas-dominated, while previous work in the literature involved N-body simulations of gas-free, bound clusters. Since our simulations run only until gas expulsion occurs, our results neglect longer scale dynamical interactions, which could occur in the gas-free star clusters after star formation ceases.

\subsection{Mass segregation}

One key question of cluster formation is whether clusters form mass segregated or become so through dynamical interactions. In our simulations star formation sites often host a single or several massive stars ($M_{\rm \star}>5\,M_{\odot}$) at their center, making them mass segregated (note that $\MSR$ is not defined for clusters with <5 massive stars). These small clusters merge to form larger structures, which in turn inherit the centrally condensed, mass-segregated substructures. Due to dynamical interactions these substructures strip stars from each other while merging. Over time dynamical processes cause the dense centers, which also host the massive stars, to sink to the center of the larger merged structure. While these processes are taking place the substructures continue to grow, forming new stars as well as continuing to accrete gas. Thus we find that whether a cluster is considered mass-segregated depends greatly on the definitions of ``cluster" and ``mass segregation," neither of which have one accepted definition in the literature (see \citealt{krumholz_2019_cluster_review} for discussion). If one defines mass segregation as any stellar configuration where \myquote{massive stars are distributed differently} than lower mass stars (as in \citealt{krumholz_2019_cluster_review}), then star clusters start out mass segregated regardless of how cluster membership is assigned, since clusters contain substructures that host a single or a few massive stars at their respective centers. However, if mass-segregation on the cluster scale occurs when \myquote{massive stars are preferentially at the center} of the cluster, as in \citet{de_Grijs_2002_LMC_mass_segregation} and \citet{krause2020}, then the cluster definition determines whether the stars are mass segregated. Choosing a method that picks out structures containing several star formation sites (like the one we used) leads to no initial mass segregation (see the scenario in Figure \ref{fig:MSO_subcluster} and the evolution in Figure \ref{fig:cluster_basic_evol_sphere}). If, however, a cluster definition picks out individual star formation sites (e.g., by defining a smaller characteristic length or by requiring that a cluster be centrally condensed), then mass segregation will appear primordial regardless of metric (see Table \ref{tab:mass_segregation} for a summary). 

Most observers define clusters as pc-sized objects with many stars \citep{Kirk_Myers_2011_mass_segregaton} and use mass segregation metrics that are insensitive to the mass segregation of any substructures within the cluster (e.g., \citealt{de_Grijs_2002_LMC_mass_segregation}). In this framework, clusters in the simulation are initially not mass segregated and become so through dynamical interactions. The process takes several Myr, which is enough time for feedback from massive stars to expel gas but not enough time for the cluster to reach a fully relaxed state. %\SO{I'm not sure I agree with this sentence -- E.g., in figure 5, the MSR metric is not mass segregated only for $<$ 0.5 Myr -- not several Myr. Before that it is undefined.}\DG{This statement is on whether the cluster reached a relaxed state (i.e. MSR would flatten). MSR>1 for a lot of time, but that does not mean that the massive stars hae all sunk to the center}

\begin{table}
    \setlength\tabcolsep{2.0pt} %compress table
	\centering
	\begin{tabular}{ | c | c | c |  }
	\hline
    \multirow{2}{*}{\textbf{Cluster definition}} & \multicolumn{2}{c|}{\bf Is there primordial mass segregation?}\\
	\cline{2-3}
	 & \myquote{Massive stars at the center} & \myquote{Distributed differently}  \\
	\hline
	No substructure & Y & Y  \\
	\hline
	With substructure & N & Y  \\
	\hline
    \end{tabular}
        \vspace{-0.1cm}
 \caption{Summary of our results regarding the initial mass segregation of newly formed clusters. We find that the answer depends on the definition of a cluster, i.e., whether the selected objects have substructure, and whether we define mass segregation as \myquote{massive stars at the center} (like \citealt{Gouliermis_2009_cluster_segregation}) or as \myquote{massive stars are distributed differently} (like \citealt{krumholz_2019_cluster_review}).}
 \label{tab:mass_segregation}\vspace{-0.5cm}
\end{table}

\subsection{Caveats}

While the simulations presented here represent the current state-of-the-art for simulating star-forming clouds, STARFORGE employs a large number of approximations and assumptions to make the simulations computationally tractable like other simulations in the literature (see \citetalias{grudic_starforge_methods} for detailed discussions). In particular, there are significant caveats when applying STARFORGE to model star cluster formation. First, the runs have a $\sim 30\,\mathrm{AU}$ Jeans-resolution, i.e., fragmentation on scales smaller than this are not resolved.  This resolution in the ideal-MHD limit effectively suppresses the formation of protostellar disks. Consequently, there is no disk fragmentation, causing the simulation to potentially miss closely formed binaries. The simulations presented here also use a gravitational softening length of $\sim 20\,\mathrm{AU}$, making gravitational interactions below this scale inaccurate and suppressing the formation of binaries with separations smaller than this value. As a result the long-term accuracy of N-body interactions is significantly lower than in pure N-body simulations. It should be noted, however, that we run our simulations until gas expulsion, corresponding to a relatively short time after star formation starts (< 5 Myr). Also, our clusters are gas-dominated for most of their lifetime and do not achieve high stellar densities, %SO I'm not sure what accretion has to do with the N-body accuracy DG: For N-body sims. they ignore the gas or have it as a simple potential, we do it self-consistently. Since the stars are actively accreting and are in a gas rich environment, their evolution is mainly governed by the gas potential, so there is a lower chance of close encounters
making close encounters rarer and lessening the effects of the gravitational softening on stellar interactions. Since we terminate the simulations soon after cloud disruption, we cannot predict the ultimate fate of the clusters, bound mass fraction and cluster mass function. We will investigate the long-term evolution and fate of the STARFORGE clusters in a follow-up paper.

% %%%%%%%%%%%%%%%%%%%%%%%%%%%%%%%%%%%%%%%%%%%%%%%%%% 
\section{Conclusions}\label{sec:conclusion}

In this work we analyze the star cluster assembly process in the STARFORGE radiation-magnetohydrodynamic simulations. These simulations follow the evolution of a mid-sized molecular cloud ($M=20000\,\msun$, $\Sigma\sim 60\,\msun/\pc^2$) taking into account gravity, gas thermodynamics, turbulence, magnetic fields, and radiation as well as stellar feedback processes (jets, radiation, winds, SNe). 

%SO Got to here
%\alr{Maybe start with that SF is distributed throughout the cloud at locations of fliaments and where filaments met to form condensations of stars, then go into how these star formation sites merge to form larger clusters via hierarchical assemply as the cGMC collapses.}
%We find that star formation is distributed throughout the cloud at locations of filaments and where filaments intersect. \SO{This is only the second time you mention filaments in the paper; the first time was inside a (). I think you should either remove this here or make a bigger note of it earlier. I think it makes more sense to focus more on filaments in the IMF paper and/or Mike's paper.}
Star clusters assemble through a series of mergers, whereby accreting star-formation sites come together to form larger structures. This hierarchical assembly continues until most stars end up in one or a few large, gas-dominated clusters. 
Once stellar feedback expels the gas they become unbound and the stars disperse.
During the assembly process clusters eject a small fraction of their members ($<$10\%).  We find no significant difference between the mass distribution of the ejected stellar population and that of the overall stellar mass spectrum of the simulation. 

%SO This is background that I don't think is necessary here, but could go in the introduction:
% \cut{Observed clouds show significant variations in their surface density and levels of turbulence (i.e., virial parameter), so} \SO{I think you should mention these in the introduction explicitly, i.e., to motivate why these two parameters are explored.}
We investigate the effect of surface density and turbulence on the cluster formation process. We find that while the initial surface density and level of turbulence significantly affect the star formation history of the cloud, they do not qualitatively affect the cluster formation process. 

We also investigate the effects of different initial cloud geometries and turbulent driving. We find that turbulent driving and a periodic \myquote{Box} geometry significantly reduces clustering and suppresses cluster mergers. This is caused by weaker gravitational focusing, as periodic boundaries lead to a shallower gravitational potential, while maintaining turbulence reduces collapse. 

We consider two different definitions for mass segregation. In all simulations, small forming groups of stars are initially mass segregated with one or a few massive stars at their center. As these structures  
merge, they (at first) become mass-segregated substructures within the newly formed cluster. Thus massive stars are not initially in the center of merged clusters. Through dynamical interactions they relax to a centralized configuration, similar to that of observed clusters. 
%SO I moved this below. I think these two paragraphs should be together.
%but %the process 
%\SO{dynamical evolution} is still ongoing at the time of gas expulsion.
%SO This is context, which seems out of place in a conclusion
%\cut{Determining whether mass segregation is primordial is a key question of star cluster formation. Although above we described the cluster assembly process,} 
We find that whether clusters are quantitatively considered to be mass segregated depends greatly on how one defines a \emph{cluster} and \emph{mass segregation}. If clusters are defined as structures that include many star-formation sites distributed throughout the GMC \emph{and} mass segregation requires massive stars to be at the center (both of these are true for most definitions used in observations), then there is no primordial mass segregation. Rather, mass segregation results from dynamical interactions. On the other hand, massive stars are usually centrally located within bound sub-groups of stars, such that they are \emph{distributed differently} with respect to low-mass stars. Thus, a definition of mass segregation that does not require massive stars to be globally centralized, concludes that clusters start out mass segregated (see Table \ref{tab:mass_segregation} for a summary).

In the simulations, dynamical evolution is still ongoing at the time of gas expulsion. Future work will investigate the evolution of the stellar distribution over 100 Myr timescales and determine the survival rate of the star clusters we identify here. 

% %%%%%%%%%%%%%%%%%%%%%%%%%%%%%%%%%%%%%%%%%%%%%%%%%% 
 
% %%%%%%%%%%%%%%%%%%%% Acknowledgements and Data Availability %%%%%%%%%%%%%%%%%%

\section{Data availability}
The data supporting the plots within this article are available on reasonable request to the corresponding authors. A public version of the {\small GIZMO} code is available at \url{http://www.tapir.caltech.edu/~phopkins/Site/GIZMO.html}.

\section*{Acknowledgements}

We would like to thank Sinan Deger for his thoughtful comments.

DG is supported by the Harlan J. Smith McDonald Observatory Postdoctoral Fellowship and the Cottrell Fellowships Award (\#27982) from the Research Corporation for Science Advancement. 
Support for MYG was provided by NASA through the NASA Hubble Fellowship grant \#HST-HF2-51479 awarded  by  the  Space  Telescope  Science  Institute,  which  is  operated  by  the   Association  of  Universities  for  Research  in  Astronomy,  Inc.,  for  NASA,  under  contract NAS5-26555
Support for PFH was provided by NSF Collaborative Research Grants 1715847 \&\ 1911233, NSF CAREER grant 1455342, and NASA grants 80NSSC18K0562 \&\ JPL 1589742.
SSRO and CM are supported by NSF Career Award AST-1748571 and by a Cottrell Scholar Award from the Research Corporation for Science Advancement. 
CAFG was supported by NSF through grants AST-1715216, AST-2108230,  and CAREER award AST-1652522; by NASA through grant 17-ATP17-0067; by STScI through grant HST-AR-16124.001-A; and by the Research Corporation for Science Advancement through a Cottrell Scholar Award.
ALR  acknowledges support from Harvard University through the ITC Post-doctoral Fellowship.
This work used computational resources provided by XSEDE allocation AST-190018, the Frontera allocation AST-20019, and additional resources provided by the University of Texas at Austin and the Texas Advanced Computing Center (TACC; http://www.tacc.utexas.edu).

% %%%%%%%%%%%%%%%%%%%%%%%%%%%%%%%%%%%%%%%%%%%%%%%%%% 
 
% %%%%%%%%%%%%%%%%%%%% REFERENCES %%%%%%%%%%%%%%%%%%

% % The best way to enter references is to use BibTeX:

 \bibliographystyle{mnras}
 \bibliography{bibliography} % if your bibtex file is called example.bib

% %%%%%%%%%%%%%%%%%%%%%%%%%%%%%%%%%%%%%%%%%%%%%%%%%%

% %%%%%%%%%%%%%%%%% APPENDICES %%%%%%%%%%%%%%%%%%%%%

\appendix

%\section{Dependence of the IMF on initial conditions}\label{sec:scaling_dependencies}
%

%%%%%%%%%%%%%%%%%%%%%%%%%%%%%%%%%%%%%%%%%%%%%%%%%%

% Don't change these lines
\bsp	% typesetting comment
\label{lastpage}
\end{document}